\documentclass[preprint, nofootinbib,superscriptaddress]{revtex4}
\usepackage{amsmath,amssymb}
\usepackage{amsthm}
\usepackage{ascmac}
\usepackage{float}
\usepackage{comment}
\RequirePackage{ifpdf}
\ifpdf
  \usepackage[pdftex]{graphicx}
\else
  \usepackage[dvipdfmx]{graphicx}
\fi
\usepackage[all]{xy}

\allowdisplaybreaks

\numberwithin{equation}{section}

\begin{document}

\title{Entanglement Spreading and Oscillation}
~\\
\preprint{EFI-17-29}
\\
\preprint{OU-HET-955}
\author{Mitsuhiro Nishida}
\email[]{mnishida@gist.ac.kr }
\affiliation{\small School of Physics and Chemistry, Gwangju Institute of Science and Technology, Gwangju 61005,
Korea}
\author{Masahiro Nozaki}
\email[]{masahiron@uchicago.edu }
\affiliation{\small Kadanoff Center for Theoretical Physics, University
  of Chicago, Chicago, IL 60637, USA}
\author{Yuji Sugimoto}
\email[]{sugimoto@het.phys.sci.osaka-u.ac.jp}
\affiliation{\small Department of Physics, Graduate School of Science, Osaka University, Toyonaka, Osaka 560-0043, Japan}
\author{Akio Tomiya}
\email[]{akio.tomiya@mail.ccnu.edu.cn}
\affiliation{\small Key Laboratory of Quark \& Lepton Physics (MOE) and Institute of Particle Physics, Central China Normal University, Wuhan 430079, China
\vspace{0.5cm}}

\begin{abstract}
We study dynamics of quantum entanglement in smooth global quenches with a finite rate, by computing the time evolution of entanglement entropy in $1+1$ dimensional free scalar theory with  time-dependent masses  which start from a nonzero value at early time and either crosses or approaches zero. The time-dependence is chosen so that the quantum dynamics is exactly solvable.
If the quenches asymptotically approach a critical point at late time, the early-time and late-time entropies are proportional to the time and subsystem size respectively. Their proportionality coefficients are determined by scales: in a fast limit, an initial correlation length; in a slow limit, an effective scale defined when adiabaticity breaks down.
If the quenches cross a critical point, the time evolution of entropy is characterized by the  scales: the initial correlation length in the fast limit and the effective correlation length in the slow limit. 
The entropy oscillates, and the entanglement oscillation comes from a coherence between right-moving and left-moving waves if we measure the entropy after  time characterized by the quench rate.
The periodicity of the late-time oscillation is consistent with the periodicity of the oscillation of  zero modes which are zero-momentum spectra of two point functions of a fundamental field and its conjugate momentum.
\end{abstract}


\maketitle


\section{Introduction and Summary}
\subsection*{Introduction}
The behavior of quantum entanglement in dynamical systems has been a subject of great interest because the entanglement plays an important role in holography \cite{rt1,rt2,VR1,VR2,BS1,t1,t2} and thermalization \cite{ca1, L1, L2, L3, MM1, MM2,HM1,ca3}.

Two kinds of protocol have been often used in order to study the dynamics of quantum entanglement. One of them is a local quench, and another is a global quench. 
In the local quenches, a state is excited by adding a local interaction to a Hamiltonian  suddenly \cite{ca2, t3, MR1, UT1, yy, edm1}, or acting by local operators \cite{m1, t4, m2, m3}. 
Entanglement entropy characterizes the resulting quantum state.
Thus, the entropy in the local quenches might shed light on the properties of holographic theories which has a gravity dual. 

In the global quenches, parameters of a Hamiltonian depend only on time, $t$. 
 The dynamics of entanglement has been well-studied by studying entanglement entropy in the global quenches where the parameters of Hamiltonian change suddenly \cite{ca1,L1,L2,MM1,HM1}.
Entanglement entropy for an excited state in the sudden global quench increases linearly with respect to $t$ in a window, $\xi_c \ll t < \frac{l}{2}$, where $\xi_c$ is a characteristic scale and $l$ is a subsystem size. After $t \sim \frac{l}{2}$, entanglement entropy is proportional to the subsystem size, $l$, so called $volume~ law$. Thus, entanglement entropy in late time becomes a ``thermal" entropy.
The time evolution of entropy for an interval can be interpreted in terms of relativistic propagation of quasi-particles. The quasi-particle interpretation can be applied to the time evolution of other quantum measures such as logarithmic negativity and mutual information \cite{ln1,mi1}. However, the authors in \cite{sc1, sc2} has pointed out that the time evolution of entropy for two or more disjoint intervals in holographic conformal field theories can not be interpreted in terms of the relativistic propagation of quasi-particles.  

It is well known that the late-time entanglement entropy in the sudden global quenches has a thermal entropy-like property.
However, very little is known about how the dynamics of entanglement entropy depends on the quench rate for a smooth quench. 
In fact for smooth quenches \cite{d1,d2,d3,d4,d5,sm1,sm2,sm3,sm4,sm5,smq1,smq2,smq3,smq4,smq5,smq6,smq7,smq8,smq9}, local quantities such as two point functions are well studied, but non-local quantities such as entanglement entropy have not been studied in much detail. The scaling behavior of the entanglement entropy in early time has been studied recently in \cite{d2}. In this paper, we concentrate on its time evolution. Authors in \cite{mspee} have studied momentum-space entanglement entropy in the similar situation to ours.
\begin{figure}[htbp]
 \begin{minipage}{0.45\hsize}
  \begin{center}
   \includegraphics[width=70mm]{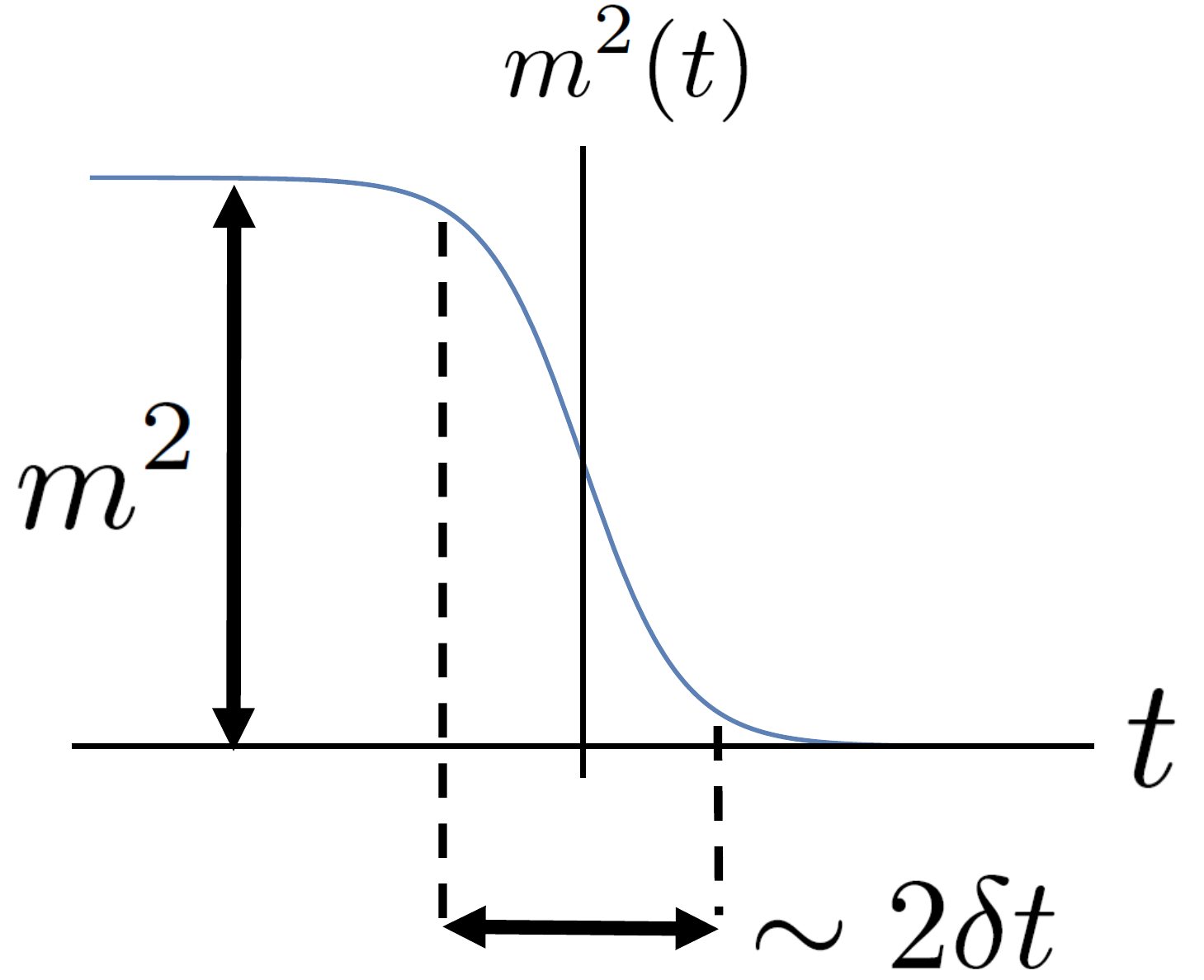}
  \end{center}
 \end{minipage}
 \begin{minipage}{0.45\hsize}
  \begin{center}
   \includegraphics[width=70mm]{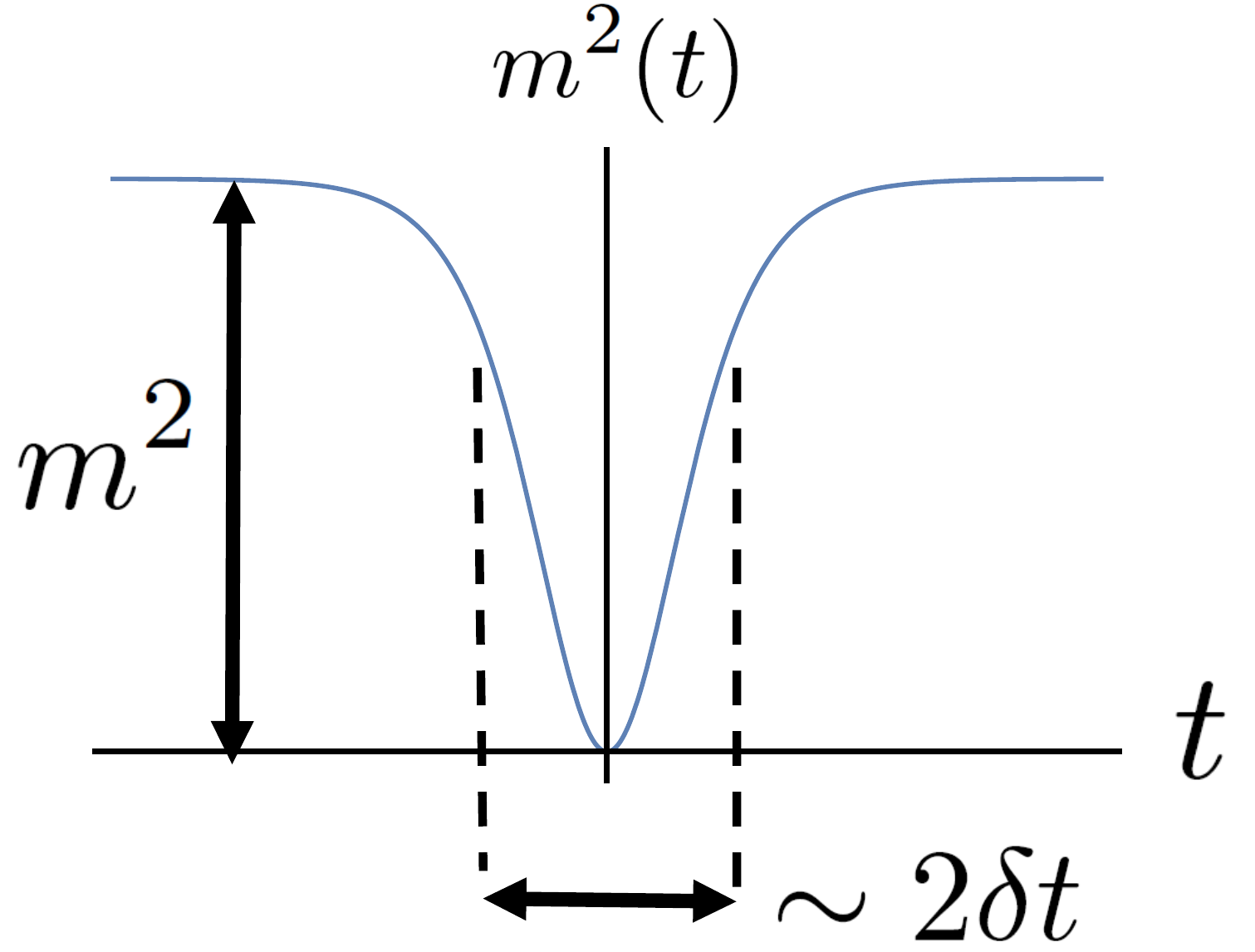}
  \end{center}
 \end{minipage}
\caption{A schematic picture of potentials: the left panel is for ECP-type mass, and the right panel is for CCP-type mass.\label{potential}}
\end{figure}
\subsection*{Summary}
In this paper, we study the dynamics of quantum entanglement in global quenches with a finite quench rate, $\delta t^{-1}$ in a $1+1$ dimensional free scalar field theory living on an infinite spatial lattice with a time-dependent masses, which are $m=\frac{1}{\xi}$ initially.
We study the  change of entanglement entropy in a quenched state at time $t$ from that in an initial state at $t=-\infty$, $\Delta S_A$\footnote{In this paper, we assume that the area law term in a cutoff expansion is independent of time. The authors in \cite{tarea} show that the terms in $3$ or more dimensions can be dependent of time. }.
These masses (which we will also call potentials) are called end-critical-point (ECP) type and cross-critical-point (CCP) type potentials.
 In the ECP-type potential we consider (a left panel in Figure \ref{potential}), the Hamiltonian at early time is a massive one.
The mass parameter gradually decreases and asymptotically vanishes at late time.
In the CCP-type potential considered (the right panel in Figure \ref{potential}), the mass is finite (and equal) in the early and late times and vanishes smoothly at an intermediate time which we choose to be $t=0$. 

We study $\Delta S_A$ in these protocols in two extreme limits: fast and slow limits.
In the slow limit, for most of the time the process of system is adiabatic. However, near the critical point adiabaticity is broken.
In the fast limit, the late-time behavior of the system is expected to be qualitatively similar to a sudden quench protocol. However, at early time, the physics is quite different, and interesting universal scaling laws emerge \cite{d1,d2,d3,d4,d5,sm1,sm2,sm3,sm4,sm5,smq1,smq2,smq3,smq4,smq5,smq6,smq7,smq8,smq9}.

The outline of the time evolution of $\Delta S_A$ is in Table \ref{t1}.
 \begin{table}[htb]
  \begin{center}
    \caption{Brief summary of the time evolution of $\Delta S_A$\label{t1}.}
    \begin{tabular}{|c|c|c|} \hline
     Protocol & Early time & Late time     \\ \hline \hline
    ECP in the fast limit &$\Delta S_A\propto t$& $\Delta S_A \propto l$; its coefficient is set by $\xi$.   \\ 
   ECP in the slow limit&$\Delta S_A\propto t$& $\Delta S_A \propto l$; its coefficient is set by $E_{kz}^{-1}$.  \\ 
     \hline
 CCP in the fast limit  &After $t=2\xi$, $\Delta S_A$ oscillates.& The periodicity of oscillation is determined by \\ 
 CCP in the slow limit & After $t=2\xi_{kz}$, $\Delta S_A$ oscillates.& the periodicity of coherent oscillation of zero modes.   \\ 
     \hline
            \end{tabular}
  \end{center}
  \end{table}

\subsection*{Time evolution in ECP-type potential}
$\Delta S_A$ in the fast limit, $\omega = \frac{\delta t}{\xi} \ll 1$, in the ECP-type potential is proportional to $t$ when $\xi \ll t < \frac{l}{2}$ ($early~ time$). 
Here, $l$ is the subsystem size and $\xi $ is an initial correlation length.
$\Delta S_A$ is proportional to $l$ (subsystems are thermalized) when $\frac{l}{2} \ll t$ ($late~ time$). 
The coefficients of $t$ and $l$ are set by $\xi$.

$\Delta S_A$ in the slow limit, $\omega \gg 1$, is proportional to $t$ when $t_{kz} \ll t < \frac{l}{2}+t_{kz}$ ($early~ time$).
Here, $t_{kz}$ is the time adiabaticity breaks down. 
 $\Delta S_A$ is proportional to $l$ when $\frac{l}{2}+t_{kz} \ll t$ ($late~ time$). 
The coefficients of $t$ and $l$ are set by the scale at $t=t_{kz}$, $E_{kz}$.  $E_{kz}$ is an energy scale at $t =t_{kz}$. 
The time evolution of $\Delta S_A$ in both limits is interpreted in terms of the propagation of entangled particles. 

\subsection*{Time evolution in CCP-type potential}

The time evolution of $\Delta S_A$ in the fast and slow limits in the CCP-type mass is characterized by the scale of the initial correlation length, $\xi$, in the fast limit and of the effective correlation length, $\xi_{kz}$, in the slow limit. $\Delta S_A$ increases monotonically before $t\simeq \xi $ or $\xi_{kz}$, and decreases monotonically from $t\simeq \xi$ or $\xi_{kz}$ to $t\simeq 2 \xi$ or $2\xi_{kz}$. After that, $\Delta S_A$ for the large subsystem increases independently of size with an oscillation, and the time evolution of $\Delta S_A$ in the fast limit depends on the subsystem after $t\simeq\frac{l}{2}$. On the other hand, $\Delta S_A$ in the slow limit depends on $l$ in very late time.

The $l$-dependence of $\Delta S_A (t=2\xi)$ in the fast limit shows it approaches a negative constant if $l >4\xi$. The negative constant shows we can define an effective correlation length, which is smaller than the initial one. We expect the effective correlation length to be related to the distance between the entangled particles created around $t=0$. On the other hand, the $l$-dependence of  $\Delta S_A(t=2\xi_{kz})$ in the slow limit shows that it approaches a positive constant if $l>6\xi_{kz}$. There is an effective correlation length, which is larger than the initial length, and this effective correlation length is expected to be associated with the distance between the entangled particles created around $t=-\xi_{kz}$.

The time evolution of $\Delta S_A$ in the fast limit is interpreted in terms of the propagation of the entangled particles. The late-time $\Delta S_A$ in both limits can be fitted by a linear function of $l$, whose proportionality depends on $\xi$ and $\delta t$.
Two point functions in $t \gg \delta t$, composed of a fundamental field and its conjugated momentum, oscillate, and the oscillation comes from the coherent between right-moving and left-moving waves. Since entanglement entropy for a Gaussian state in weakly coupled theories is related to the two point functions, the entanglement oscillation comes from the oscillation of two point functions. The periodicity of oscillation in late time is set by $\xi$ which is related to the coherent oscillation of  zero modes which are zero-momentum spectra of two point functions for a fundamental field and its conjugate.

\subsubsection*{Organization}
Our paper is organized as follow.
In section $2$, we explain our setup: the definition of $S_A$, smooth quenches and how to compute $\Delta S_A$.
In section $3$, we explain the detail of our results.
In section $4$, we conclude our results and discuss a few future directions.
\section{Our setup}
In this paper, we study the time evolution of quantum entanglement by measuring a change of entanglement entropy in global quenches with finite quench rate (smooth quenches).
The change is defined by subtracting the entropy of an initial state from the entropy at $t$. 
We will explain the definition of entanglement entropy, smooth quenches and how to compute the change in this section. In our notation, $t, l, \delta t$ and $\xi$ are dimensionless.

\subsection{Definition of entanglement entropy}
Here, we will explain how we define entanglement entropy in quantum field theories.
We divide a total Hilbert space into two subsystems, $A$ and $B$, geometrically as follows:
\begin{equation}
\mathcal{H}_{total} = \mathcal{H}_A+\mathcal{H}_B,
\end{equation}
where $\mathcal{H}_{A,B}$ are Hilbert spaces of $A$ and $B$.
Entanglement entropy can measure the quantum entanglement between $A$ and $B$.
The entropy is given by the von Neumann entropy $S_A$ for a reduced density matrix $\rho_A$:
\begin{equation}
S_A = -\text{Tr}_A\rho_A\log{\rho_A},
\end{equation}
where $\rho_A$ is given by
\begin{equation}
\rho_A= \text{Tr}_B \rho.
\end{equation}
Thus, $S_A$ is defined by ignoring the degrees of freedom in $B$.
\subsubsection{A change of entanglement entropy}
Here, we study the dynamics of quantum entanglement by computing the change of entanglement entropy defined by subtracting the entropy for the initial state from the entropy for the state at $t$:
\begin{equation}
\Delta S_A(t) = S_A(t) -S_A(t_{in}),
\end{equation}
where $S_A(t_{in})$ is the entropy for the initial state which is defined at $t=-\infty$. Since the time-dependent mass in this paper change slowly in the very early time, the entropy for the initial state is approximated by the one in a massive theory.
\subsection{Smooth quenches}
A Hamiltonian in this paper has a time-dependent mass which has two tunable parameters: $\delta t$ and $m$. 
There are two kinds of potential in this paper.
 One of them is an ECP-type mass, and the other is a CCP-type. The parameter, $\delta t$, determines a time scale of the potential as in Figure \ref{potential}.
 On the other hand, $m$ determines the correlation length, $\xi$, for the initial state.

\subsubsection{ECP-type potential}
The ECP-type potential in this paper is given by 
\begin{equation} \label{mecp}
m^2(t)= \frac{m^2}{2}\cdot\left(1-\tanh{\left(\frac{t}{\delta t}\right)}\right),
\end{equation}
where $m$ is the inverse of the correlation length for the initial state. As in the left side of Figure \ref{potential}, the Hamiltonian before $t\simeq-\delta t$ is the one with the finite mass, $m$.
The time-dependent mass $m(t)$ decreases monotonically with respect to $t$. Thus, the system approaches a critical point asymptotically.

\subsubsection{CCP-type potential}
The CCP-type mass is given by
\begin{equation}\label{mccp}
m^2(t)=m^2 \tanh^2{\left(\frac{t}{\delta t}\right)}.
\end{equation}
In this potential, the initial and final Hamiltonians ($t \rightarrow \pm \infty$) are approximated by the Hamiltonian with a finite mass $m$, as in the right side of Figure \ref{potential}.
Before $t=0$, the mass monotonically decreases with respect to $t$ and vanishes at $t=0$. After $t=0$, it monotonically increases with respect to $t$. 

\subsection{Two extreme limits}
The time-dependent mass has two parameters which we can tune in order to change the quench rate. We can change the time evolution of $\Delta S_A$ by tuning these parameters. In this paper, we take two extreme limits: the slow and fast limits. We will explain these limits.

\subsubsection{Slow limit}
In the slow limit, the system is expected to time-evolve adiabatically when it is far from the critical point.
Whether the time evolution of system is adiabatic is determined by Landau criteria. We define a function, $C_L(t)$:
\begin{equation}
C_L(t)=\left|\frac{1}{m^2(t)}\times \frac{d m(t)}{dt}\right|.
\end{equation}
If $C_L(t) \ll 1$, the local quantities such as two point functions can be approximated by the leading term in the adiabatic expansion. Thus, the local quantities is computed adiabatically.
When the system approaches the critical point, an adiabaticity breaks down: $C_L(t) \simeq \mathcal{O}(1)$. 
The {\it Kibble--Zurek time}, $t_{kz}$, is the time $C_L(t) \simeq 1$.
In the ECP case, the adiabaticity breaks down after $t\simeq t_{kz}$. On the other hand, the adiabaticity breaks down from $t\simeq-t_{kz}$ to $t\simeq t_{kz}$ in the CCP case.
By tuning the parameters, the condition $C_L(t) \simeq 1$ is satisfied only near the critical point (the slow limit). 
\subsubsection*{ECP-type Potential}
Here, we consider the ECP-type mass in (\ref{mecp}).
$C_L(t)$ for (\ref{mecp}) is given by
\begin{equation}
C_L(t)=\frac{1}{\sqrt{2}m \delta t e^{-\frac{3t}{2\delta t}}\sqrt{\cosh{\left(\frac{t}{\delta t}\right)}}}.
\end{equation}
The breaking of adiabaticity near the critical point is that  $C_L(t_{kz}) \sim 1$ when  $e^{\frac{t_{kz}}{\delta t}} \gg 1$. It means that
\begin{equation}
\begin{split}
&C_L(t_{kz}) \sim \frac{e^{\frac{t_{kz}}{\delta t}}}{m \delta t} \sim 1\\
&\Rightarrow e^{\frac{t_{kz}}{\delta t}} \sim \delta t m. \\
\end{split}
\end{equation}
Thus, $\omega =\delta t m \gg 1$.
A {\it Kibble--Zurek energy}, $E_{kz}$, is given by
\begin{equation}
E_{kz}=m(t_{kz}) \sim \frac{1}{\delta t}.
\end{equation}
A Kibble--Zurek time can be defined by
\begin{equation}
t_{kz}\sim \delta t \log{\omega}.
\end{equation} 
\subsubsection*{CCP-type Potential}
 If mass profile  is (\ref {mccp}), $C_L(t)$ is given by
\begin{equation}
C_L(t) =\frac{1}{m(t)\delta t \tanh{\left(\frac{t}{\delta t}\right)} \cosh^2{\left(\frac{t}{\delta t}\right)}}.
\end{equation}
The breaking of the adiabaticity near the critical point means $C_L(t=-t_{kz}) \simeq 1$ for $\frac{t_{kz}}{\delta t} \ll 1$. When the adiabaticity breaks down around the critical point, $C_L(t=-t_{kz}) \simeq 1$ means that
\begin{equation} \label{tkz1}
\begin{split}
&C^2_L(-t_{kz})=\frac{1}{m^2\delta t^2  \sinh^4{\left(\frac{t}{\delta t}\right)}}\sim1\\
&\Rightarrow t_{kz } \sim \left(\frac{\delta t}{m}\right)^{\frac{1}{2}},
\end{split}
\end{equation}
where we use $\frac{t_{kz}}{\delta t} \ll 1$ in the second line in (\ref{tkz1}).
Then,
\begin{equation}
\begin{split}
&\frac{t_{kz}}{\delta t} \ll 1 \\
&\Rightarrow \omega = m \delta t \gg 1,
\end{split}
\end{equation}
where $\xi_{kz} \equiv \frac{1}{m(-t_{kz})}=t_{kz}$.
Thus, the limit, $\omega \gg 1 $, is the slow limit.
If $\frac{t_{kz}}{\delta t} \ll 1$ (the slow quenches), Kibble and Zurek conjectured that the dynamics is frozen, which means that  local quantities between $t\simeq-t_{kz}$ and $t\simeq t_{kz}$ might be approximated by the leading terms in the adiabatic expansion  at $t\simeq-t_{kz}$.

\subsubsection{Fast limit}
The fast limit is the opposite of the slow limit:
\begin{equation}\label{fast}
\omega \ll 1,
\end{equation}
where the system is expected to be similar to the one in sudden quenches.
\subsection{How to Compute}
Here, we explain how to compute entanglement entropy. 
A well-known and powerful method to compute the entropy is a {\it replica method} \cite{rp1, rp2}, where 
R$\acute{e}$nyi entanglement entropy is given by a ``free energy'' from a partition function on a replica space. In this paper, we compute entanglement entropy by a method, {\it correlator method}, where the entropy is given by two point functions. It is the powerful method to compute entanglement entropy numerically in a lattice theory \cite{MM1,d2,ln1, cme}. 

\subsubsection{Setup}
We put a dynamical system on the lattice as follows.
A time-dependent Hamiltonian is given by
\begin{equation}
H(t)= \frac{1}{2}\int dx \left[\pi(x)\pi(x)+\left(\nabla \phi(x)\right)^2+\tilde{m}(t)^2\phi(x)^2\right],
\end{equation}
where the time-dependent mass $\tilde{m}(t)$ is dimensionful.
A discrete Hamiltonian on a circle with the circumference $L=N \epsilon$ is obtained by replacing $\int dx \rightarrow \epsilon \sum_{k=0}^{N-1}$ and $\phi \rightarrow q_k, \pi \rightarrow p_k/\epsilon, \tilde{m}(t) \rightarrow m(t)/\epsilon$ as well as $\nabla \phi \rightarrow (q_{k+1}-q_k)/\epsilon$,
\begin{equation}
H(t)=\frac{1}{2\epsilon}\sum_{k=0}^{N-1}\left[p_k^2+(2+ m(t)^2)q_k^2-2 q_{k+1}q_k\right],
\end{equation}
where we assume that $N=2m+1$ ($m \in {\bf Z}$). Dimensionful parameters  $\tilde{t}, \tilde{l}, \delta \tilde{t}$ and $\tilde{\xi}$ are given by 
\begin{align}
\tilde{t}=\epsilon t,~~ \tilde{l}=\epsilon l,~~ \delta \tilde{t}=\epsilon \delta t,~~ \tilde{\xi}=\epsilon\xi.
\end{align}
We impose om $q_k$ and $p_k$ a periodic boundary condition:
\begin{equation}
p_0=p_{N-1}, ~q_0=q_{N-1},
\end{equation}
and a canonical commutation relation:
\begin{equation}
[q_a, p_b]= i \delta_{ab}, ~~[q_a, q_b]= [p_a, p_b]=0.
\end{equation}
We use a discrete Fourier transform:
\begin{equation}
\begin{split}
q_k=\frac{1}{\sqrt{N}}\sum_{l=-\frac{N-1}{2}}^{\frac{N-1}{2}}e^{i \frac{2\pi k l}{N}}\tilde{q}_l, \\
p_k=\frac{1}{\sqrt{N}}\sum_{l=-\frac{N-1}{2}}^{\frac{N-1}{2}}e^{i \frac{2\pi k l}{N}}\tilde{p}_l, \\
\end{split}
\end{equation}
where $\tilde{q}^{\dagger}_l=\tilde{q}_{-l}$ and $\tilde{p}^{\dagger}_l=\tilde{p}_{-l}$ because $p_k$ and $q_k$ are real.
In terms of variables in the momentum space, the Hamiltonian can be written by
\begin{equation}
H(t)=\frac{1}{2\epsilon}\sum_{k=-\frac{N-1}{2}}^{\frac{N-1}{2}}\left[\tilde{p}_k \tilde{p}^{\dagger} _k+ \left(4\sin^2{\left(\frac{\pi k}{N}\right)}+ m^2(t)\right)\tilde{q}_k\tilde{q}^{\dagger}_k\right],
\end{equation}
where $\tilde{p}$ and $\tilde{q}$ satisfy that $\left[\tilde{q}_a, \tilde{p}_b \right]=i\delta_{a, -b}$ and  $\left[\tilde{p}_a, \tilde{p}_b\right]=\left[\tilde{q}_a, \tilde{q}_b\right]=0$.
$\tilde{p}$ and $\tilde{q}$ are written by
\begin{equation}\label{fs}
\begin{split}
&\tilde{q}_k =f_k(t) a_k + f_{-k}^*(t) a^{\dagger}_{-k}, \\
&\tilde{p}_k =\dot{f}_k(t) a_k + \dot{f}_{-k}^*(t) a^{\dagger}_{-k},
\end{split}
\end{equation}
where $\left[a_k, a_l\right]=\left[a^{\dagger}_k, a^{\dagger}_l\right]=0$ and $\left[a_k, a^{\dagger}_l\right]=\delta_{k,l}$. We impose a condition, $f_k(t)=f_{-k}(t)$, on $f_k$. Then, $f_k(t)$ satisfies the following Wronskian condition:
\begin{equation}
\begin{split}
&\left[\tilde{q}_L, \tilde{p}_{-l}\right]=\left( f_l(t)\dot{f}^*_l(t)-\dot{f}_l(t)f^*_l(t)\right)\delta_{l,L}=i \delta_{l ,L}. \end{split}
\end{equation}
After replacing $\frac{\pi k}{N} \rightarrow k/2$, $f_k(t)$ obeys the following equation of motion:
\begin{equation}
\frac{d^2 f_k(t)}{dt^2}+\left[4 \sin^2{\left(\frac{k}{2}\right)}+ m^2(t)\right]f_k(t)=0.
\end{equation}
\subsubsection{Correlator method}
If we take a limit where $N\rightarrow \infty$, and $L \rightarrow \infty$, but $\epsilon$ is finite, the coordinates and its conjugate momenta are given by
\begin{equation}
\begin{split}
q_n=X_n =\int^{\pi}_{-\pi} \frac{dk}{\sqrt{2\pi}}\tilde{q}_k e^{i k n}, \\
p_n=P_n= \int^{\pi}_{-\pi} \frac{dk}{\sqrt{2\pi}}\tilde{p}_k e^{i k n}, \\
\end{split}
\end{equation} 
where $\tilde{p}$ and  $\tilde{q}$ are defined in (\ref{fs}).

We can use the correlator method for a Gaussian state in a weakly coupled theory when we compute R$\acute{e}$nyi entanglement entropy. 
In the method, $n$-th R$\acute{e}$nyi entanglement entropy is given by
\begin{equation}
S_A^{(n)}=\sum_{r=1}^{l}\frac{1}{n-1}\log{\left[\left(\gamma_r+\frac{1}{2}\right)^n-\left(\gamma_r-\frac{1}{2}\right)^n\right]},
\end{equation}
where $l$ is the number of subsystem site, and $\gamma_r$ are positive eigenvalues of a $2l \times 2l$ matrix $\mathcal{M}$:
\begin{equation}
\begin{split}
&\mathcal{M}= i J \Gamma, \\
&J=\begin{bmatrix}
0 & I_{l \times l} \\
-I_{l \times l} & 0 \\
\end{bmatrix}, \\
&\Gamma=\begin{bmatrix}
X_{ab}(t) & D_{ab}(t) \\
D_{ab}(t) & P_{ab}(t) \\
\end{bmatrix}, \\
\end{split}
\end{equation}
where $X_{ab}(t), P_{ab}(t)$ and $D_{ab}(t)$ are given by
\begin{equation}\label{spect}
\begin{split}
&X_{ab}(t)=\left\langle X_a (t)X_b(t) \right \rangle =\int^{\pi}_{-\pi} \frac{dk}{2\pi}X_k\cos{\left(k\left|a-b\right|\right)} =\int^{\pi}_{-\pi} \frac{dk}{2\pi}\left|f_k(t)\right|^2\cos{\left(k\left|a-b\right|\right)}, \\
&P_{ab}(t)=\left\langle P_a (t)P_b(t) \right \rangle = \int^{\pi}_{-\pi} \frac{dk}{2\pi}P_k\cos{\left(k\left|a-b\right|\right)} = \int^{\pi}_{-\pi} \frac{dk}{2\pi}\left|\dot{f}_k(t)\right|^2\cos{\left(k\left|a-b\right|\right)}, \\
&D_{ab}(t)=\frac{1}{2}\left\langle \left\{X_a (t), P_b(t) \right\} \right \rangle=\int^{\pi}_{-\pi} \frac{dk}{2\pi}D_k\cos{\left(k\left|a-b\right|\right)} =\int^{\pi}_{-\pi} \frac{dk}{2\pi}Re\left[\dot{f}^*_k(t)f_k(t)\right]\cos{\left(k\left|a-b\right|\right)}. \\
\end{split}
\end{equation}
In this paper, we compute $\Delta S_A$ numerically. The detail of numerical computation is explained in Appendix \ref{sec:appendix_numerical_calculation} and \ref{sec:error_examination}.
\section{Time evolution of $\Delta S_A$}
Here, we study the dynamics of quantum entanglement in the smooth quenches by measuring time evolution of entanglement entropy. Since we are interested in how the entanglement structure changes in the smooth quenches, we define a change of entanglement entropy, $\Delta S_A(t)$, by subtracting $S_A(t_{in})$ for an initial state from $S_A(t)$ for the smooth-quenched state.  First, we study the time evolution of $\Delta S_A$ in the ECP-type potential.
Second, we study the time evolution of $\Delta S_A$ in the CCP-type potential. 
\subsection{Time evolution of $\Delta S_A$ in ECP-type protocol}
We study the time evolution of $\Delta S_A$ in both fast and slow limits in ECP-type potential and interpret it in terms of entangled particles.
\subsubsection{Fast limit}
Here, we study the time evolution of $\Delta S_A$ in the fast limit, $\omega \ll 1$, in the ECP-type potential. 
$\Delta S_A$ with several parameters is shown in Figure \ref{xi100dt5}. If parameters $(\xi, \delta t, l)$ are sufficiently large, we expect $\Delta S_A$ to depend on ratios of physical length scales, and the lattice spacing drops out.  Furthermore, Figure \ref{prop_ECP_fast} shows the data for $\Delta S_A$ as a function of $t/\xi$ for different values of $l,\xi,\delta t$ with the same value of $l/\xi$ and $\omega$. All the points lie on the same curve, showing that the function $\Delta S_A$ is of the form
\begin{equation} \label{sce}
\Delta S_A\sim\Delta S_A \left(\frac{l}{\xi}, \frac{t}{\xi}, \omega \right).
\end{equation}
Figure~\ref{ECP_fast_vol_dep} shows the $l$-dependence of $\Delta S_A$.
We also find that Figure~\ref{xi100dt5}, \ref{prop_ECP_fast} and \ref{ECP_fast_vol_dep} show the following properties of the time evolution:
\begin{itemize}
\item[(1)] The entanglement structure does not change at $t<0$.
\item[(2)] $\Delta S_A$ starts to increase around $t=0$.
\item[(3)] If the subsystem size $l$ is sufficiently larger than the initial correlation length, $\xi$, the time evolution of $\Delta S_A$ does not depend on $l$ before $t \simeq\frac{l}{2}$. When $\xi \ll t <\frac{l}{2}$,  $\Delta S_A$ is fitted by a linear function of $t$.
\item[(4)] $\Delta S_A$ depends on $l$ after $t\simeq\frac{l}{2}$. A liner function of $l$ fits $\Delta S_A$.
\end{itemize}

\subsubsection*{Time growth}
\begin{figure}[htbp]
 \begin{minipage}{0.2\hsize}
 \begin{center}
   \includegraphics[clip,width=2.5cm]{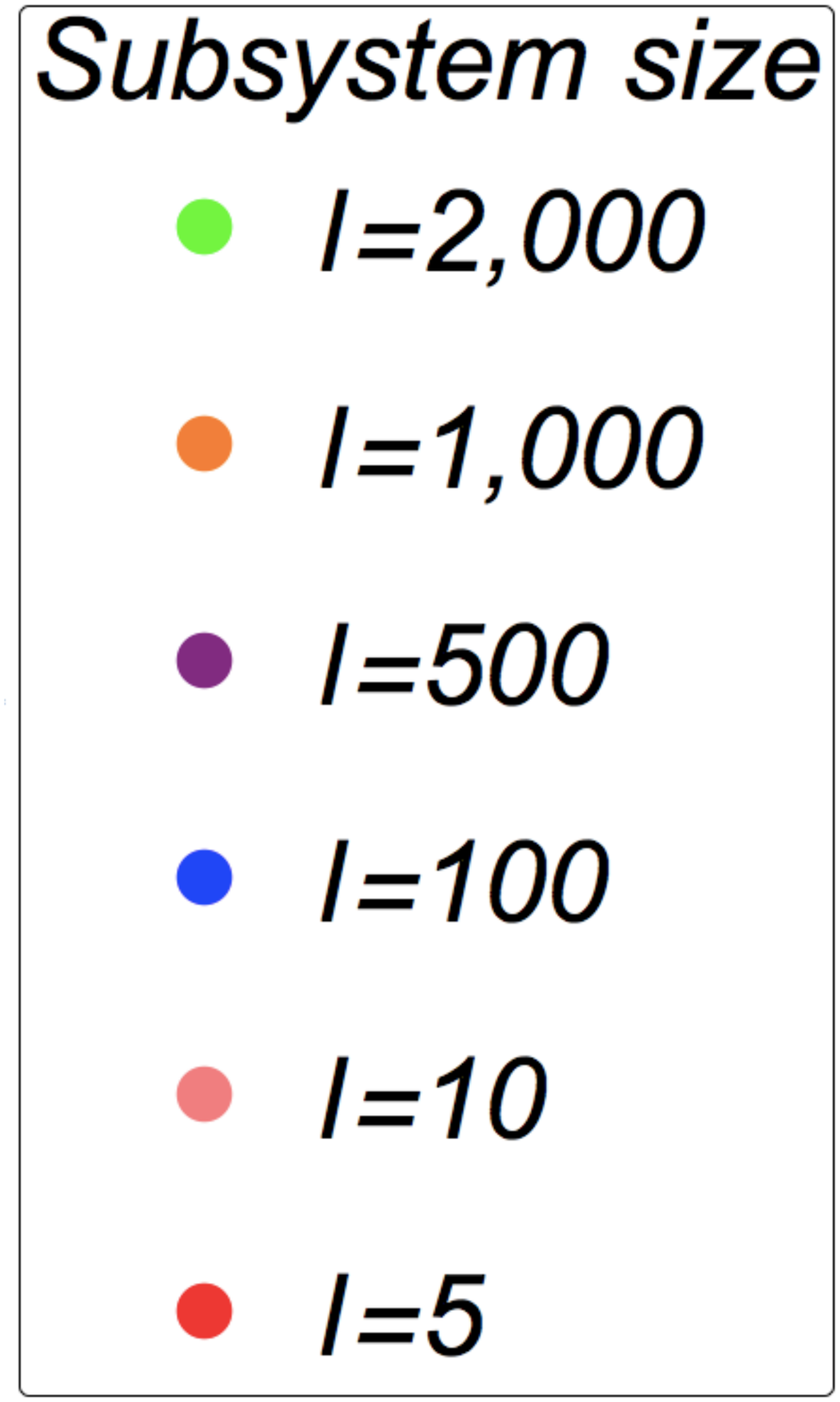}
      \end{center}
  \end{minipage}
 \begin{minipage}{0.25\hsize}
 \begin{center}
    \includegraphics[clip,width=4cm]{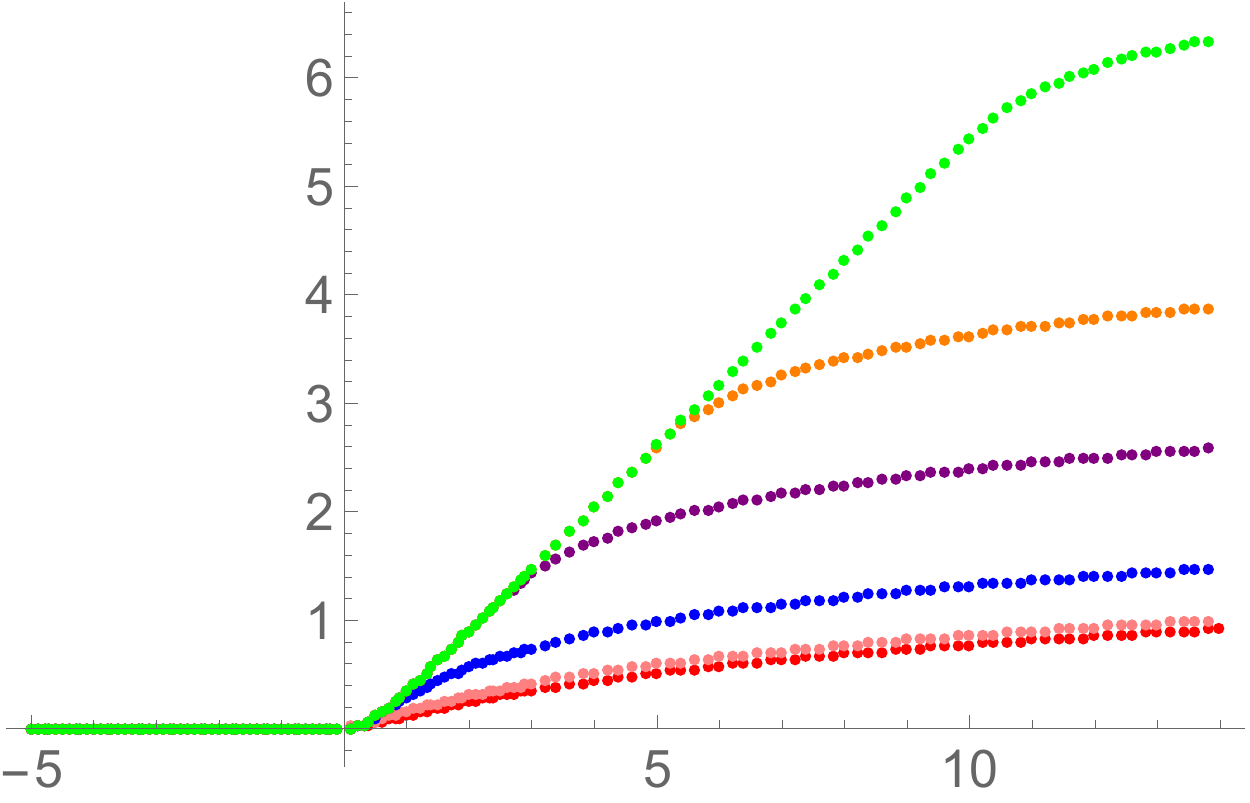}
   \put(-10,-10){$t/\xi$}
   \put(-100,80){$\Delta S_A$}
   \end{center}
  \end{minipage}
 \begin{minipage}{0.25\hsize}
 \begin{center}
     \includegraphics[clip,width=4cm]{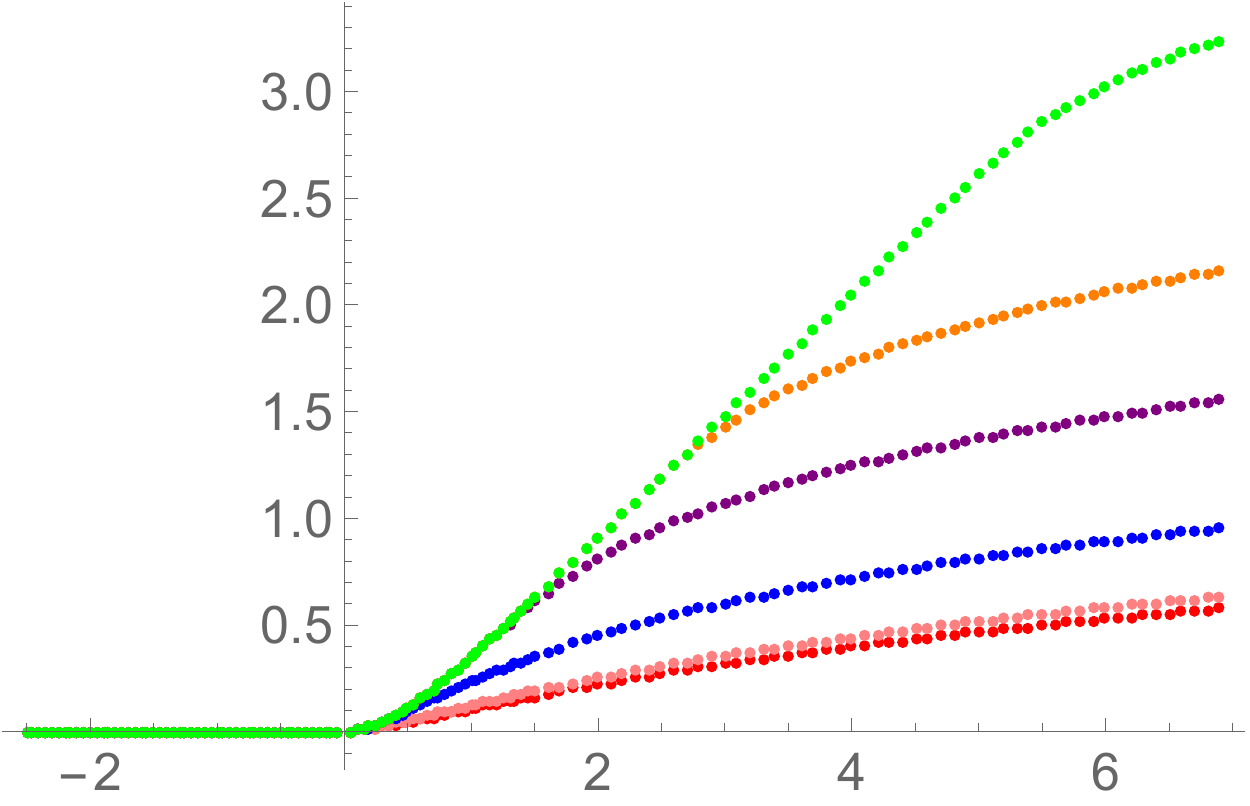}
       \put(-10,-10){$t/\xi$}
    \put(-100,80){$\Delta S_A$}
  \end{center}
  \end{minipage}
   \begin{minipage}{0.25\hsize}
 \begin{center}
    \includegraphics[clip,width=4cm]{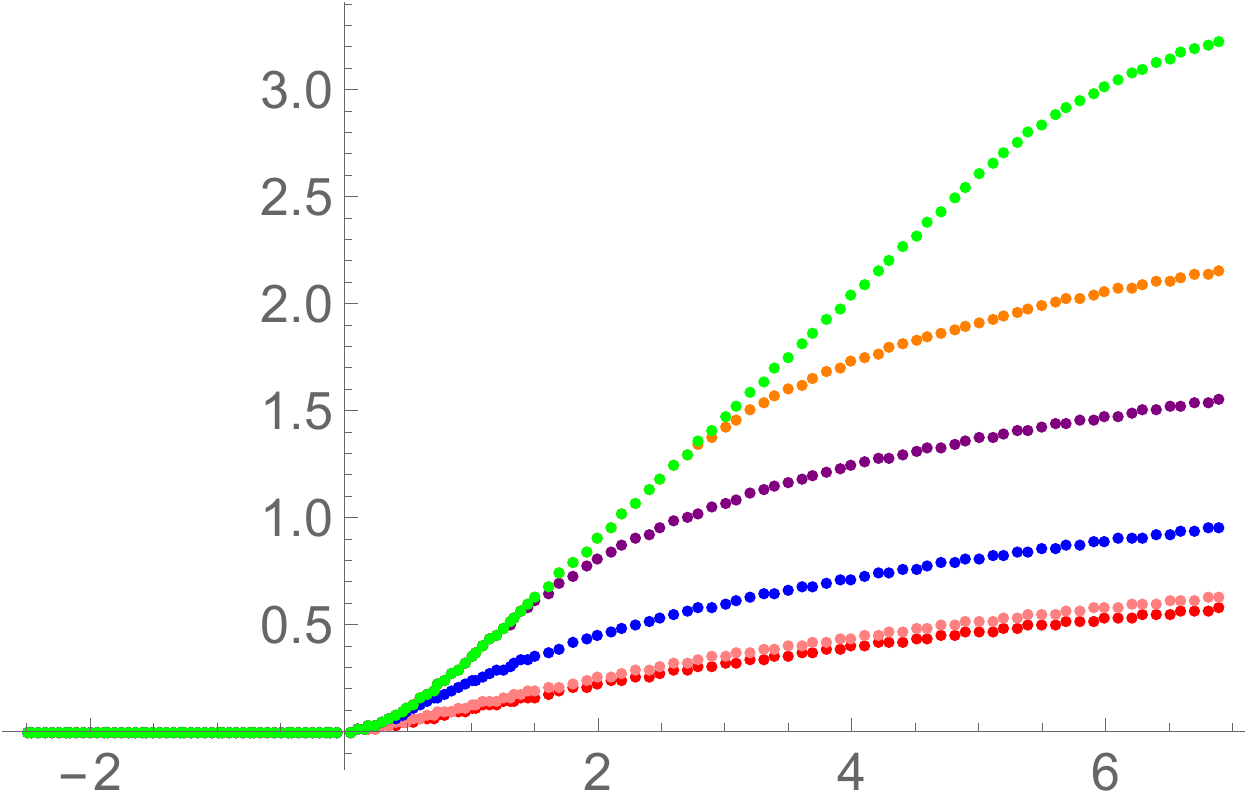}
         \put(-10,-10){$t/\xi$}
    \put(-105,80){$\Delta S_A$}
  \end{center}
  \end{minipage}
    \caption{The time evolution of the entanglement entropy. We plot $\Delta S_A$ with $(\xi,\delta t)=(100,5)$ in the left panel, $(\xi,\delta t)=(200,5)$ in the middle panel and $(\xi,\delta t)=(200,10)$ in the right panel.   \label{xi100dt5}} 
     \begin{minipage}{0.2\hsize}
 \begin{center}
   \includegraphics[clip,width=2.5cm]{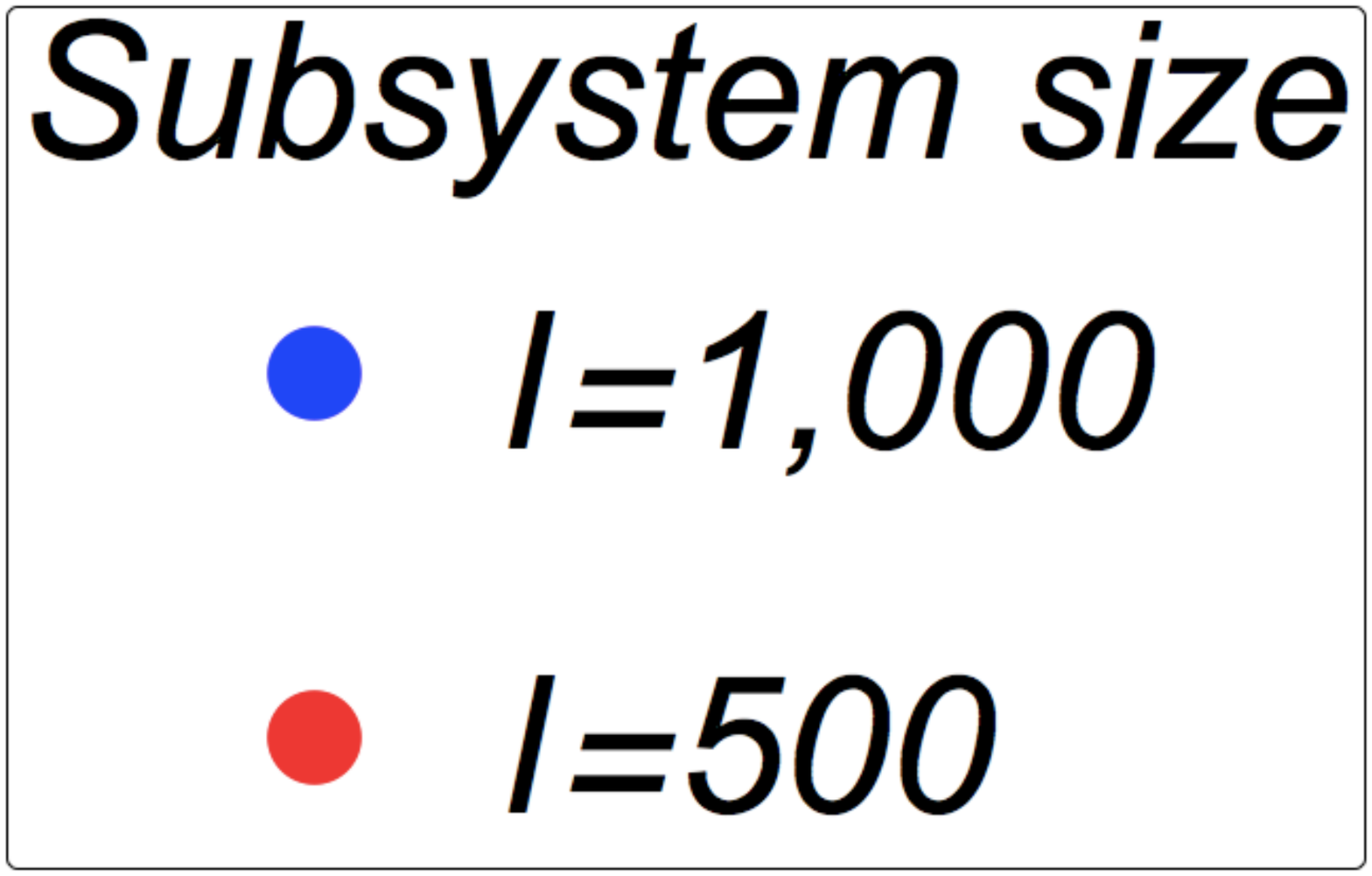}
      \end{center}
  \end{minipage}
   \begin{minipage}{0.45\hsize}
  \begin{center}
   \includegraphics[width=65mm]{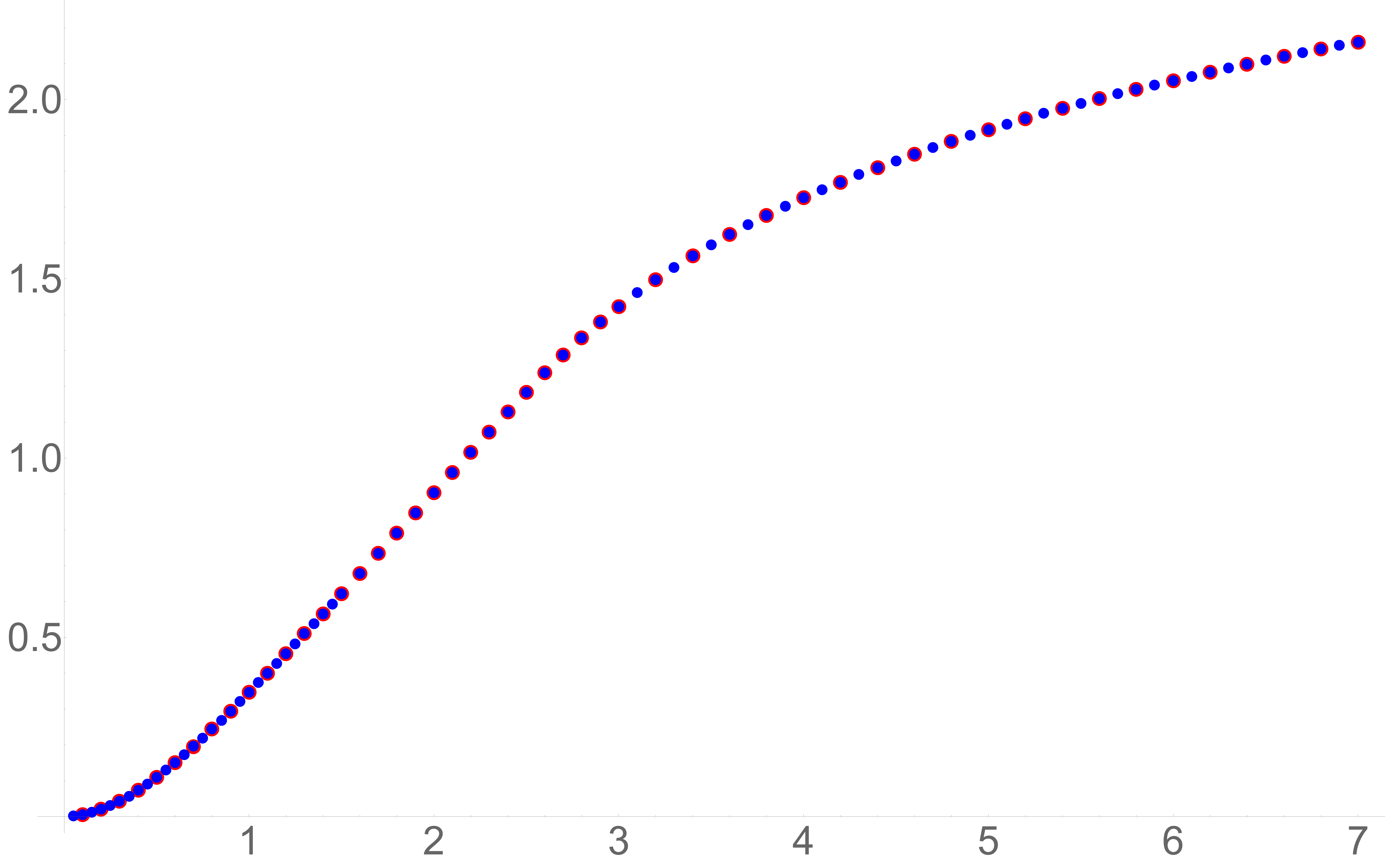}
         \put(5,5){$t/\xi$}
    \put(-190,120){$\Delta S_A $}
  \end{center}
 \end{minipage} 
 \caption{The properties in the time-dependence of the entanglement entropy. We plot $\Delta S_A$ with $\xi=100, \delta t=5$ corresponding to red plots and $\xi=200, \delta t=10$ corresponding to blue plots.    \label{prop_ECP_fast} }
\end{figure}

In the window (3),  $\xi \ll t <\frac{l}{2}$, the time evolution of $\Delta S_A$ is fitted by a linear function of $t$ (in Figure \ref{xi100dt5}):
\begin{equation} \label{tl1}
\Delta S_A \simeq a_1 \frac{t}{\xi} + a_2,
\end{equation}
where $a_1$ and $a_2$ are in Table \ref{ta1}. It  shows that 
\begin{equation} \label{tl2}
a_1\simeq 0.57, ~~ a_2\simeq -0.23.
\end{equation}
(\ref{tl1}) and  (\ref{tl2}) show that the linear growth of $\Delta S_A $ is determined by the initial correlation length.
\begin{table}[htb]
  \begin{center}
          \caption{Fit results for Figure \ref{xi100dt5} in the window (3), $\xi \ll t <\frac{l}{2}$. \label{ta1}}
  \begin{tabular}{|c|c|c|c|c|} \hline
    $\delta t$ & $\xi$ &$l$& Fit Result & Fit Range    \\ \hline \hline
        $5$& $100$&$1,000$ &$\Delta S_A=-0.231489 + 0.567424  \frac{t}{\xi}$ & $2.5 \le \frac{t}{\xi} \le 5$ \\ \hline
           $5$& $200$&$2,000$ &$\Delta S_A=-0.232435 + 0.569479  \frac{t}{\xi}$ & $2.5 \le \frac{t}{\xi} \le 5$ \\ \hline
        $10$& $200$&$2,000$ &$\Delta S_A=-0.231518 + 0.567467 \frac{t}{\xi}$ & $2.5 \le \frac{t}{\xi} \le 5$ \\ \hline
    \end{tabular}
  \end{center}
  \end{table}

\subsubsection*{Late time behavior}
\begin{figure}[htbp]
     \begin{minipage}{0.2\hsize}
 \begin{center}
   \includegraphics[clip,width=3cm]{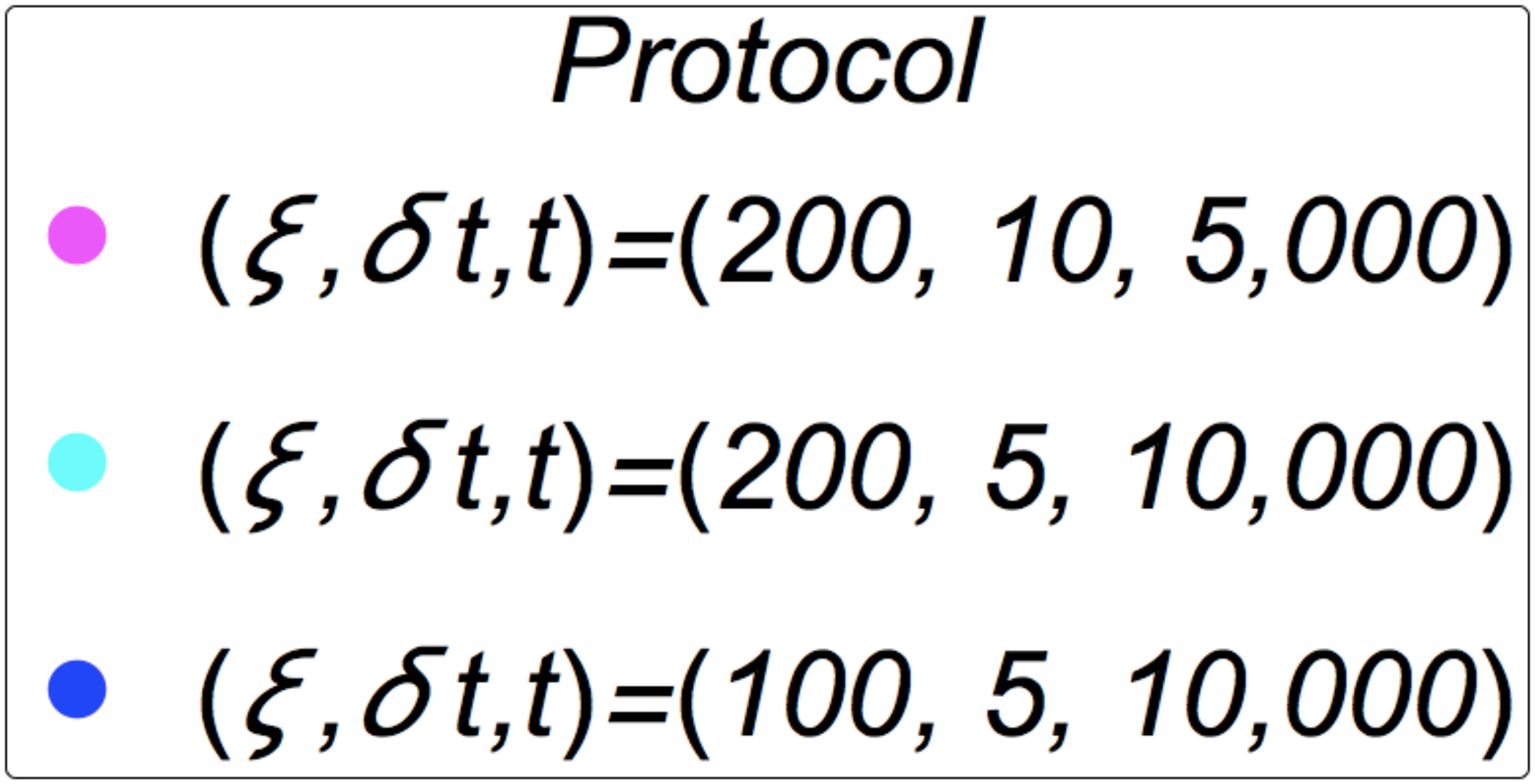}
      \end{center}
  \end{minipage}
  \begin{minipage}{0.4\hsize}
  \begin{center}
  \includegraphics[width=6cm]{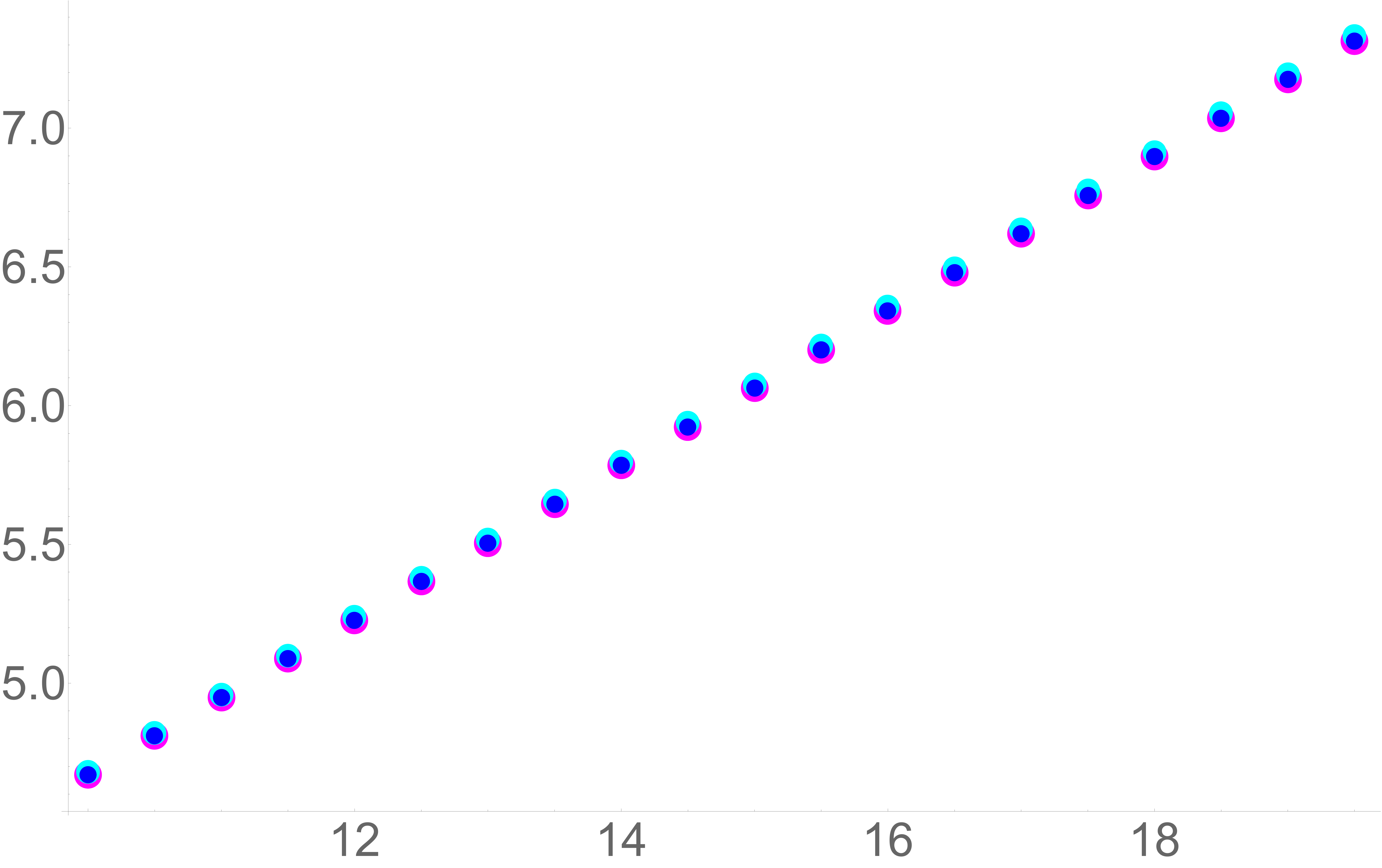}
             \put(5,5){$l/\xi$}
    \put(-180,110){$\Delta S_A $}
      \end{center}
    \end{minipage}
\caption{The volume-dependence of the entanglement entropy. Three results are on top of each other.\label{ECP_fast_vol_dep}}
\end{figure}

In the window (4), $t \ge \frac{l}{2}$, $\Delta S_A$ depends on the subsystem size $l$ as in Figure~\ref{ECP_fast_vol_dep}.  $\Delta S_A$ is fitted by
\begin{equation}\label{v1}
\Delta S_A \simeq b_1 \frac{l}{\xi}+ b_2 
\end{equation}
where $b_1, b_2$ are in Table \ref{ta2}. 
Table \ref{ta2} shows that $b_1$ and $b_2$ are given by
\begin{equation}\label{v2}
b_1 \simeq 0.28, b_2\simeq 1.89.
\end{equation}
(\ref{v1}) and (\ref{v2}) show that $\Delta S_A$ is proportional to the volume of subsystem.
Thus, entanglement entropy appears to be thermal entropy:
\begin{equation} \label{teff}
\Delta S_A \sim T_{eff} \cdot l,\hspace{0.8mm}T_{eff} \sim \frac{1}{\xi}.
\end{equation}
$T_{eff}$ is set by the initial correlation length. $\Delta S_A$ even in late time increases slowly, unlike the result in \cite{ca1}. The authors in \cite{MM1} have found that $\Delta S_A$ in late time in a lattice theory increases logarithmically, $\Delta S_A \sim \frac{1}{2}\log[t]$. Table \ref{log} shows that $\Delta S_A$ even in this setup appears to be $\Delta S_A \sim \frac{1}{2}\log[t]$. It might come from a discretization effect as we will explain later.  

\begin{table}[htb]
  \begin{center}
                \caption{Fit results for Figure \ref{ECP_fast_vol_dep} in the window (4), $t \gg \frac{l}{2}$. \label{ta2}}
    \begin{tabular}{|c|c|c|c|c|c|} \hline
         $\delta t$ &  $\xi$ &$t$& Fit Result & Fit Range    \\ \hline \hline  
       $5$   &  $100$ &   $5,000$ & $\Delta S_A=1.88818 + 0.278238 \frac{l}{\xi} $ & $10 \le \frac{l}{\xi} \le 20$ \\ \hline
       $5$   &  $200$ & $10,000$ & $\Delta S_A=1.88808 + 0.279255 \frac{l}{\xi} $ & $10 \le\frac{l}{\xi}  \le 20$ \\ \hline      
       $10$ &  $200$ & $10,000$ & $\Delta S_A=1.88822 + 0.278254 \frac{l}{\xi} $ & $10 \le \frac{l}{\xi}  \le 20$ \\ \hline      
            \end{tabular}
  \end{center}
  \end{table}

  \begin{table}[htb]
  \begin{center}
        \caption{Fit results for Figure \ref{xi100dt5} in very late time. \label{log}}
    \begin{tabular}{|c|c|c|c|c|} \hline
      $\delta t$ & $\xi$ &$l$& Fit Result & Fit Range    \\ \hline \hline
       $5$& $200$&$2,000$ &$\Delta S_A=-0.633481 + 0.571142 \log [t]$ & $2,000 \le t \le 100,000$ \\ \hline
       $5$& $200$&$2,000$ &$\Delta S_A=-0.0697051 + 0.516526 \log [t]$ & $10,000 \le t \le 100,000$ \\ \hline
       $5$& $200$&$2,000$ &$\Delta S_A=0.0496387 + 0.505738 \log [t]$ & $60,000 \le t \le 100,000$ \\ \hline
    \end{tabular}
  \end{center}
\end{table}
\subsubsection{Slow limit}
\begin{figure}[htbp]
 \begin{minipage}{0.2\hsize}
 \begin{center}
     \includegraphics[clip,width=2.5cm]{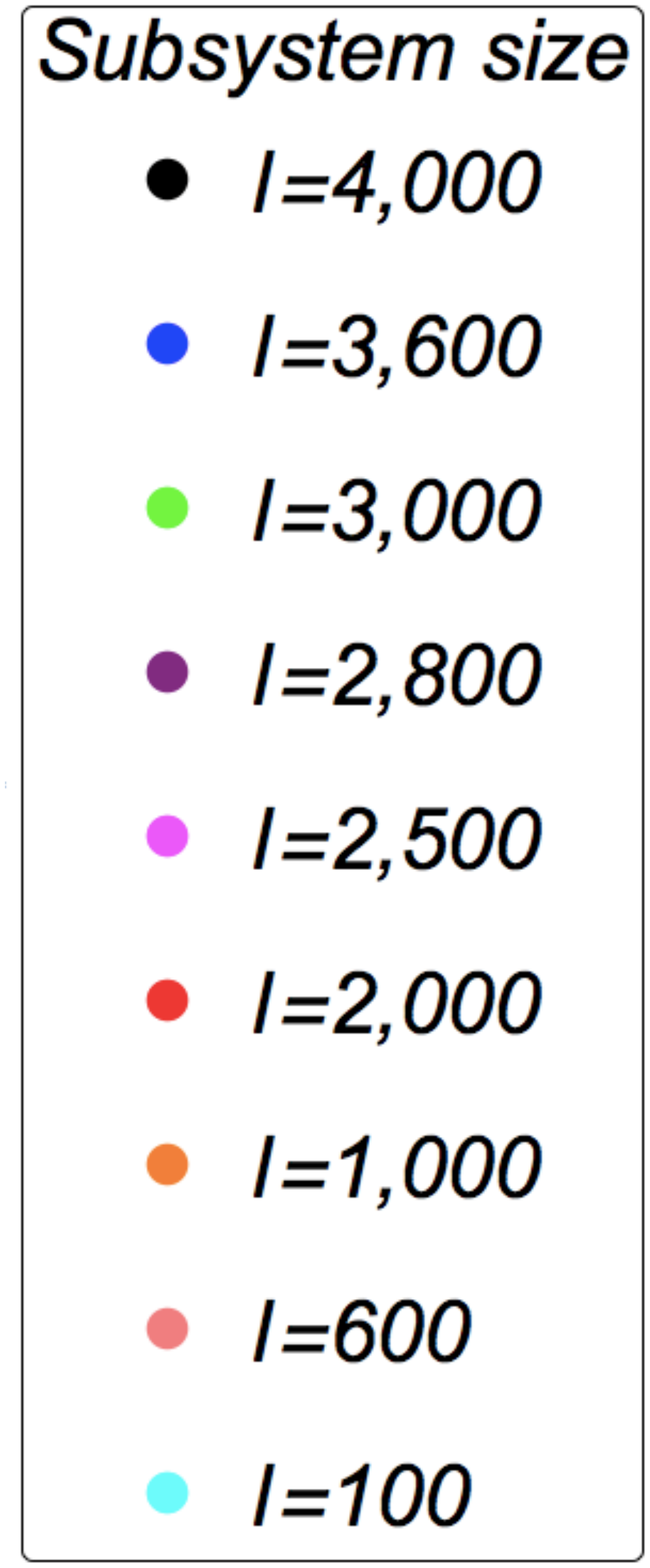}
   \end{center}
\end{minipage}
\begin{minipage}{0.25\hsize}
 \begin{center}  
    \includegraphics[clip,width=3.8cm]{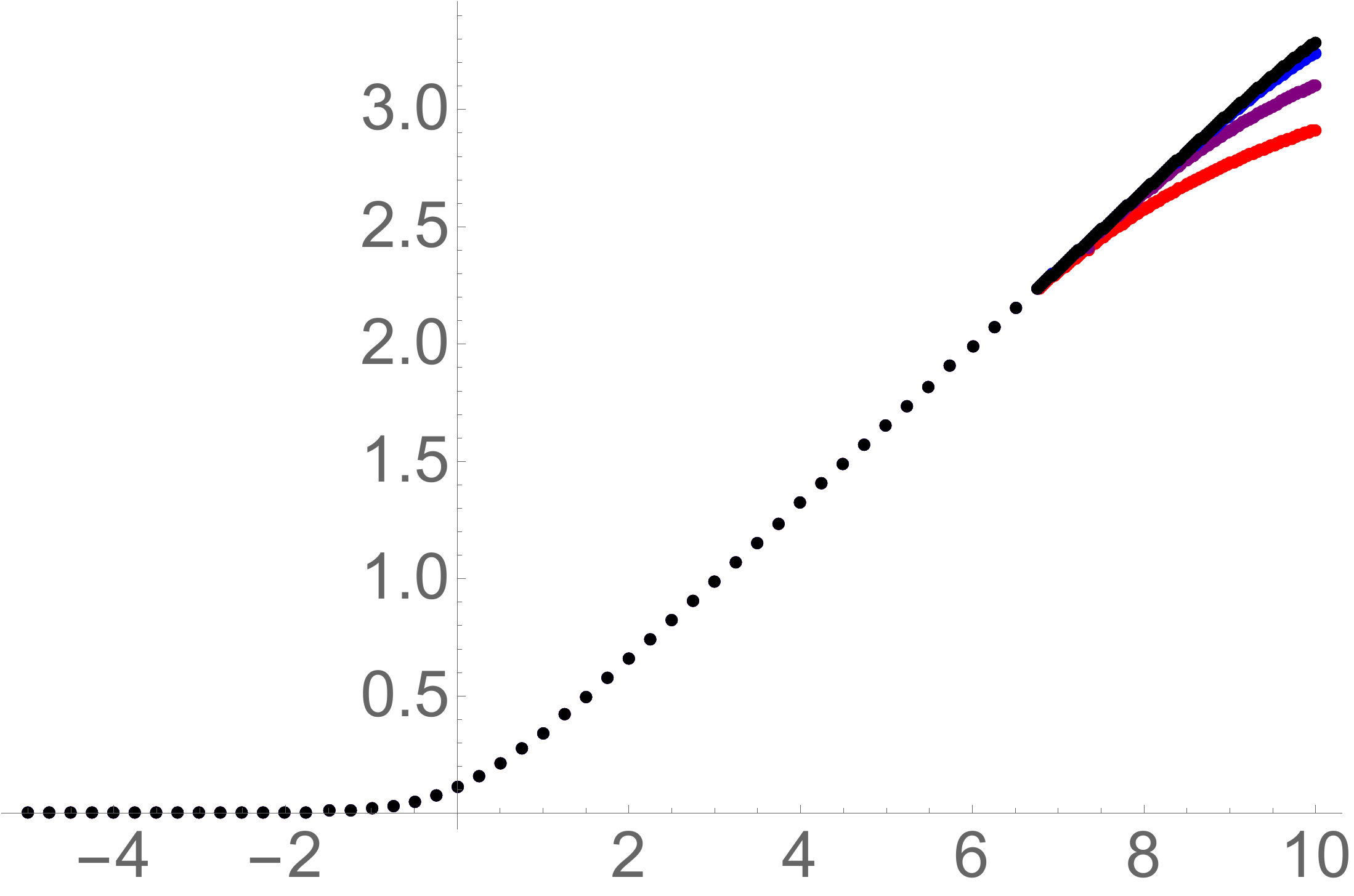}
   \put(-25,-10){$t\cdot E_{kz}$}
   \put(-95,75){$\Delta S_A$}
   \end{center}
\end{minipage}
\begin{minipage}{0.25\hsize}
 \begin{center}
     \includegraphics[clip,width=3.8cm,bb=0mm 0mm 130mm 84mm]{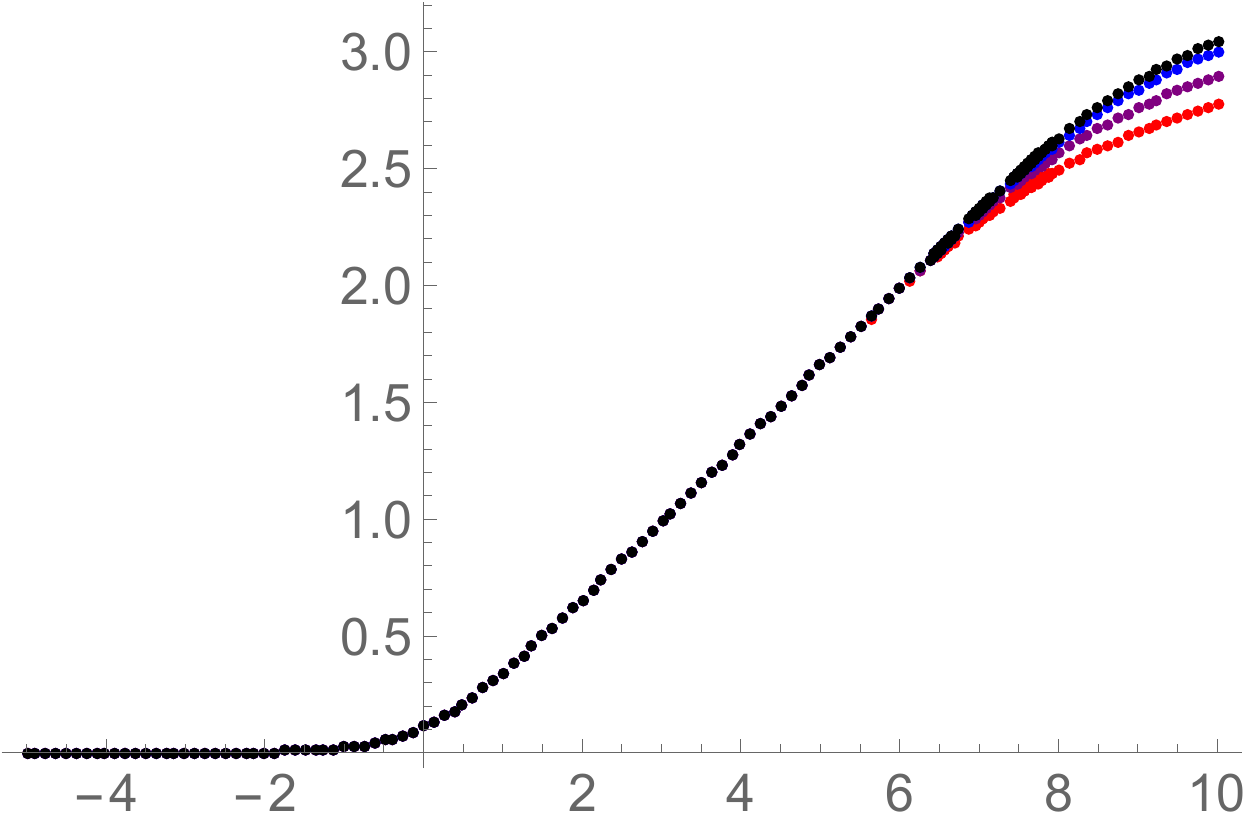}
       \put(-25,-10){$t\cdot E_{kz}$}
    \put(-95,75){$\Delta S_A$}
  \end{center}
  \end{minipage}
   \begin{minipage}{0.25\hsize}
 \begin{center}
    \includegraphics[clip,width=3.8cm, bb=0mm 0mm 130mm 84mm]{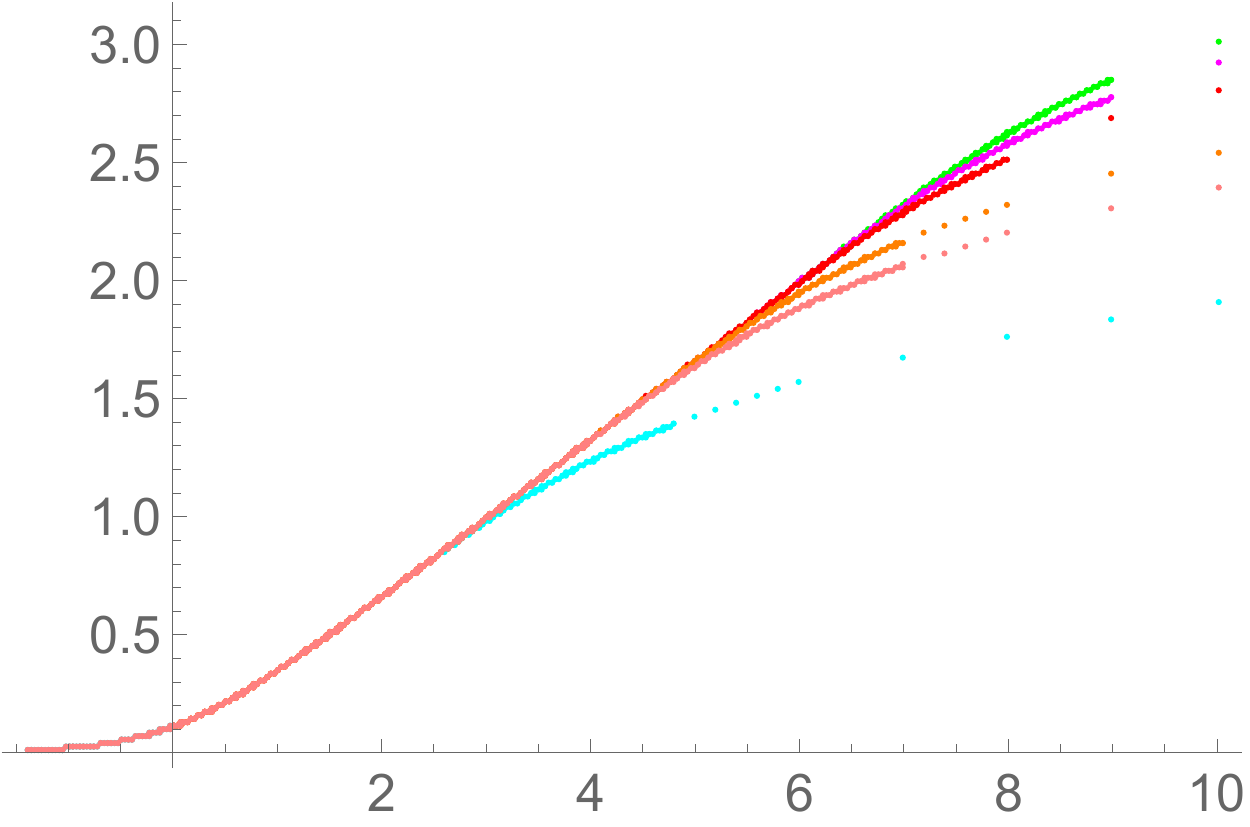}
         \put(-20,-10){$t\cdot E_{kz}$}
    \put(-105,75){$\Delta S_A$}
  \end{center}
  \end{minipage}
\caption{The time evolution of the entanglement entropy in the slow limit. We plot $\Delta S_A$ with $(\xi,\delta t)=(4,400)$ in the left panel, $(\xi,\delta t)=(4,800)$ in the middle panel, and $(\xi,\delta t)=(5,500)$ in the right panel.
\label{ECP_slow}}
\end{figure}

\begin{figure}[htbp]
 \begin{minipage}{0.2\hsize}
 \begin{center}
     \includegraphics[clip,width=2.5cm]{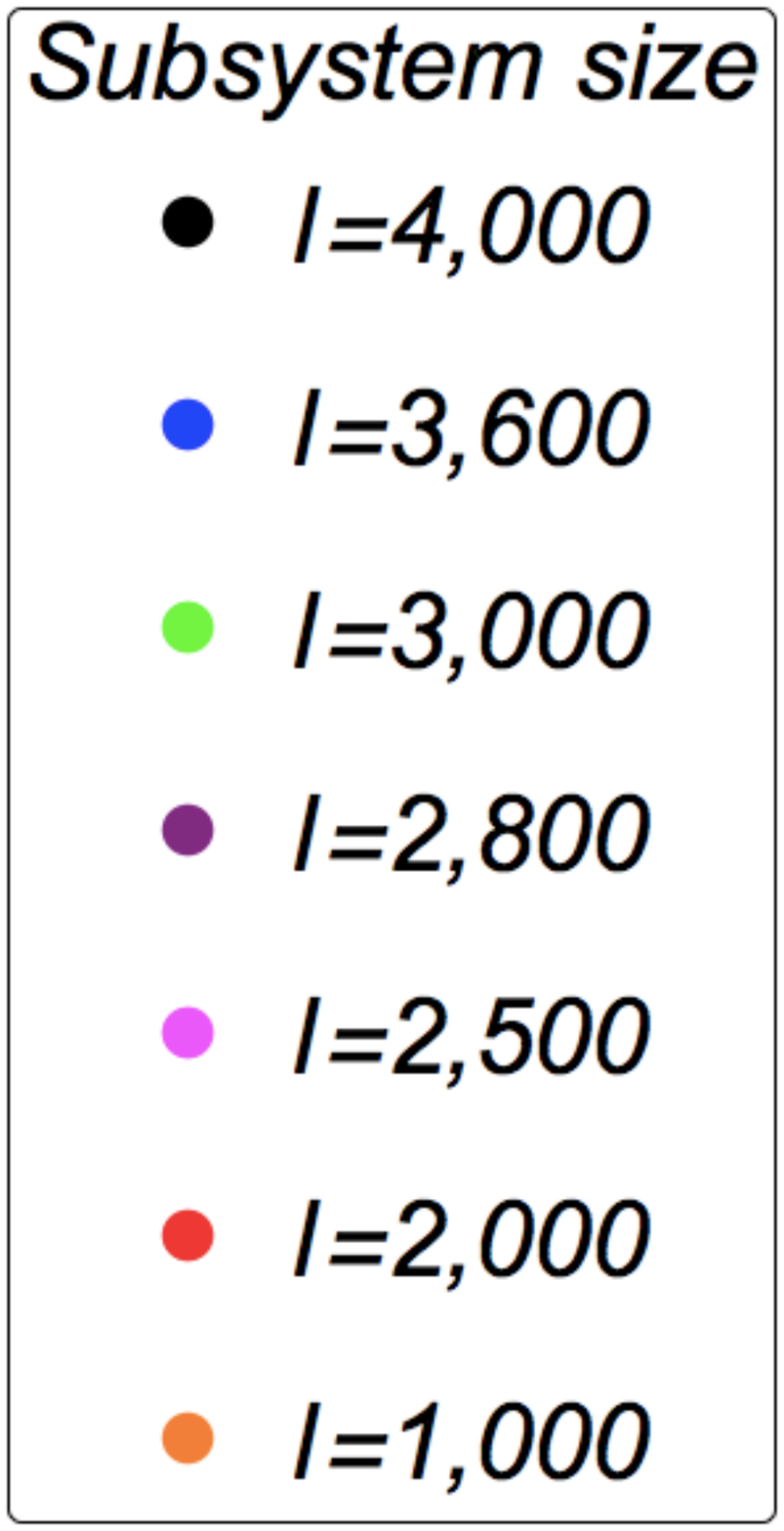}
   \end{center}
\end{minipage}
\begin{minipage}{0.25\hsize}
 \begin{center}  
    \includegraphics[clip,width=3.8cm, bb=0mm 0mm 130mm 84mm]{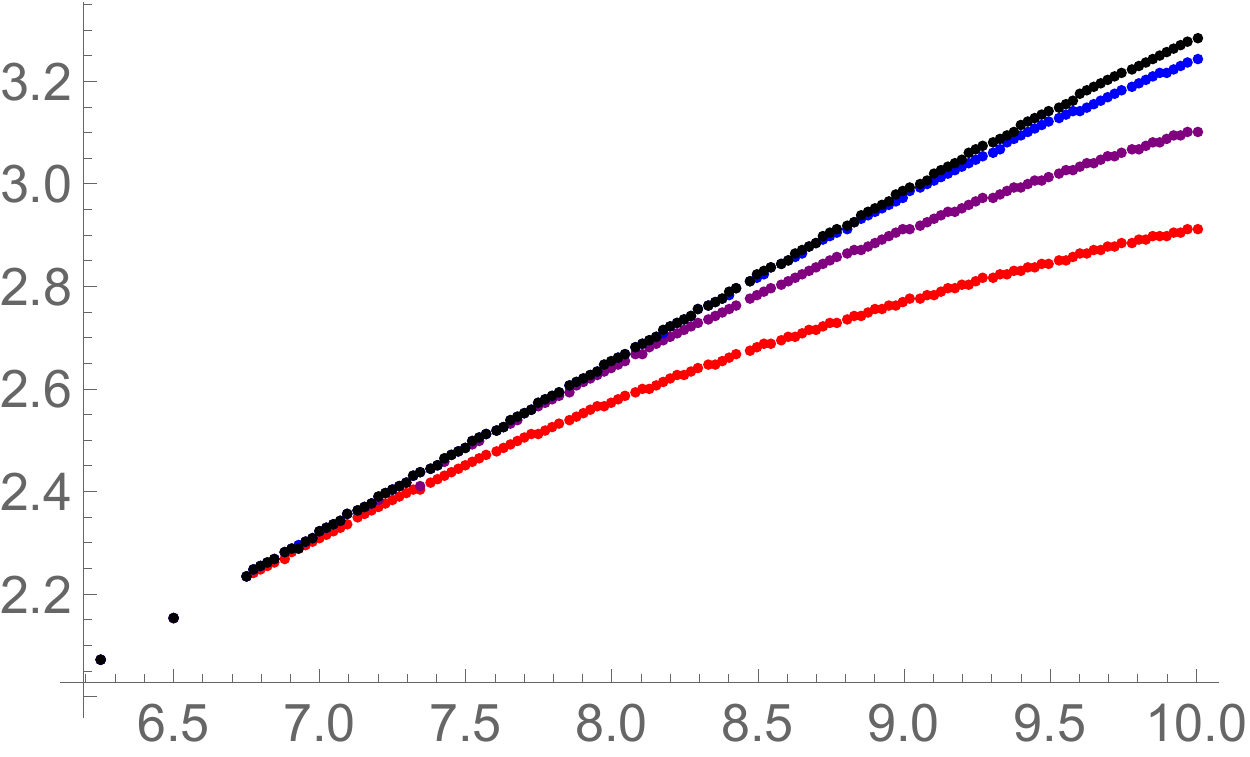}
   \put(-20,-10){$t\cdot E_{kz}$}
   \put(-115,70){$\Delta S_A$}
   \end{center}
\end{minipage}
\begin{minipage}{0.25\hsize}
 \begin{center}
     \includegraphics[clip,width=3.8cm,bb=0mm 0mm 130mm 84mm]{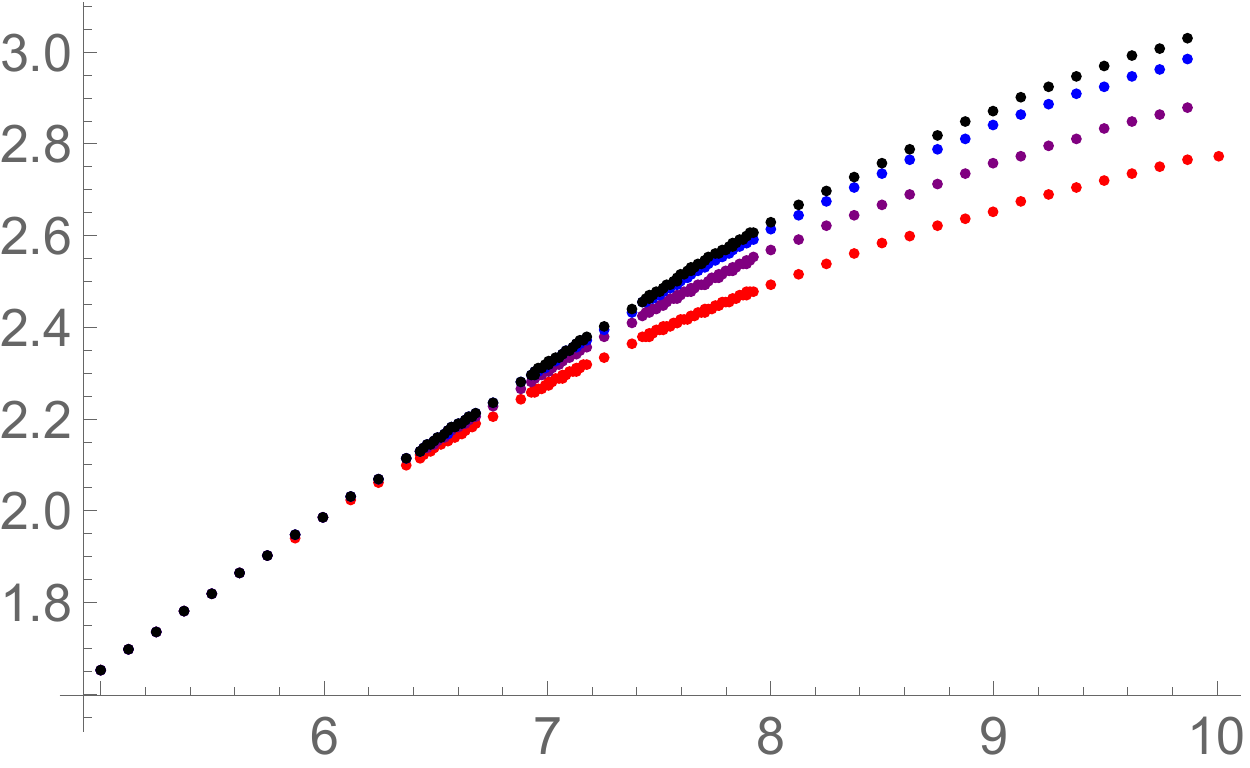}
       \put(-20,-10){$t\cdot E_{kz}$}
    \put(-115,70){$\Delta S_A$}
  \end{center}
  \end{minipage}
   \begin{minipage}{0.25\hsize}
 \begin{center}
    \includegraphics[clip,width=3.8cm, bb=0mm 0mm 130mm 84mm]{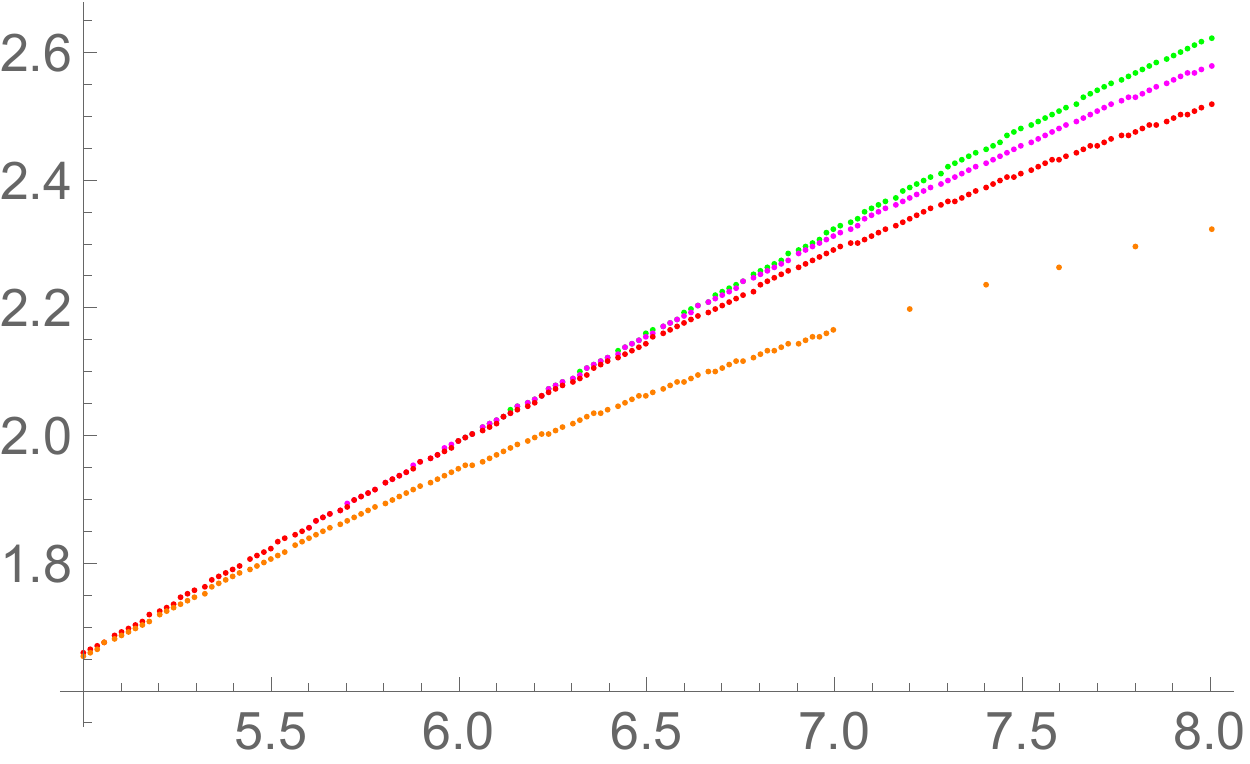}
         \put(-10,-10){$t\cdot E_{kz}$}
    \put(-115,70){$\Delta S_A$}
  \end{center}
  \end{minipage}
\caption{The time that the behavior of the entanglement entropy changes. This figure is same as Figure \ref{ECP_slow}, but enlarged. In this case, the Kibble--Zurek times are $t_{kz}\cdot E_{kz}\simeq 4.605$ in the left panel, $t_{kz}\cdot E_{kz}\simeq 2.648$ in the middle panel, and $t_{kz}\cdot E_{kz}\simeq 4.605$ in the right panel.
\label{ECP_enlarged}}
\end{figure}

Here, we study the time evolution of $\Delta S_A$ in the slow limit, $\omega \gg 1$, in the ECP-type potential\footnote{We expect  $\Delta S_A$ with the parameters sufficiently larger than the lattice spacing to be independent of the spacing. $\Delta S_A$ is given by a scaling function of $E_{kz} \cdot l,$ $E_{kz}\cdot t$ and $\omega$:
 \begin{equation} \Delta S_A\simeq \Delta S_A(E_{kz} \cdot l, E_{kz}\cdot t, \omega). \end{equation}}.
Figure \ref{ECP_slow} and \ref{ECP_enlarged} show the time evolution of $\Delta S_A$ whose behavior is as follows:
\begin{itemize}
\item[(1)] The entanglement structure changes around $t=0$.
\item[(2)] If the subsystem size, $l$, is efficiently larger than Kibble--Zurek time, $t_{kz}$, the time evolution of $\Delta S_A$ does not depends on $l$ before $t \simeq t_{kz}+\frac{l}{2}$. When $t_{kz} \ll t < t_{kz}+\frac{l}{2}$,  $\Delta S_A$ is fitted by a linear function of $t$.
\item[(3)] $\Delta S_A$ depends on $l$ if $t>t_{kz}+\frac{l}{2}$, and $\Delta S_A$ in the window, $t \gg \frac{l}{2}+t_{kz}$, is fitted by a linear function of $l$.
\end{itemize}


\subsubsection*{Time growth}
\begin{figure}
  \begin{center}
     \includegraphics[width=25mm]{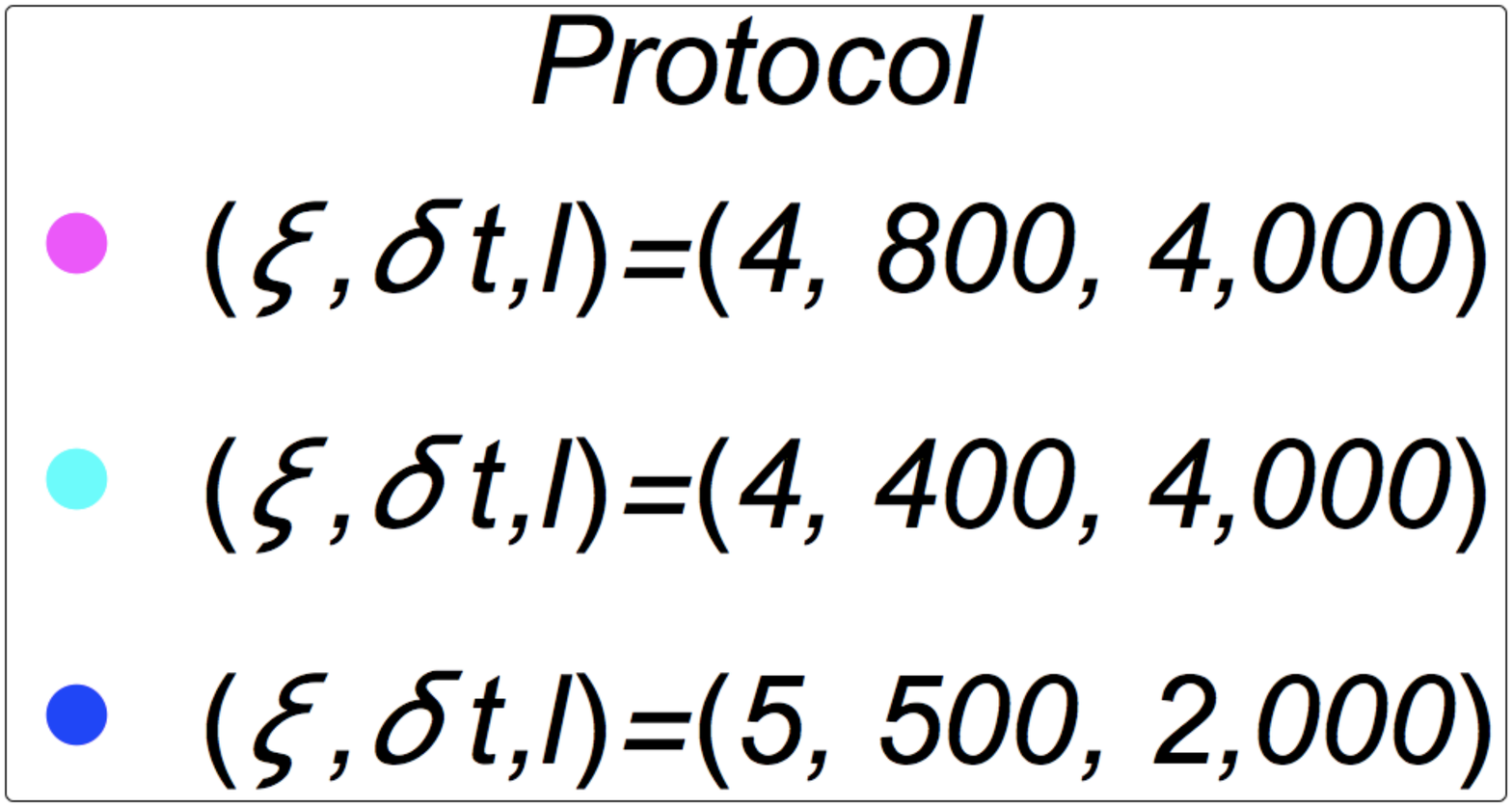}
   \includegraphics[width=55mm]{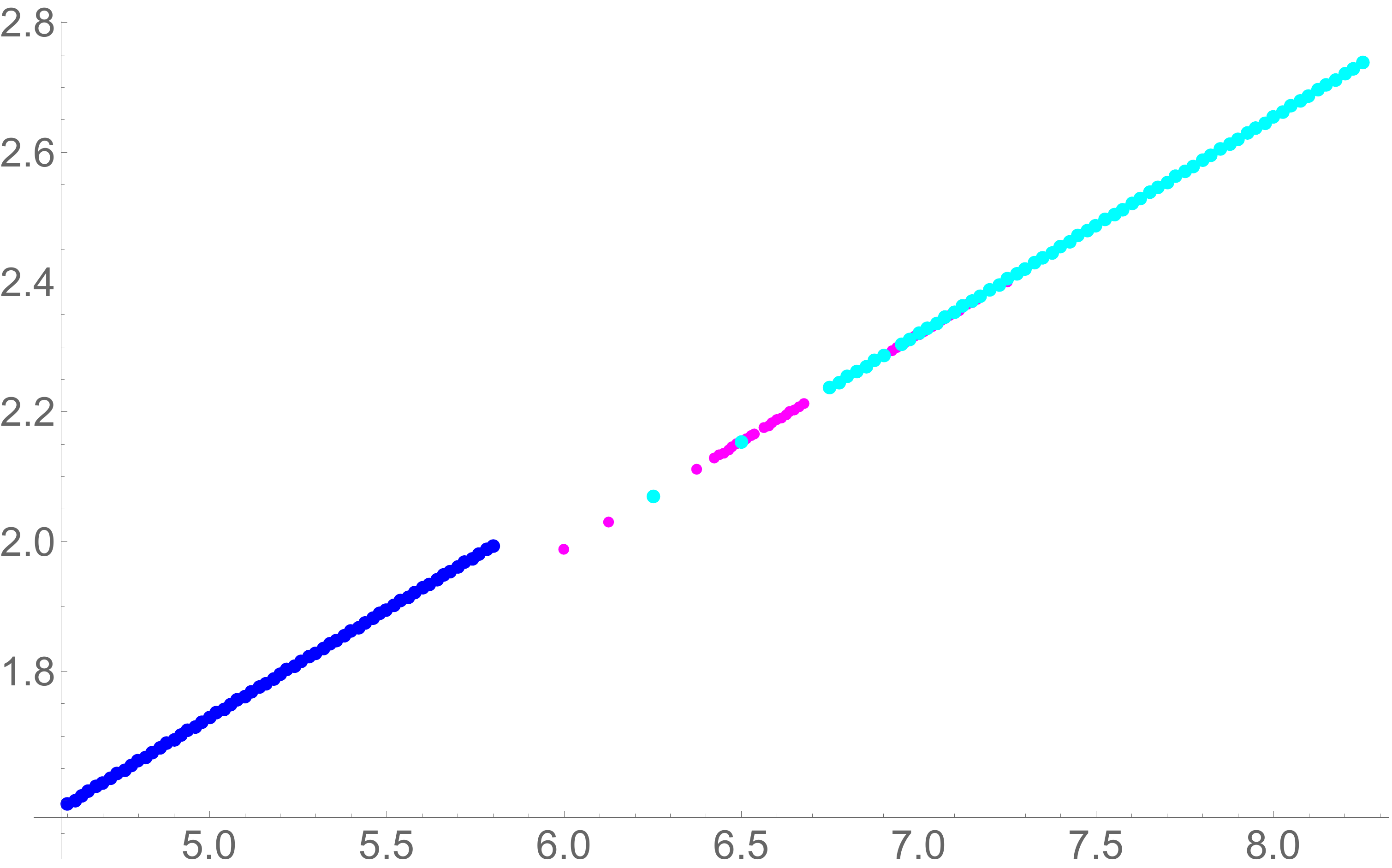}
         \put(5,5){$t\cdot E_{kz}$}
    \put(-170,105){$\Delta S_A $}
  \end{center}
  \caption{The time-dependence of $\Delta S_A$ for  $t_{kz} \ll t \ll t_{kz}+ \frac{l}{2}$.\label{ECP_slow_time_dep} }
\end{figure}

If $t_{kz} \ll t < t_{kz}+ \frac{l}{2}$, $\Delta S_A$ is a linear function of $t$ as in Figure \ref{ECP_slow_time_dep}:
\begin{equation}
\Delta S_A \sim c_1 E_{kz} t+c_2,
\end{equation}
where $c_1$ and $c_2$ are in Table \ref{t3}\footnote{The result of the protocol, $(\delta t,\xi)=(100,1)$, might strongly depend on the lattice spacing since the lattice spacing highly affects the results when $\xi$ approaches $1$.}.
Thus, $c_1$ is given by
\begin{equation}
c_1 \simeq 0.33.
\end{equation}
A parameter, $c_2$ depends on a fit range. 
The proportionality coefficient of $t$ is determined by $E_{kz}$. 
Thus, how entanglement entropy increases is determined by the scale, $E_{kz}$, which is an energy scale when adiabaticity breaks down. 

\begin{table}[htb]
  \begin{center}
        \caption{Fit results for Figure \ref{ECP_slow_time_dep}.\label{t3}}
    \begin{tabular}{|c|c|c|c|c|} \hline
     $\delta t$ & $\xi$ &$l$& Fit Result & Fit Range    \\ \hline \hline
        $100$& $1$&$2,000$ &$\Delta S_A=0.333333 E_{kz}\cdot t-0.111814 $ & $8 \le t \cdot E_{kz} \le 12$ \\ \hline
        $100$& $1$&$2,000$ &$\Delta S_A=0.333161E_{kz}\cdot t-0.109879   $ & $10 \le t \cdot E_{kz} \le 14$ \\ \hline

        $500$& $5$&$2,000$ &$\Delta S_A=0.332814  E_{kz}\cdot t-0.00621652$ & $4.6 \le t \cdot E_{kz} \le 5.8$ \\ \hline
        $500$& $5$&$2,000$ &$\Delta S_A=0.326995 E_{kz}\cdot t+0.0258795$ & $5.2 \le t \cdot E_{kz} \le 6.4$ \\ \hline

        $400$& $4$&$4,000$ &$\Delta S_A=0.333327E_{kz}\cdot t-0.0127876$ & $6.25 \le t \cdot E_{kz} \le 8.25$ \\ \hline
        $400$& $4$&$4,000$ &$\Delta S_A=0.3333E_{kz}\cdot t-0.0125858$ & $6.75 \le t \cdot E_{kz} \le 8.5$ \\ \hline
        
        $800$& $4$&$4,000$ &$\Delta S_A=0.331369 E_{kz}\cdot t-0.000153294$ & $6 \le t \cdot E_{kz} \le 7.25$ \\ \hline
        $800$& $4$&$4,000$ &$\Delta S_A=0.331066E_{kz}\cdot t+0.00206931$ & $6.5 \le t \cdot E_{kz} \le 7.125$ \\ \hline
  \end{tabular}
  \end{center}
  \end{table}

\subsubsection*{Late time behavior}
\begin{figure}
\begin{minipage}{0.7\hsize}
  \begin{center}
     \includegraphics[width=25mm]{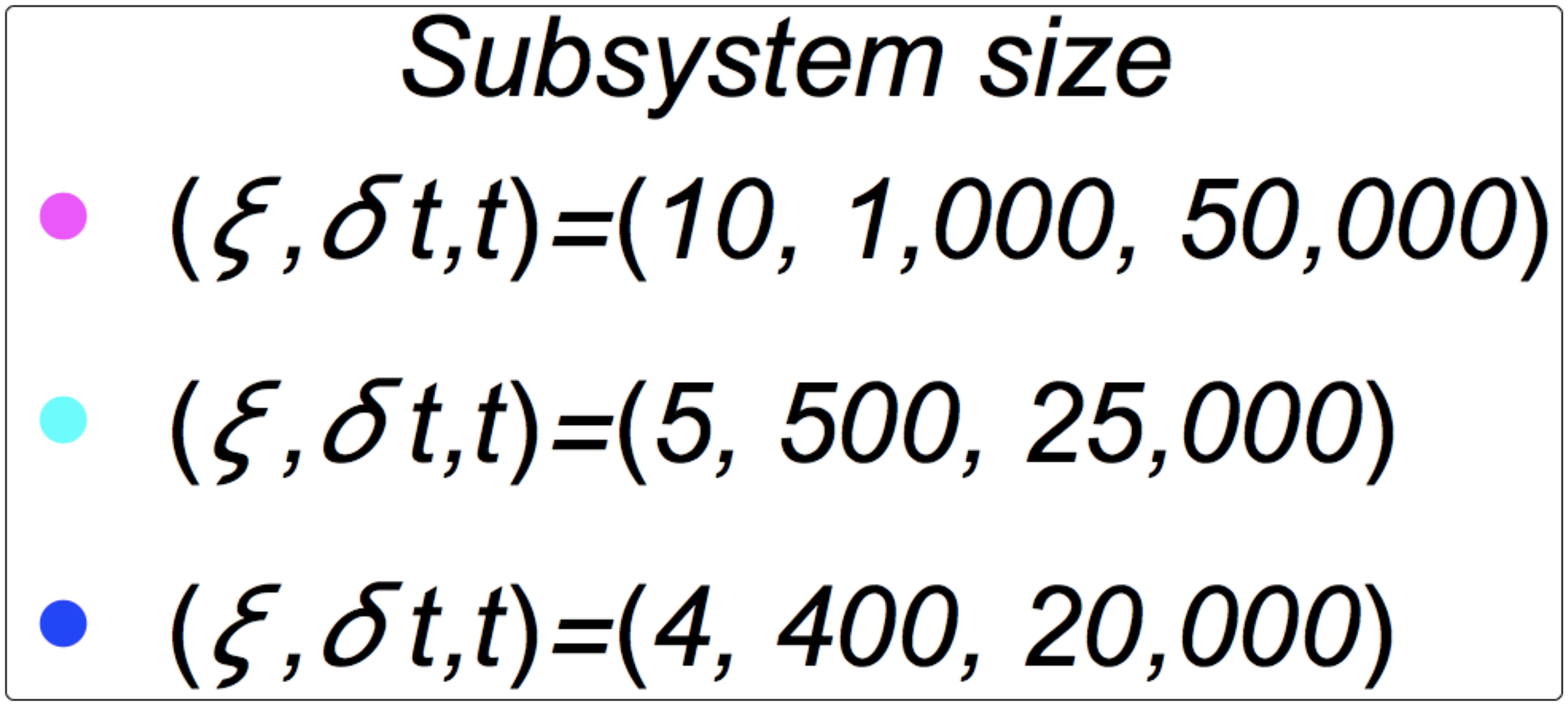}
   \includegraphics[width=65mm]{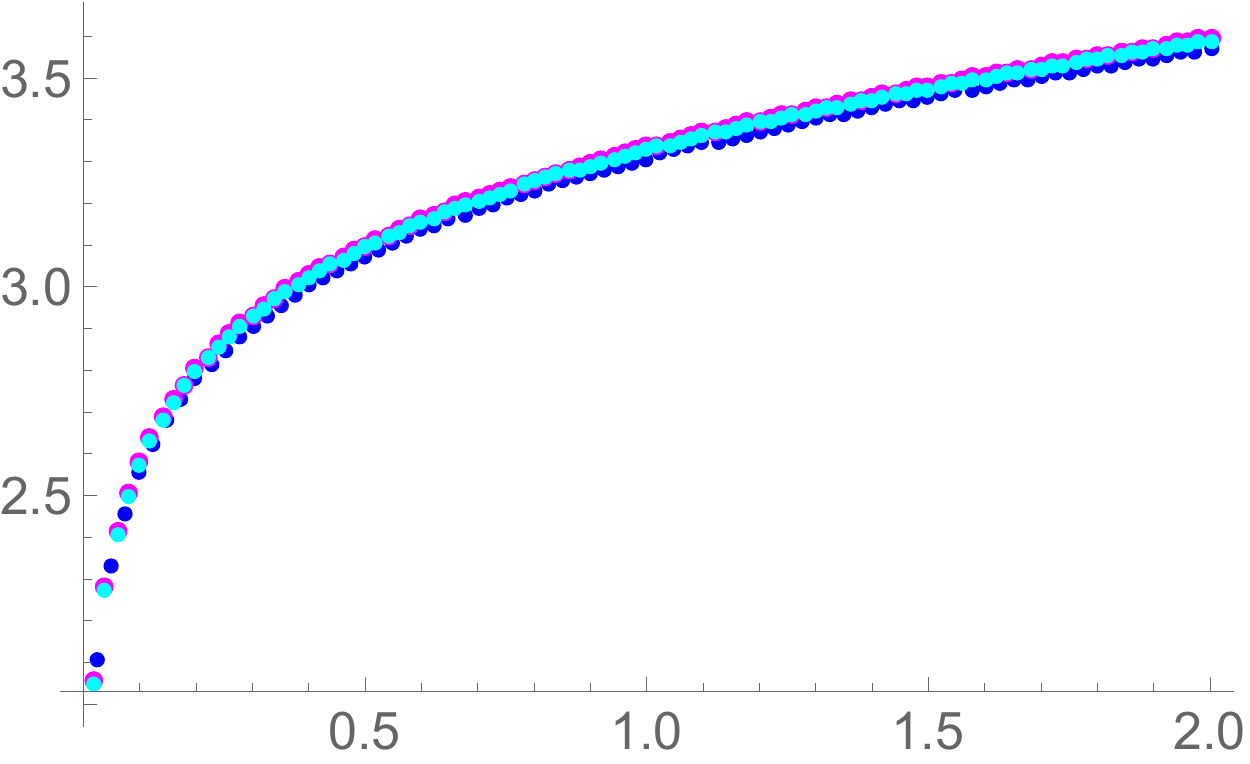}
         \put(5,5){$l\cdot E_{kz}$}
    \put(-190,120){$\Delta S_A $}
  \end{center}
\end{minipage}
  \caption{The size-dependence of $\Delta S_A $. We can see that  $\Delta S_A $ in the late time grows linearly on time, and three results are on top of each other.\label{volume_xi4_dt800_t40000} }
\end{figure}

Figure \ref{volume_xi4_dt800_t40000} shows the size-dependence of $\Delta S_A$ in the late time, $t \gg t_{kz}+\frac{l}{2}$. $\Delta S_A$ is fitted by
\begin{equation}
\Delta S_A \sim d_1 E_{kz} l+d_2,
\end{equation}
where $d_1$ and $d_2$ are in Table \ref{VIta}.
It shows that
\begin{equation}
d_1 \simeq  0.16^{+0.01}_{-0.00},
\end{equation}
and $d_2$ depends on the fit range. 
$\Delta S_A$ is proportional to the subsystem size, $l$, whose coefficient is set by $E_{kz}$. Thus, the effective temperature is given by $T_{eff} \sim E_{kz}$. 
\begin{table}[htb]
  \begin{center}
      \caption{Fit results for Figure \ref{volume_xi4_dt800_t40000}. \label{VIta}}
    \begin{tabular}{|c|c|c|c|c|} \hline
     time $t$  & $\xi$ &$dt$& Fit Result & Fit Range    \\ \hline \hline
$20,000$ & $4$& $400$ &$\Delta S_A=0.167555 E_{kz}\cdot l+3.26166  $ & $3 \le  l\cdot E_{kz} \le 5$ \\ \hline  
$20,000$ & $4$& $400$ &$\Delta S_A=0.164994 E_{kz}\cdot l+3.27151 $ & $3 \le  l\cdot E_{kz} \le \frac{25}{4}$ \\ \hline  
$40,000$ & $4$& $800$ &$\Delta S_A= 0.168396E_{kz}\cdot l+3.497$ & $3 \le  l\cdot E_{kz}  \le 5$ \\ \hline
$40,000$ & $4$& $800$ &$\Delta S_A= 0.166204 E_{kz}\cdot l+3.5049$ & $3 \le  l\cdot E_{kz}  \le 6$ \\ \hline
$25,000$ & $5$& $500$ &$\Delta S_A= 0.16115 E_{kz}\cdot l+ 3.31292$ & $5 \le l\cdot E_{kz}  \le 10$ \\ \hline    
\end{tabular}
\end{center}
\end{table}
~~\\




\subsection{Physical interpretation}

\begin{figure}[htbp]
  \begin{center}
  \includegraphics[width=10cm]{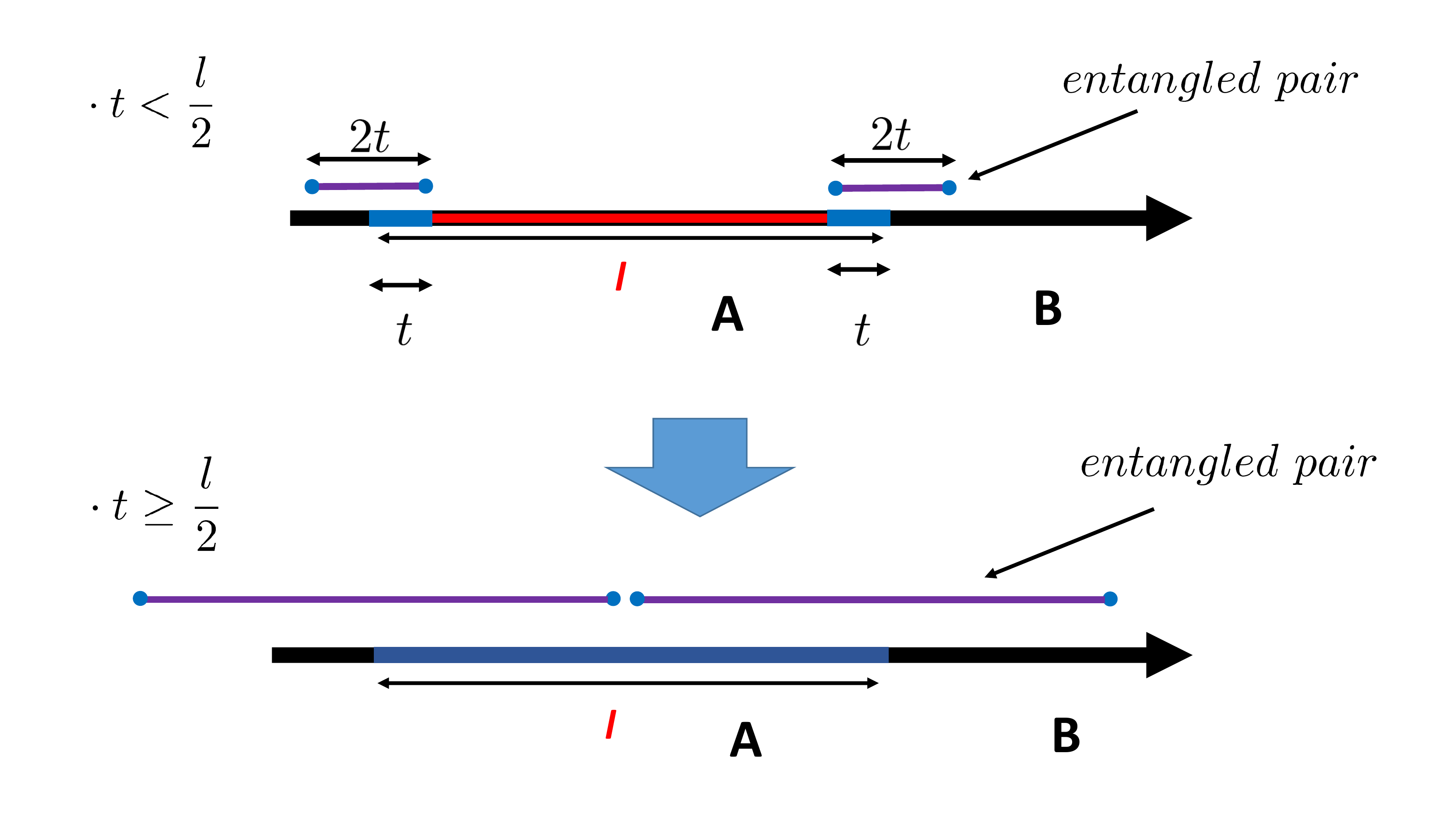}
\caption{A picture of entangled pair.\label{pep}}
  \end{center}
\end{figure}

We interpret the time evolution of $\Delta S_A$ in both limits in the ECP-type potential in terms of entangled particles because the time evolution of $\Delta S_A$ is similar to the time evolution in the sudden global quenches \cite{ca1} .
The time evolution of $\Delta S_A$ in \cite{ca1} is explained in terms of relativistic propagation of quasi-particles as in Figure \ref{pep}.
In sudden quenches, entangled pairs are created everywhere when parameters in a Hamiltonian are suddenly changed. Here, we assume that the parameters change at $t=0$. 
In two-dimensional quantum field theory, the pair is constructed of two entangled particles which propagate in right and left directions with the speed of light.  
Here, the total system is divided into $A$ and $B$. 
If a particle of the pair is  in $A$, and the other is in $B$, the entanglement between entangled particles contributes to $\Delta S_A$. 
Therefore, if the subsystem size is $l$, entangled particles in the blue region can contribute to $\Delta S_A$, before  $t \le \frac{l}{2}$. The number of pairs contributing to $\Delta S_A$ is proportional to $t$. Thus, $\Delta S_A$ is proportional to $t$.
After $t=\frac{l}{2}$, whole region in $A$ is entangled with $B$ because the distance between the two entangled particles is larger than the subsystem size. $\Delta S_A$ is proportional to the subsystem size.

In the ECP-type mass, the mass profile in the very early time, $t \ll -\delta t$, changes slowly. Therefore, we expect the entanglement structure of state to change adiabatically.
The structure is expected to change drastically because the profile in the window, $-\delta t < t < \delta t$, changes drastically in the fast ECP-type quench\footnote{ In this paper, we assume that $\delta t$ in the fast limit is not so large , $\delta t \sim \mathcal{O}(10)$.}. 
Here, we assume that entangled pairs whose velocities, $v_k$, depend on momenta, $k$, as in \cite{MM1, ln1},  are created around $t=0$. Since $\xi$ is large enough, the maximal velocity in $0 \le t$ will be the speed of light, $v^{max}_k \simeq \pm 1$.  The time evolution of $\Delta S_A$ except for the late-time logarithmic growth can be explained by the relativistic propagation of entangled particles. 
The late-time logarithmic growth of $\Delta S_A$ in $t\gg \delta t$ will be explained by the contribution from the particles with small velocity. 
Since the mass potential in $t\gg \delta t$ changes slowly, a dispersion relation $\omega_k$ can be approximated by
\begin{equation}
\omega_k \simeq \sqrt{4\sin^2{\left(\frac{k}{2}\right)}+m^2} \label{eq:dispersion},
\end{equation}
where the velocity, $v_k$, can be defined by
\begin{equation}
v_k=\frac{d\omega_k}{dk}.
\end{equation}
Since the velocity for $k\simeq \pm\pi, 0$ is small, these slow-moving particles contribute to the late-time $\Delta S_A$. 
However, the particles with $k\simeq \pm \pi$ suffer from lattice artifacts.
In this paper, unfortunately, we do not find whether the logarithmic growth comes from the physical mode, $k=0$.
This dispersion \eqref{eq:dispersion} is strongly related to how to discretize the space, $k^2 \rightarrow 4\sin{\left(\frac{k^2}{2}\right)}$, and, unphysical modes might vanish when we employ
another discretization scheme like the symmetric difference instead of the forward one.

The time evolution of $\Delta S_A$ even in the slow ECP-type mass can be interpreted in terms of the momentum-dependent propagation of entangled particles created around $t=t_{kz}$ . Therefore, we expect the entangled particles to be created when adiabaticity breaks down. $\Delta S_A$ in late time is proportional to $l$ due to the contribution of entangled particles.


\subsection{Time evolution of $\Delta S_A$ in CCP-type protocol}
Here, we study $\Delta S_A$ at the fast and the slow limits in the CCP-type potential. We study the time evolution and the $l$-dependence of $\Delta S_A$ in the CCP-type potential.
\subsubsection{Fast limit}
\begin{figure}[htbp]
\begin{minipage}{0.1\hsize}
 \begin{center}
  \includegraphics[clip,width=1.5cm]{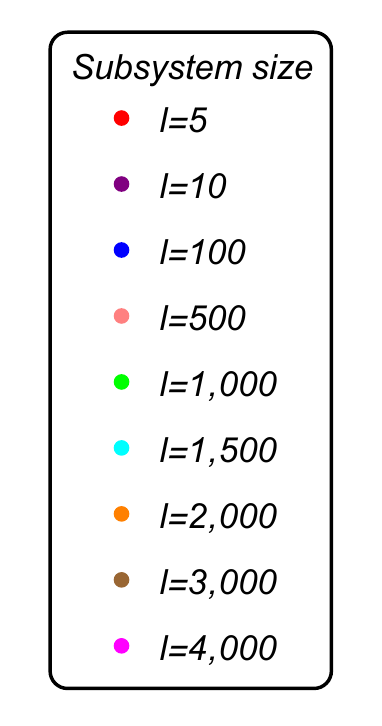}
      \end{center}
  \end{minipage}
 \begin{minipage}{0.25\hsize}
 \begin{center}
    \includegraphics[clip,width=4cm]{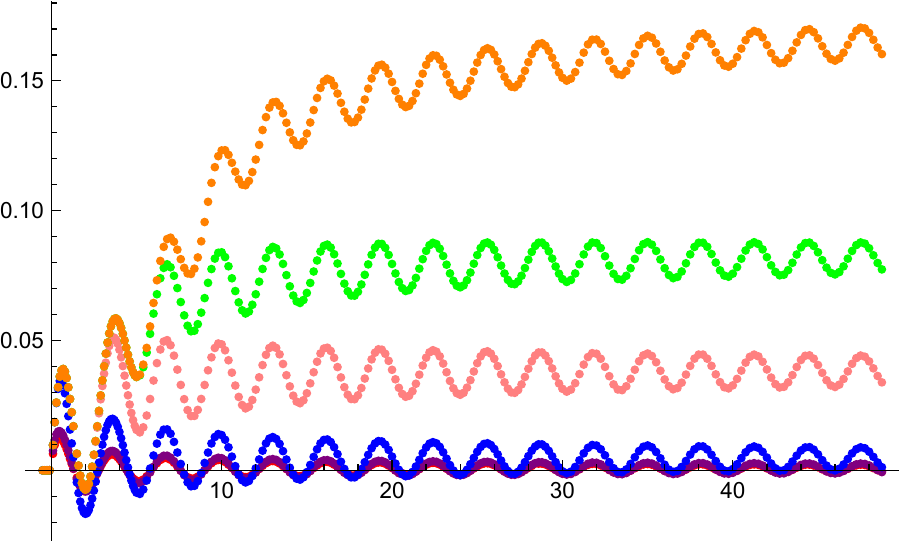}
       \put(-10,-5){$t/\xi$}
    \put(-110,70){$\Delta S_A$}
   \end{center}
  \end{minipage}
 \begin{minipage}{0.25\hsize}
 \begin{center}
     \includegraphics[clip,width=4cm]{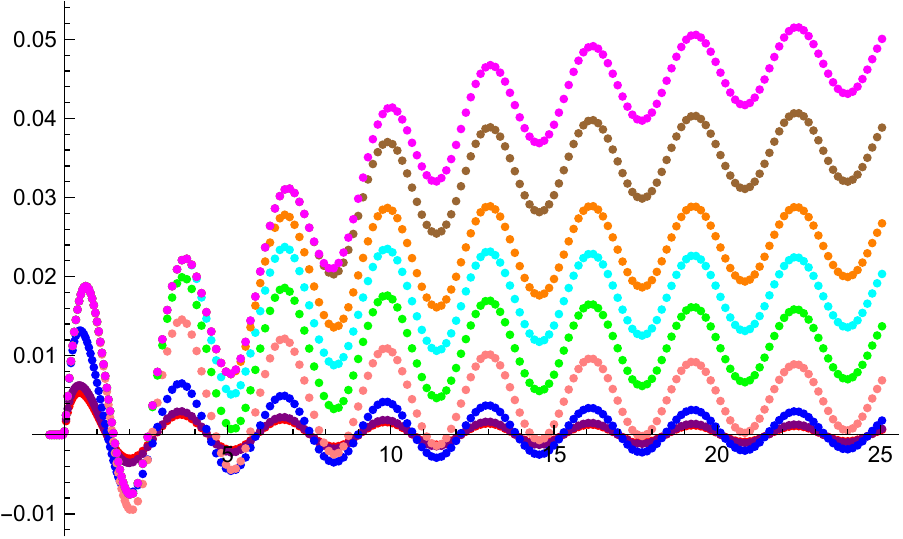}
       \put(-10,-5){$t/\xi$}
    \put(-110,70){$\Delta S_A$}
  \end{center}
  \end{minipage}
   \begin{minipage}{0.25\hsize}
 \begin{center}
    \includegraphics[clip,width=4cm]{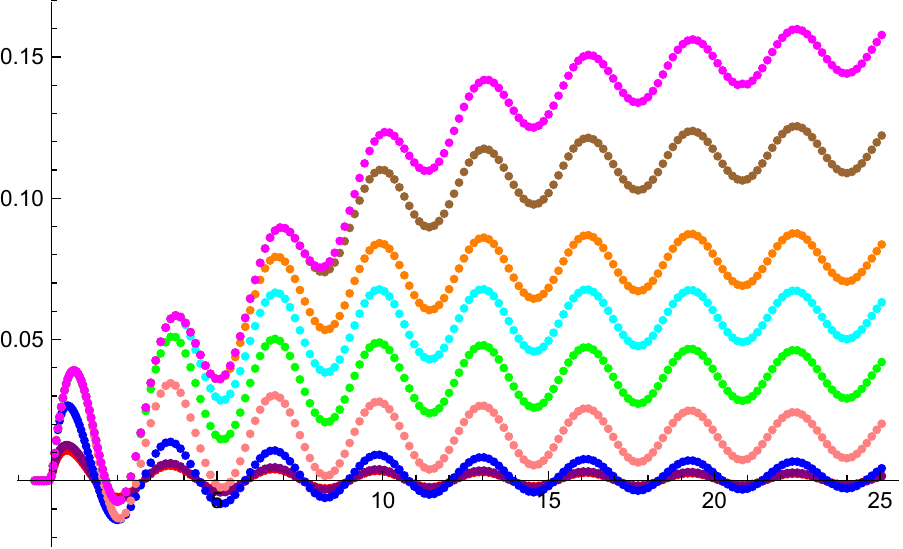}
        \put(-10,-5){$t/\xi$}
    \put(-110,70){$\Delta S_A$}
  \end{center}
  \end{minipage}
      \caption{The $t$-dependence of $\Delta S_A$. The left, middle and right panels show $\Delta S_A$ with $(\delta t=5, \xi=100)$, $(\delta t=5, \xi=200)$ and $(\delta t=10, \xi=200)$, respectively.  }
      \label{fig:CCP_tdependence}
\end{figure}

\begin{figure}[htbp]
  \begin{center}
    \includegraphics[clip,width=3.5cm]{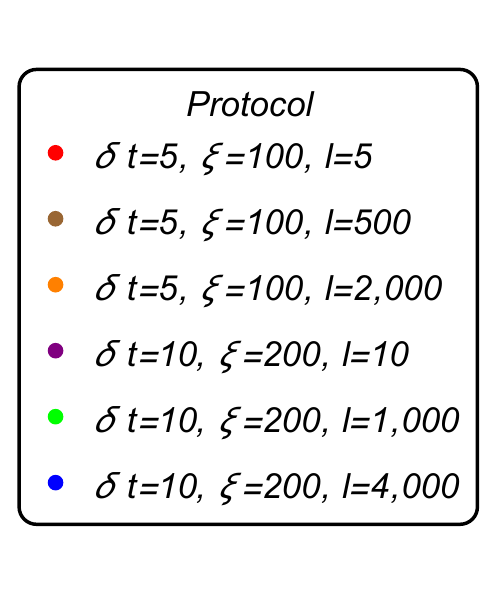}
    \includegraphics[clip,width=7.0cm]{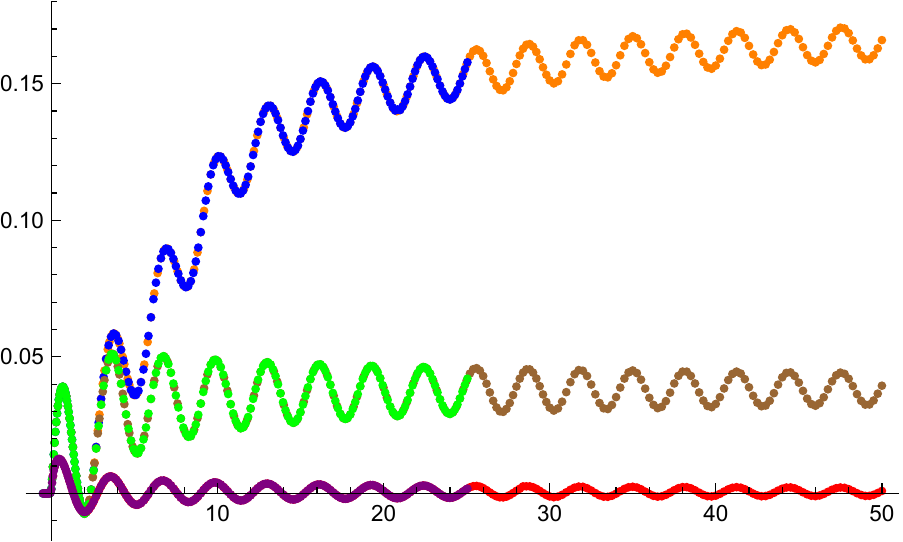}
         \put(5,8){$t/\xi$}
    \put(-220,115){$\Delta S_A$}
     \end{center}
    \caption{The $t$-dependence of $\Delta S_A$ with different parameters. As the time evolution in the ECP-type potential, the data points for $\Delta S_A$ with the same value of $\frac{l}{\xi}$, $\frac{t}{\xi}$ and $\omega$ lie on the same curves. Therefore, this panel shows the scaling law in (\ref{sce}).  }
     \label{fig:scalinglaw}
\end{figure}

Figure \ref{fig:CCP_tdependence} shows the time-dependence of $\Delta S_A$. 
If the parameters $(\xi, \delta t, l)$ are
 much larger than $1$, $\Delta S_A$ does not depend on the UV regularization. 
As in Figure \ref{fig:scalinglaw}, $\Delta S_A$ can be written by the scaling function of $\frac{t}{\xi}, \frac{l}{\xi}$ and $\omega$ in (\ref{sce}).
The time evolution of $\Delta S_A$ in the fast CCP-type potential has the following properties:
\begin{itemize}
\item[(1)] The structure of quantum entanglement starts to change around $t=0$.
\item[(2)] $\Delta S_A$ is minimized around $t=2\xi$, and oscillates after that time.
\item[(3)] If $l$ is sufficiently large, $\Delta S_A$ is independent of the subsystem size before $t \simeq \frac{l}{2}$. 
After $t \simeq \frac{l}{2}$, $\Delta S_A$ depends on the subsystem size, and oscillates around the constant which depends on $l$. 
\end{itemize}

\subsubsection*{Minimum of $\Delta S_A$}
\begin{figure}[htbp]
   \begin{center}
    \includegraphics[clip,width=4cm]{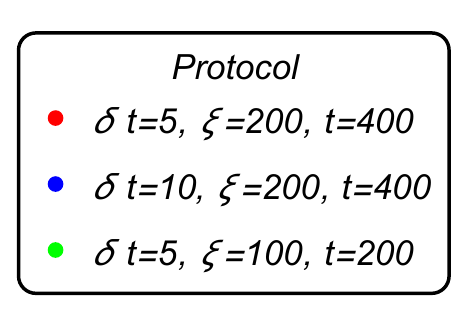}
    \includegraphics[clip,width=7.0cm]{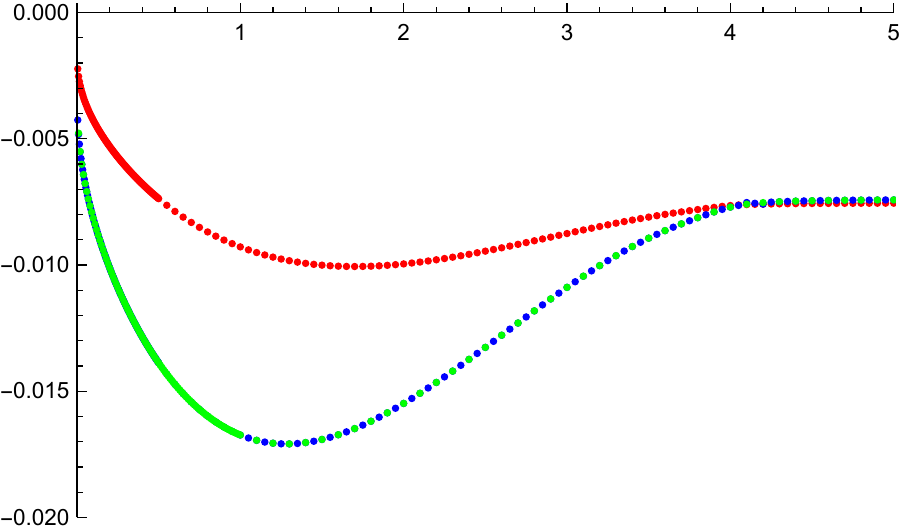}  
    \put(5,110){$l/\xi$}
    \put(-220,100){$\Delta S_A$}
    \end{center}
      \caption{The $l$-dependence of $\Delta S_A$ at $t=2\xi$.  \label{fig:plot_t2xi}}
\end{figure}

$\Delta S_A$ is minimized around $t \simeq 2 \xi$. 
The size-dependence of $\Delta S_A$ at $t=2\xi$ is shown in Figure \ref{fig:plot_t2xi}.
If $l$ is smaller than 4$\xi$, $\Delta S_A\bigl(\frac{l}{\xi}\bigr)$ depends on only $\omega$. If l is larger than $4\xi$, the entropy is a negative constant.

\subsubsection*{Oscillation of $\Delta S_A$}
\begin{figure}[htbp]
 \begin{minipage}{0.21\hsize}
  \begin{center}
    \includegraphics[clip,width=3cm]{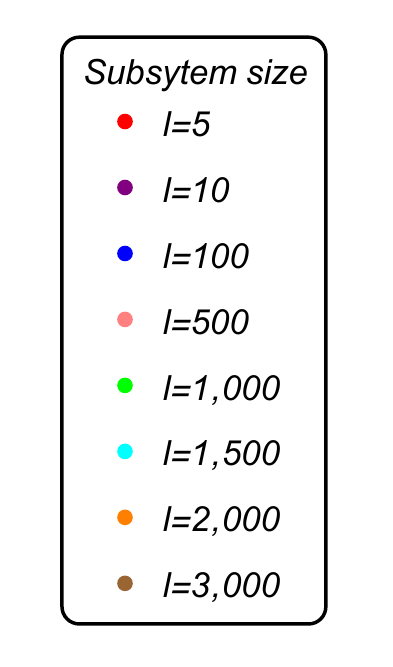}
\end{center}
 \end{minipage}
 \begin{minipage}{0.37\hsize}
  \begin{center}
     \includegraphics[clip,width=6.0cm]{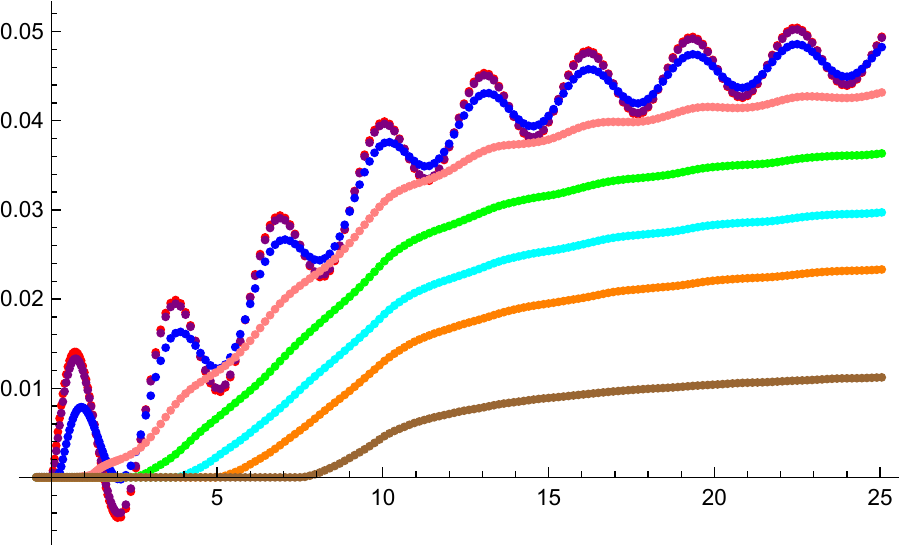}
     \put(-5, -5){$t/\xi$}
    \put(-190,110){$\delta S_A=\Delta S_A(l=4,000)-\Delta S_A(l)$}
\end{center}
 \end{minipage}
 \begin{minipage}{0.37\hsize}
  \begin{center}
    \includegraphics[clip,width=6.0cm]{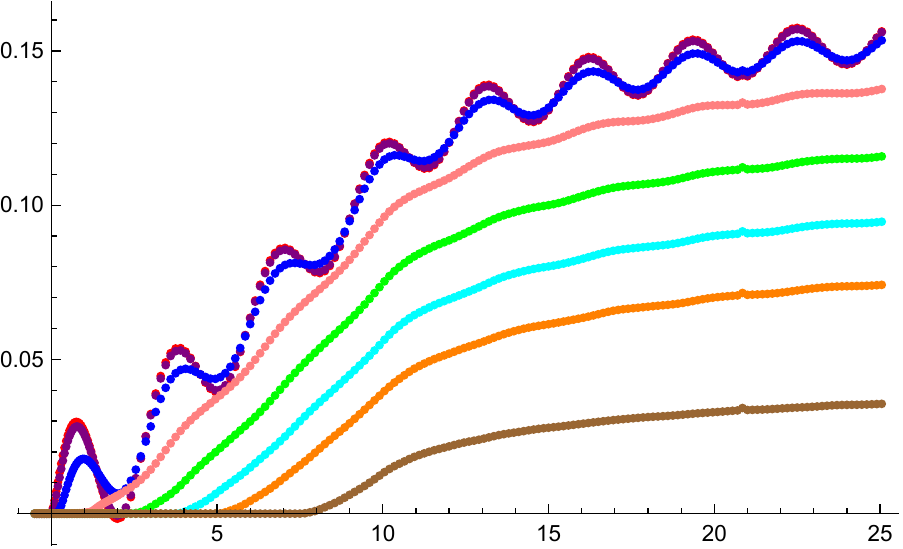}
       \put(-5, -10){$t/\xi$}
    \put(-190,110){$\delta S_A=\Delta S_A(l=4,000)-\Delta S_A(l)$}
     \end{center}
 \end{minipage}
\caption{The $t$-dependence of $\delta S_A$. The left and right panels show $\delta S_A$ with $(\delta t=5, \xi=200)$ and $\delta S_A$ with $(\delta t=10, \xi=200)$, respectively.\label{fig:dCCP_difference}}
\end{figure}

Figure \ref{fig:dCCP_difference} shows $\delta S_A$ defined by subtracting $\Delta S_A$ for $l<4,000$ from $\Delta S_A$ for $l=4,000$. The amplitude of oscillation in $\delta S_A$ becomes smaller as $l$ becomes larger.
Therefore, the amplitude of oscillation in $\Delta S_A$ appears to be independent of $l$ when $l$ is larger than $\xi$.  At late time, the periodicity of oscillation approaches $\pi \xi$. 

\subsubsection{Subsystem size dependence}
\begin{figure}[htbp]
 \begin{minipage}{0.18\hsize}
  \begin{center}
    \includegraphics[clip,width=2.5cm]{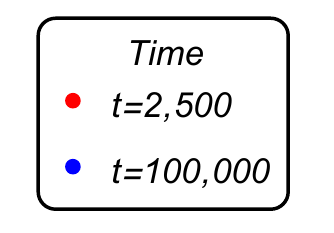}
\end{center}
 \end{minipage}
 \begin{minipage}{0.39\hsize}
  \begin{center}
   \includegraphics[width=65mm]{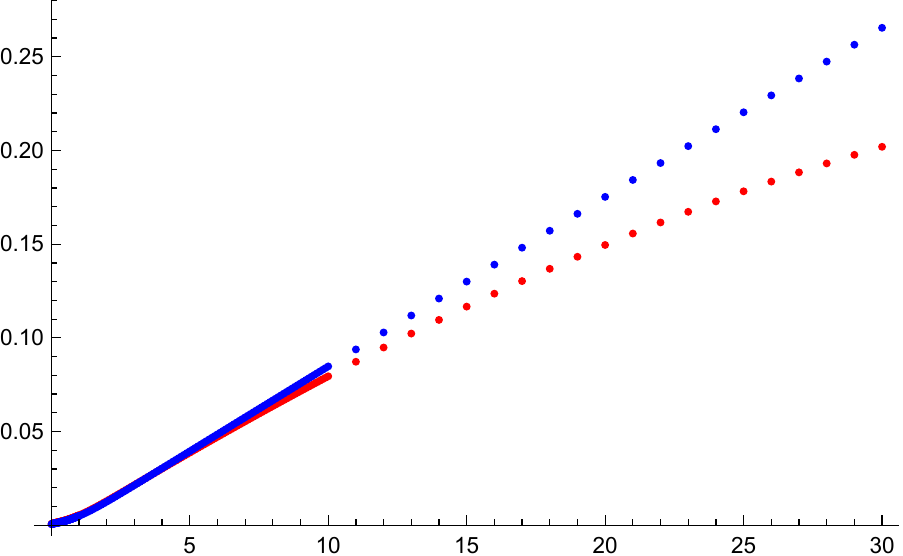}
       \put(-5, -10){$l/\xi$}
    \put(-180,115){$\Delta S_A$}
  \end{center}
 \end{minipage}
 \begin{minipage}{0.39\hsize}
  \begin{center}
   \includegraphics[width=65mm]{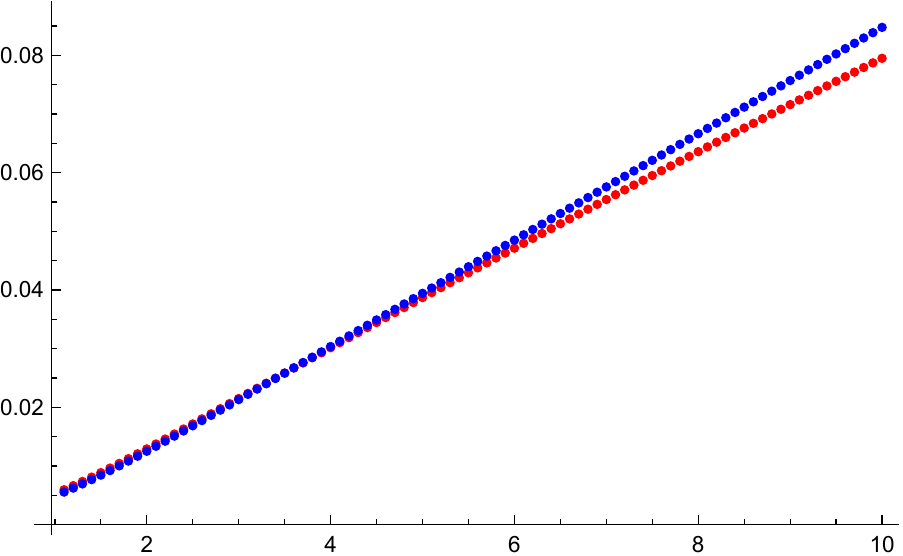}
       \put(-5, -10){$l/\xi$}
    \put(-180,115){$\Delta S_A$}
  \end{center}
 \end{minipage}
\caption{The plot of $\Delta S_A$ with $(\xi, \delta t)=(100, 5)$.   The left and right panels are for $\frac{1}{100}\le \frac{l}{\xi} \le 30$ and for $1 \le \frac{l}{\xi} \le 10$, respectively. \label{fig:four}}
\end{figure}

\begin{figure}[htbp]
 \begin{minipage}{0.21\hsize}
  \begin{center}
    \includegraphics[clip,width=3cm]{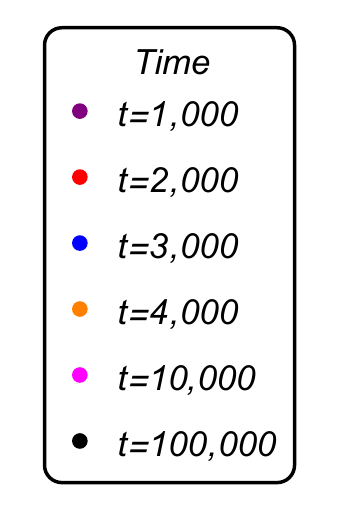}
\end{center}
 \end{minipage}
 \begin{minipage}{0.37\hsize}
  \begin{center}
   \includegraphics[width=60mm]{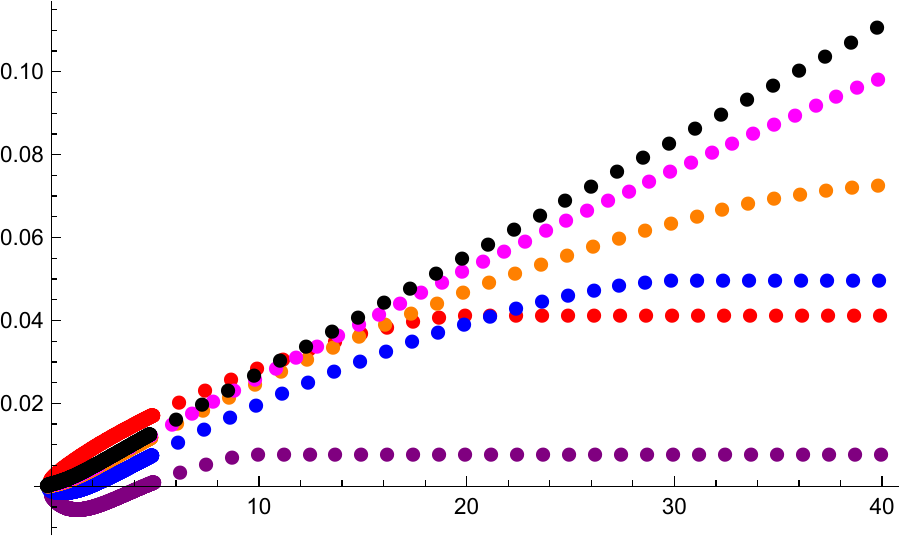}
    \put(-5, -10){$l/\xi$}
    \put(-180,105){$\Delta S_A$}
  \end{center}
 \end{minipage}
 \begin{minipage}{0.37\hsize}
  \begin{center}
   \includegraphics[width=60mm]{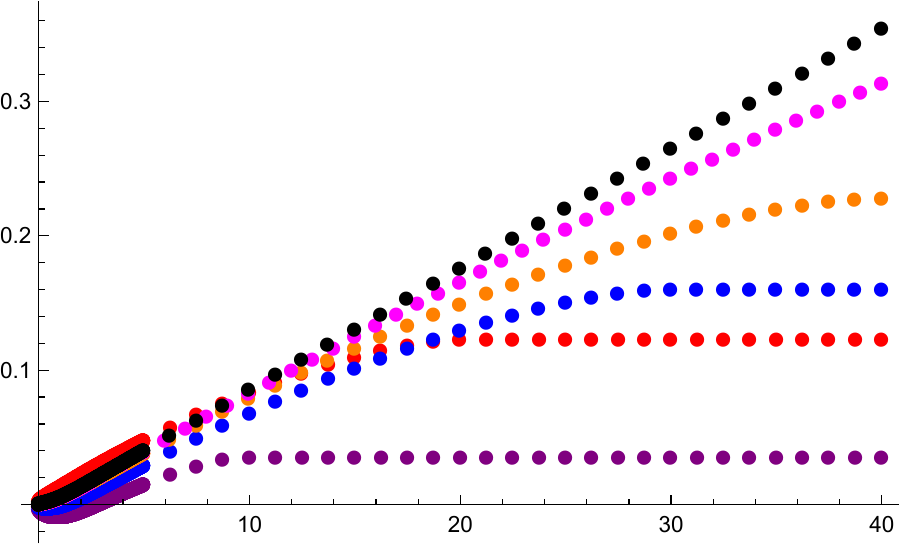}
     \put(-5, -10){$l/\xi$}
    \put(-180,105){$\Delta S_A$}
  \end{center}
 \end{minipage}
\caption{The $l$-dependence of $\Delta S_A$. The left  and right panels show $\Delta S_A$ with $(\xi, \delta t)=(200, 5)$ and $\Delta S_A$ with $(\xi, \delta t)=(200, 10)$, respectively.  Since $\Delta S_A$ in the CCP-type protocol oscillates, the values of $\Delta S_A$ with different $t$ are not the same for small $l$. \label{fig:CCPVL3}}
\end{figure}

\begin{figure}[htbp]
  \begin{center}
    \includegraphics[clip,width=7.0cm]{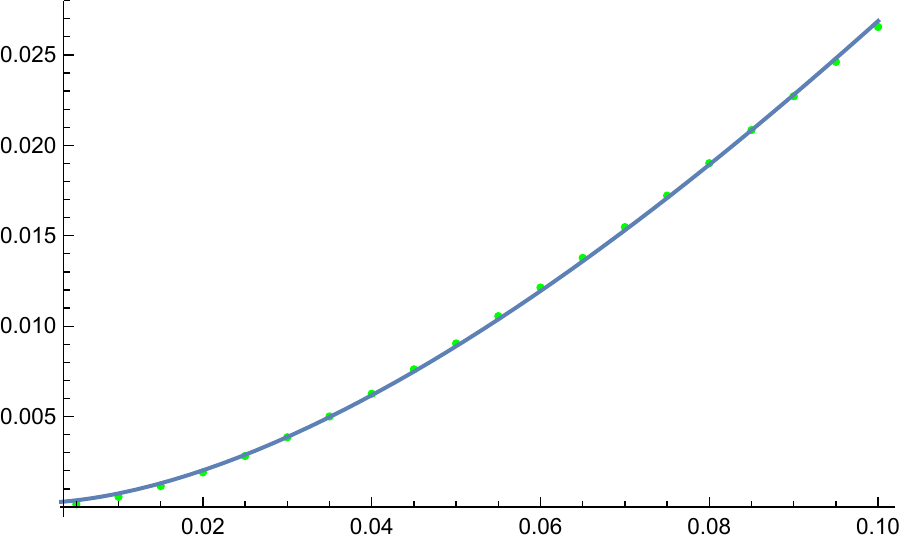}
       \put(5,5){$\omega$}
    \put(-200,125){$a$}
    \caption{The $\omega$-dependence of the coefficient $a$ in $\Delta S_A\simeq al/\xi +b$.   The curve is a nonlinear fitting in $1/200\le\omega\le20/200$.\label{fig:CCPVL5}}
  \end{center}
\end{figure}

The $l$-dependence of $\Delta S_A$  is shown in Figure~\ref{fig:four} and \ref{fig:CCPVL3}. 
 In Figure~\ref{fig:CCPVL3}, $\Delta S_A$ does not depend on $l$ when $l$ is larger than $2t$.
 In the window, $\xi\ll l/2\ll t$, $\Delta S_A$ is fitted by a linear function:
\begin{equation}
\Delta S_A \simeq a \frac{l}{\xi} + b,
\end{equation}
where $a$ and $b$ are in Table~\ref{tab:fCCP1}, which shows that $a$ and $b$ depend on $\omega$. 
Figure~\ref{fig:CCPVL5} shows how $a$ depends on $\omega$. As $\omega (\ll 1)$ decreases, $a$ decreases monotonically and is fitted by a nonlinear function of $\omega$:
\begin{equation}
a=g_1 \omega^2\log{\left[\frac{1}{\omega}\right]}+g_2,\label{fc1}
\end{equation}
where $g_1$ and $g_2$ are in Table~\ref{tab:fCCP2}. We estimate $g_1 \simeq 1.2$. Table~\ref{tab:fCCP2} shows that $g_2$ depends strongly on the fit range and becomes smaller when the upper bound of the fit range becomes smaller. Therefore, we expect $a$ to vanish around $\omega=0$.  
The effective temperature in (\ref{teff}) depends on $\omega$ since the effective temperature, $T_{eff}$ is given by
\begin{align}
\frac{a}{\xi} \sim T_{eff}.
\label{T_eff}
\end{align}

  \begin{table}[htb]
  \begin{center}
    \caption{Fit results for Figure \ref{fig:four} and \ref{fig:CCPVL3}.\label{tab:fCCP1}}
    \begin{tabular}{|c|c|c|c|c|} \hline
  $\delta t$ & $\xi$ &$t$& Fit Result & Fit Range    \\ \hline \hline
     $5$& $100$&$2500$ &$\Delta S_A=-0.00326117 + 0.00834283 \frac{l}{\xi}$ & $2 \le \frac{l}{\xi} \le 10$ \\ 
     \hline
   $5$& $100$&$100,000$ &$\Delta S_A=-0.005807 + 0.00905522 \frac{l}{\xi}$ & $2 \le \frac{l}{\xi} \le 10$ \\ 
     \hline
     $5$& $100$&$100,000$ &$\Delta S_A=-0.00567712 + 0.00904089 \frac{l}{\xi}$ & $2 \le \frac{l}{\xi} \le 30$ \\ 
     \hline
      $5$& $200$&$10,000$ &$\Delta S_A=-0.00146424 + 0.00273676    \frac{l}{\xi}$ & $2 \le \frac{l}{\xi} \le 10$ \\ \hline
         $5$& $200$&$100,000$ &$\Delta S_A=-0.00136572 + 0.00281051   \frac{l}{\xi}$ & $2 \le \frac{l}{\xi}\le 10$ \\ \hline   
      $5$& $200$&$100,000$ &$\Delta S_A=-0.00135296 + 0.00280731    \frac{l}{\xi}$ & $2 \le \frac{l}{\xi} \le 40$ \\ \hline
         $10$& $200$&$10,000$ &$\Delta S_A=-0.00465824 + 0.00878667    \frac{l}{\xi}$ & $2 \le \frac{l}{\xi} \le 10$ \\ \hline
        $10$& $200$&$100,000$ &$\Delta S_A=-0.00450837 + 0.00902687    \frac{l}{\xi}$ & $2 \le \frac{l}{\xi} \le 10$ \\ \hline
       $10$& $200$&$100,000$ &$\Delta S_A=-0.00440558 + 0.00899895    \frac{l}{\xi}$ & $2 \le \frac{l}{\xi} \le 40$ \\ \hline    
            \end{tabular}
  \end{center}
  \end{table}
  
\begin{table}[htb]
  \begin{center}
    \caption{Fit results for Figure \ref{fig:CCPVL5}.\label{tab:fCCP2}}
    \begin{tabular}{|c|c|c|c|c|} \hline
      $\xi$ &$t$& Fit Result & Fit Range    \\ \hline \hline  
    $200$&$100,000$ &$a=0.000217153 - 1.1576 \omega^2 \textrm{log}[\omega]$ & $1/200 \le \omega \le 20/200$ \\ \hline
        $200$&$100,000$ &$a=0.0000177701 - 1.20953 \omega^2 \textrm{log}[\omega]$ & $1/200 \le \omega \le 10/200$ \\ \hline      
            \end{tabular}
  \end{center}
  \end{table}


\subsubsection{Slow limit}
Here, we study the time evolution of $\Delta S_A$ {in the slow CCP-type potential.
Its time evolution is shown in Figure~\ref{fig:CCPslow_tdependence}. 
Figure~\ref{fig:scalinglaw_CCPslow} shows that $\Delta S_A$ with the large parameters is independent of the lattice spacing and is a function of $\frac{l}{\xi_{kz}}$, $\frac{t}{\xi_{kz}}$ and $\omega$: 
\begin{equation}
\Delta S_A \simeq \Delta S_A\biggl(\frac{l}{\xi_{kz}}, \frac{t}{\xi_{kz}}, \omega \biggr).\label{sce2}
\end{equation}
As in Figure~\ref{fig:CCPslow_tdependence}, the time evolution of $\Delta S_A$ has the following properties: 
\begin{itemize}
\item[(1)] If the subsystem size is larger than the initial correlation length, $\Delta S_A$ in the early time is independent of the subsystem size. 
\item[(2)] Before $t \simeq \xi_{kz}$, $\Delta S_A$ monotonically increases. From $t\simeq\xi_{kz}$ to $t\simeq2\xi_{kz}$, $\Delta S_A$ monotonically decreases.  After $t \simeq 2\xi_{kz}$, $\Delta S_A$ oscillates.
\item[(3)] $\Delta S_A$ in  late time depends on the subsystem size. $\Delta S_A$ with small $l$ decreases, and $\Delta S_A$ with large $l$ increases  when $t$  increases.
\end{itemize}

\subsubsection*{Local minimum of $\Delta S_A$}
$\Delta S_A$ is locally minimized around $t \simeq 2 \xi_{kz}$ as explained in (2). The $l$-dependence of $\Delta S_A$  at $t=2 \xi_{kz}$ is plotted in Figure~\ref{fig:CCPslow_ldependence_t2xikz}. 
After $t\simeq2\xi_{kz}$, $\Delta S_A$ oscillates. Here, $\xi_{kz}$ is an effective correlation length when adiabaticity breaks down. Thus, $\Delta S_A$ oscillates after the scale characterizes the time evolution: $2\xi$ in the fast limit and $2\xi_{kz}$ in the slow limit. 

\begin{figure}[htbp]
 \begin{minipage}{0.1\hsize}
 \begin{center}
  \includegraphics[clip,width=1.5cm]{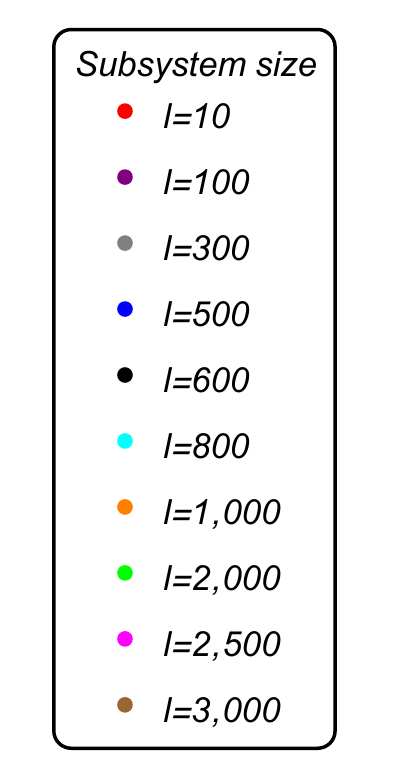}
  \end{center}
  \end{minipage}
  \begin{minipage}{0.25\hsize}
 \begin{center}
    \includegraphics[clip,width=4cm]{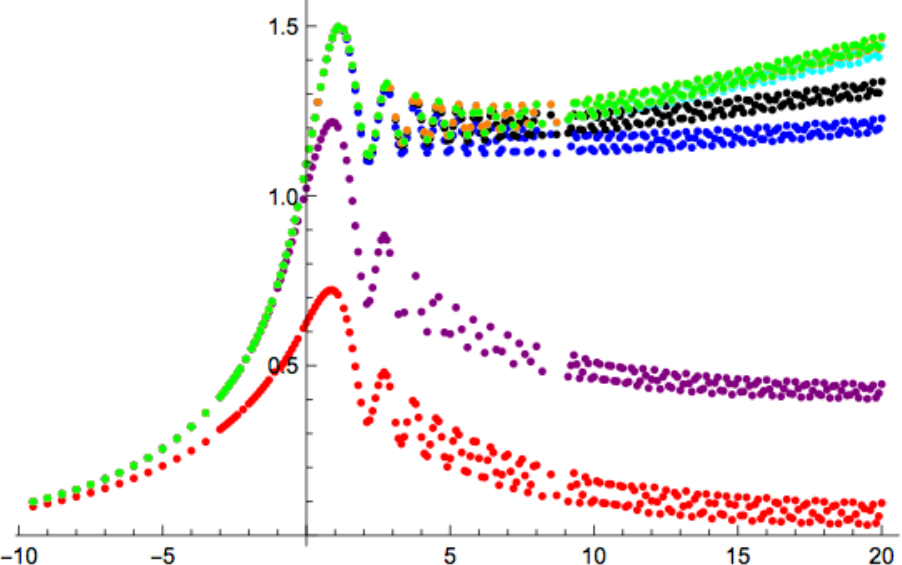}
   \put(-10,-10){$t/\xi_{kz}$}
   \put(-105,65){$\Delta S_A$}
   \end{center}
  \end{minipage}
 \begin{minipage}{0.25\hsize}
 \begin{center}
     \includegraphics[clip,width=4cm]{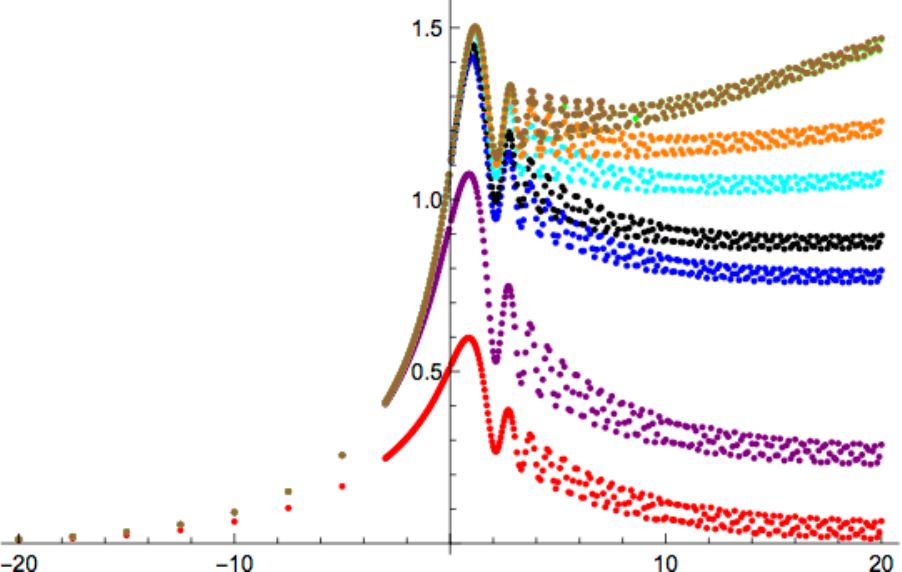}
       \put(-10,-10){$t/\xi_{kz}$}
    \put(-90,65){$\Delta S_A$}
  \end{center}
  \end{minipage}
   \begin{minipage}{0.25\hsize}
 \begin{center}
    \includegraphics[clip,width=4cm]{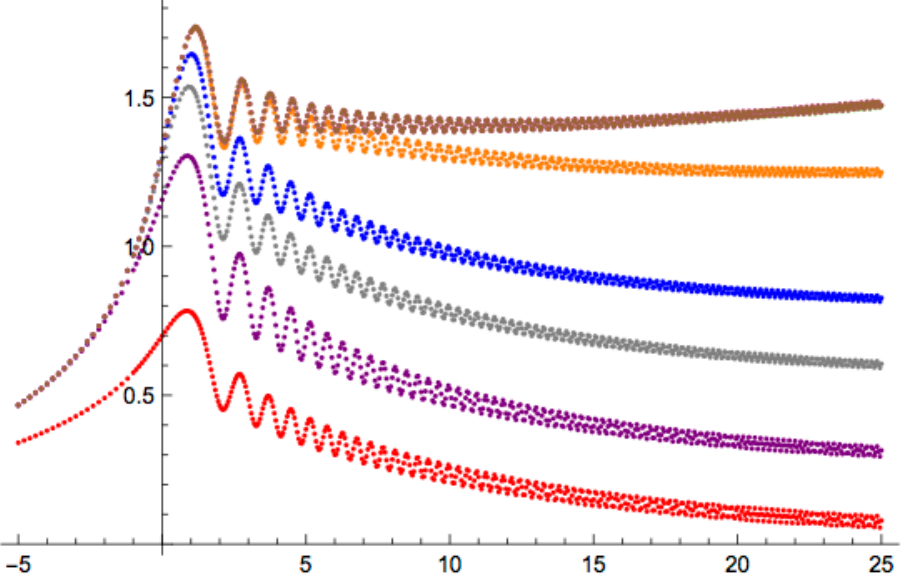}
         \put(-10,-10){$t/\xi_{kz}$}
    \put(-120,65){$\Delta S_A$}
  \end{center}
  \end{minipage}
      \caption{The $t$-dependence of $\Delta S_A$. The left, middle and right panels show $\Delta S_A$ with $(\delta t=1,000, \xi=10), (\delta t=2,000, \xi=20)$ and $(\delta t=4,000, \xi=10)$, respectively.}
    \label{fig:CCPslow_tdependence}
\end{figure}

\begin{figure}[htbp]
  \begin{center}
   \includegraphics[clip,width=6cm]{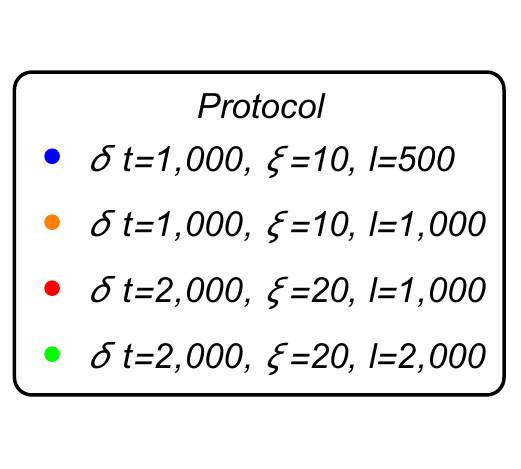}
    \includegraphics[clip,width=7.0cm]{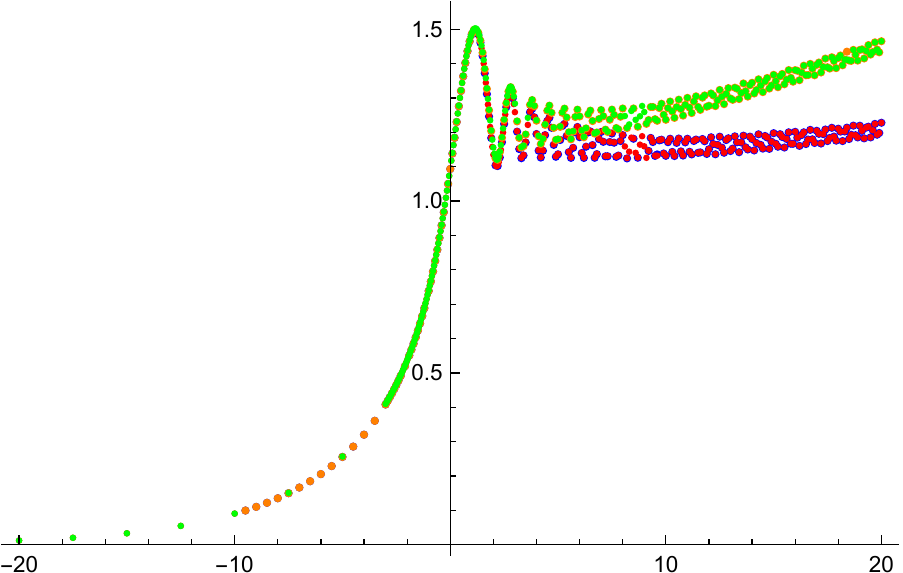}
         \put(5,5){$t/\xi_{kz}$}
    \put(-130,115){$\Delta S_A$}
    \caption{The  $t$-dependence of $\Delta S_A$ with different parameters.   This panel shows the scaling law in (\ref{sce2}).}
    \label{fig:scalinglaw_CCPslow}
  \end{center}
\end{figure}

\begin{figure}[htbp]
  \begin{center}
  \includegraphics[clip,width=6cm]{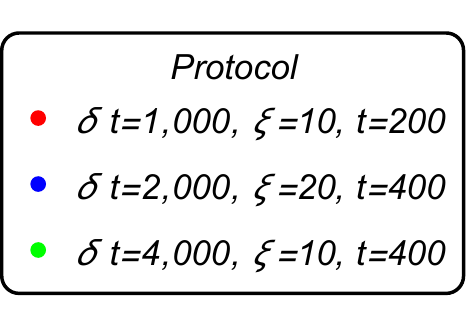}
    \includegraphics[clip,width=7.0cm]{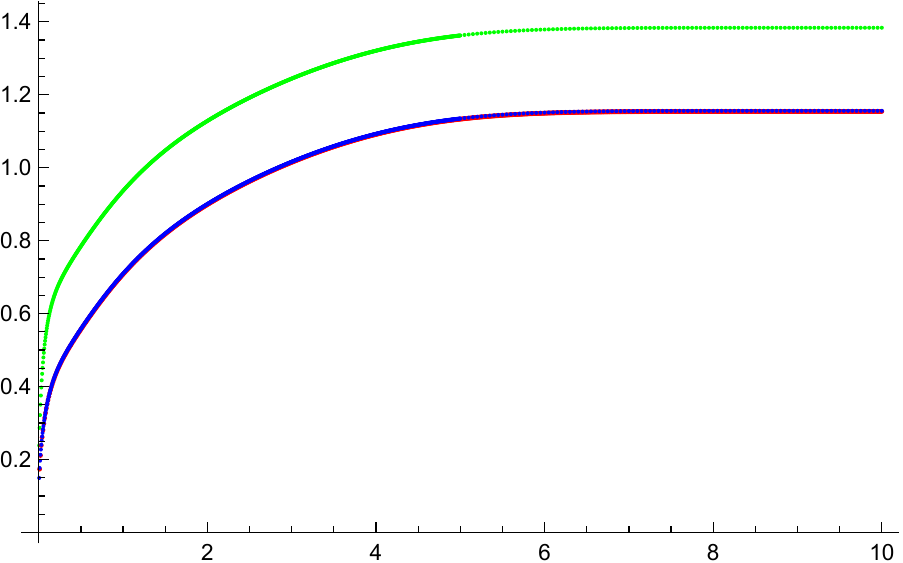}
         \put(5,5){$l/\xi_{kz}$}
    \put(-225,115){$\Delta S_A$}
    \caption{The $l$-dependence of $\Delta S_A$ at $t=2\xi_{kz}$.  }
    \label{fig:CCPslow_ldependence_t2xikz}
  \end{center}
\end{figure}

\subsubsection*{Oscillation of $\Delta S_A$}
$\Delta S_A$ in the slow CCP-type potential also oscillates. 
The periodicity of oscillation of $\Delta S_A$ asymptotically approaches $\pi \xi$ at  late time.
As we will explain later, the periodicity depends on the periodicity of oscillation of zero modes, which is zero-momentum spectra of two point functions, $X_{k=0}, P_{k=0}$ and $D_{k=0}$. 

\subsubsection*{$l$-dependence of $\Delta S_A$}
Figure~\ref{fig:ldependence_xi10_dt1000} and \ref{fig:ldependence_xi20_dt2000}  show the $l$-dependence of $\Delta S_A$. If $l$ is  sufficiently large, $\Delta S_A$ is independent of $l$. 
If $t$ is sufficiently large, a linear function of $l$ fits $\Delta S_A $ in $\xi \ll l \ll t_{*}$, although we have not found what determines $t_{*}$\footnote{If we apply the  entangled-particles interpretation where they are  created at $t=\pm t_{kz}$ to the time evolution of $\Delta S_A$ in the slow CCP-potential, the time evolution might change around $\frac{l}{2}\pm t_{kz}$. However, $t_{*}$ appears to be larger than $\frac{l}{2}\pm t_{kz}$.}.

\begin{figure}[htbp]
  \begin{center}
  \includegraphics[clip,width=4.8cm]{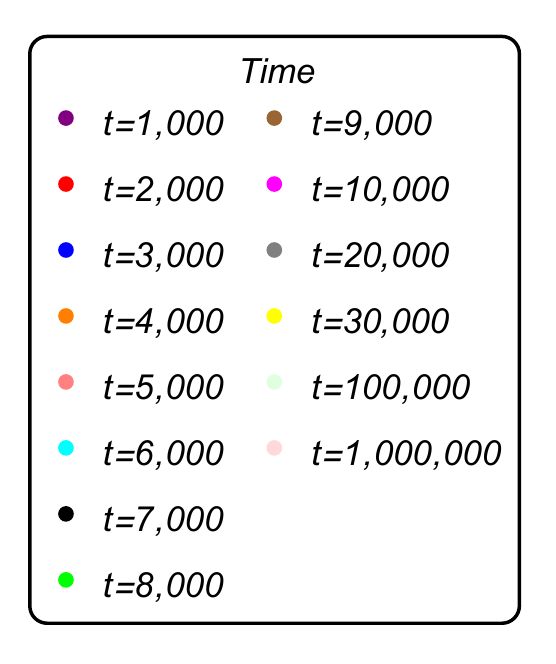}
  \hspace{1cm}
    \includegraphics[clip,width=7.0cm]{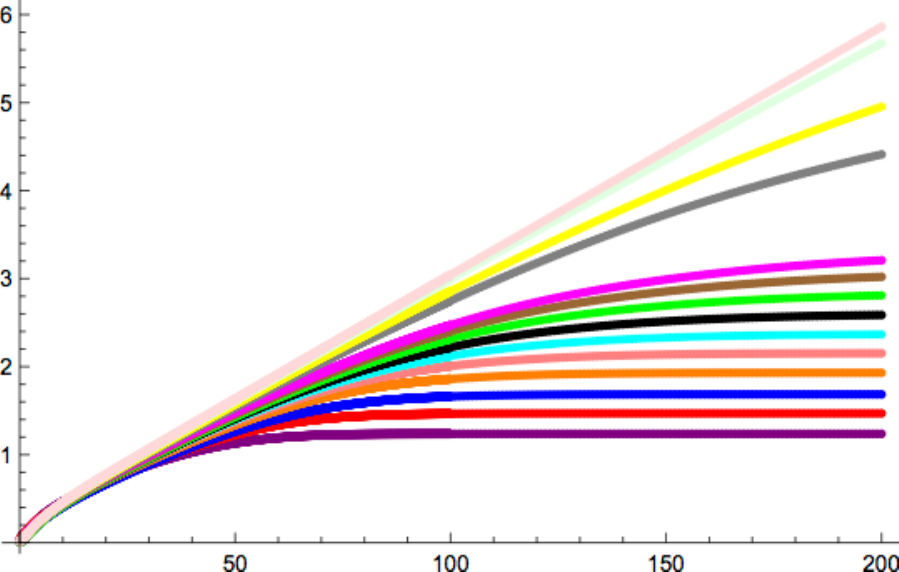}
   \put(5,0){$l/\xi$}
   \put(-225,100){$\Delta S_A$}
    \caption{The $l$-dependence of $\Delta S_A$ with $(\delta t=1,000, \xi=10)$. }
\label{fig:ldependence_xi10_dt1000}
  \end{center}
\end{figure}

\begin{figure}[htbp]
  \begin{center}
   \includegraphics[clip,width=3.8cm]{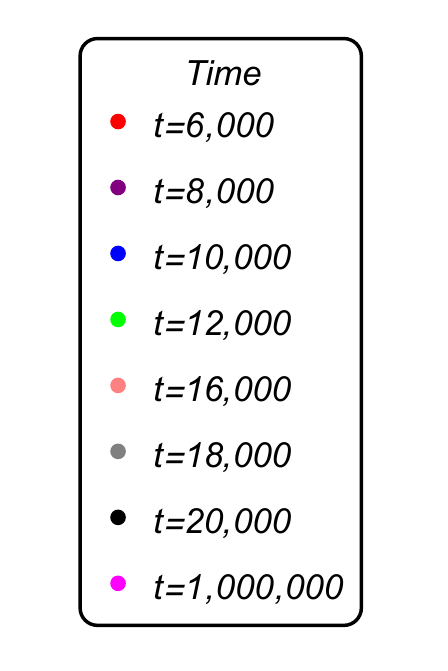}
   \hspace{1cm}
    \includegraphics[clip,width=7.0cm]{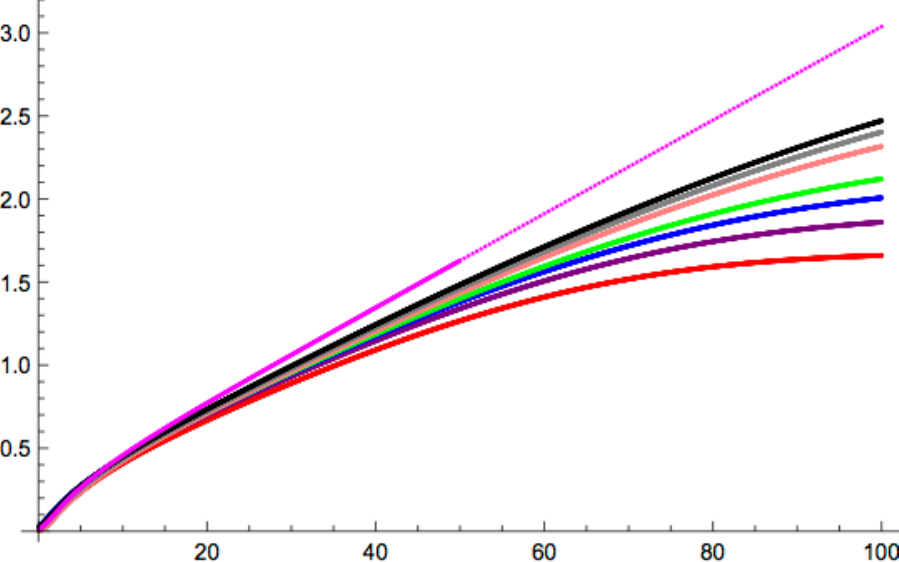}
    \put(5,0){$l/\xi$}
   \put(-225,100){$\Delta S_A$}
    \caption{The $l$-dependence of $\Delta S_A$ with $(\delta t=2,000, \xi=20)$.  }
\label{fig:ldependence_xi20_dt2000}
  \end{center}
\end{figure}

The left panel of Figure \ref{fig_fit_o} is a plot of $\Delta S_A$  in the window, $\xi \ll l \ll t_{*}$, which is fitted by a linear function of $l$:
\begin{equation} \label{fitslow}
\Delta S_A \simeq A \frac{l}{\xi} +B,
\end{equation} 
where $A$ and $B$ are in Table~\ref{tab:sccp}. In the case of $\Delta S_A$ in Figure \ref{fig_fit_o}, $A$ depends on $\omega$. As shown in the right panel of Figure~\ref{fig_fit_o}, as $\omega(\gg 1)$ increases $A $ decreases  and can be fitted by a function of $\omega$:
\begin{equation}
A \simeq 0.167802 -\frac{0.053054 \log {(\omega)}}{{\omega}^{0.121652}},\label{fc2}
\end{equation} 
where $\xi=10$ and $t=1,000,000$.
Moreover, $B$ also depends on $\omega$. The $\omega$-dependence of $B$ might be related to the  fit range.

If $t$ is large enough, $\Delta S_A$ in the slow and fast limits can be fitted by a linear function of $l/\xi$. Its proportionality coefficient $T_{eff}\cdot\xi$ depends on $\omega$ as shown in \eqref{T_eff}. In the slow (fast) limit, $T_{eff}\cdot\xi$ decreases as $\omega$ increases (decreases). 
As we will explain later, the $\omega$-dependence of $T_{eff}\cdot\xi$ is consistent with the number of particles with $\xi$ fixed.

\begin{figure}[htbp]
 \begin{minipage}{0.21\hsize}
  \begin{center}
    \includegraphics[clip,width=3cm]{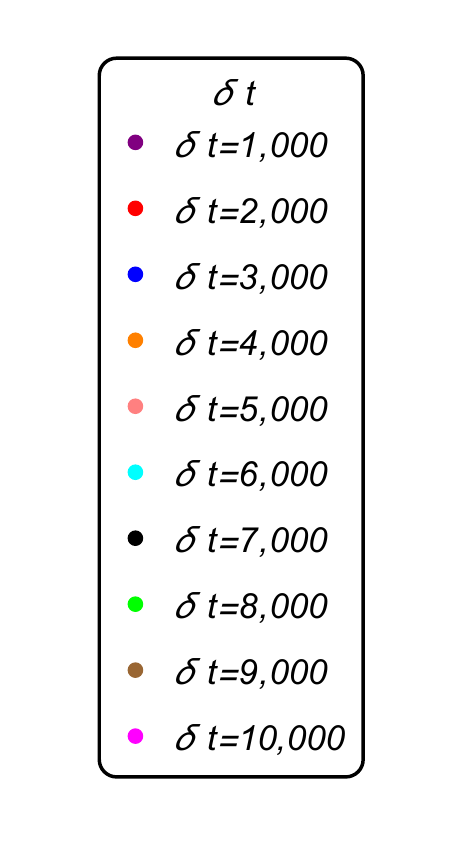}
\end{center}
 \end{minipage}
 \begin{minipage}{0.38\hsize}
   \begin{center}
    \includegraphics[clip,width=6.0cm]{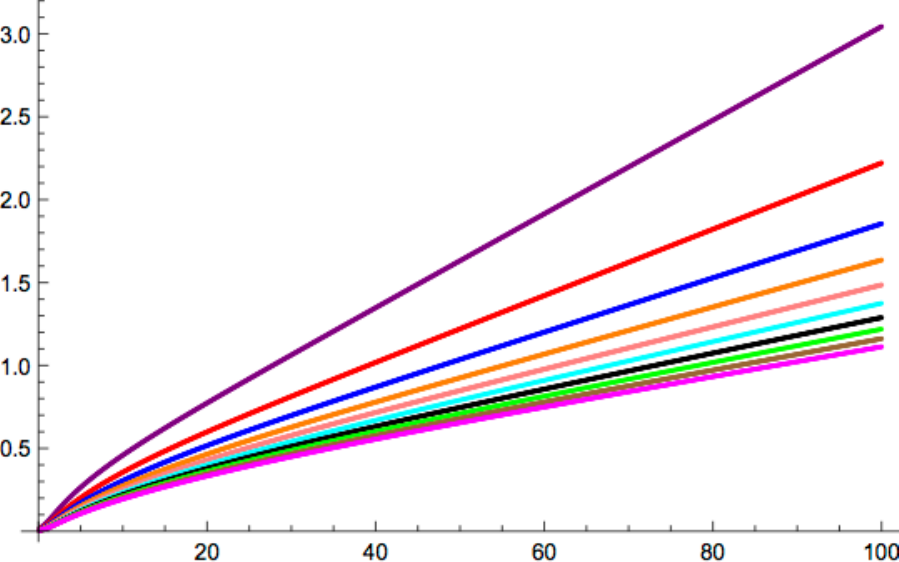}  
  \put(-10,-10){$l/\xi$}
    \put(-180,110){$\Delta S_A$}
    \end{center}
 \end{minipage}
 \begin{minipage}{0.38\hsize}
 \begin{center}
    \includegraphics[clip,width=6.0cm]{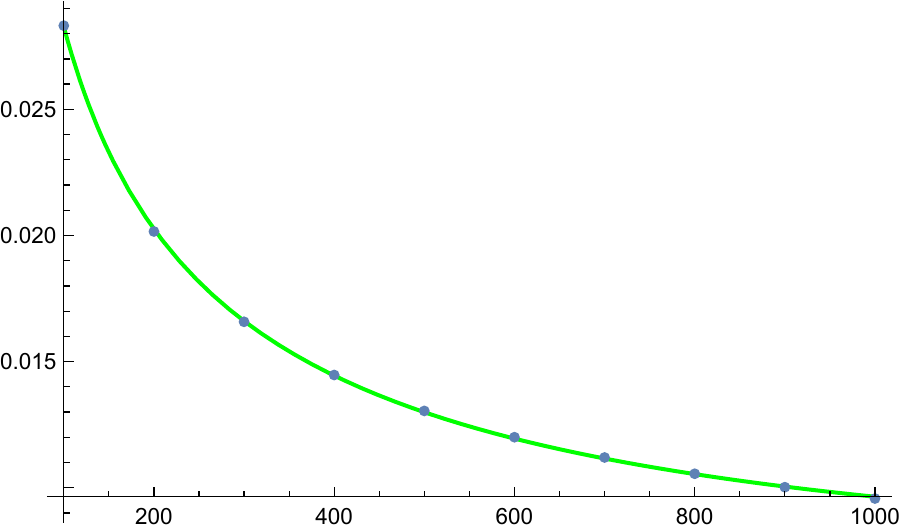} 
    \put(-10,-10){$\omega$}
    \put(-160,105){$A$}
     \end{center}
 \end{minipage}
    \caption{The left panel shows the $l$-dependence of $\Delta S_A$ with $(\xi, t)=(10, 1,000,000)$. 
The right panel shows the $\omega$-dependence of $A$ in $\Delta S_A \simeq A \frac{l}{\xi} +B$. The green curve is a plot of the fit function of $\omega$ in (\ref{fc2}).
    }
\label{fig_fit_o}
\end{figure}

\begin{table}[htb]
  \begin{center}
    \caption{Fit results for Figure \ref{fig:ldependence_xi20_dt2000} and \ref{fig_fit_o} . \label{tab:sccp}}
    \begin{tabular}{|c|c|c|c|c|} \hline
    $\delta t$ & $\xi$ &$t$& Fit Reslut & Fit Range    \\ \hline \hline
    $1,000$& $10$&$1,000,000$ &$\Delta S_A=0.214255 + 0.0283171 \frac{l}{\xi}$ & $20 \le \frac{l}{\xi} \le 100$ \\ 
     \hline
     $2,000$& $10$&$1,000,000$ &$\Delta S_A=0.209519 + 0.0201569  \frac{l}{\xi}$ & $20 \le \frac{l}{\xi} \le 100$ \\ \hline
       $3,000$& $10$&$1,000,000$ &$\Delta S_A=0.202759  + 0.0165772    \frac{l}{\xi}$ & $20 \le \frac{l}{\xi} \le 100$ \\ \hline   
      $4,000$& $10$&$1,000,000$ &$\Delta S_A=0.195988 + 0.0144678      \frac{l}{\xi}$ & $20 \le \frac{l}{\xi} \le 100$ \\ \hline
    $5,000$& $10$&$1,000,000$ &$\Delta S_A=0.189682 + 0.0130437    \frac{l}{\xi}$ & $20 \le \frac{l}{\xi} \le 100$ \\ \hline    
    $6,000$& $10$&$1,000,000$ &$\Delta S_A=0.184001 + 0.0120019   \frac{l}{\xi}$ & $20 \le \frac{l}{\xi} \le 100$ \\ \hline 
  $7,000$& $10$&$1,000,000$ &$\Delta S_A=0.178983 + 0.0111977  \frac{l}{\xi}$ & $20 \le \frac{l}{\xi} \le 100$ \\ \hline 
 $8,000$& $10$&$1,000,000$ &$\Delta S_A=0.174596 + 0.0105525 \frac{l}{\xi}$ & $20 \le \frac{l}{\xi} \le 100$ \\ \hline 
 $9,000$& $10$&$1,000,000$ &$\Delta S_A=0.170754 + 0.0100198  \frac{l}{\xi}$ & $20 \le \frac{l}{\xi} \le 100$ \\ \hline 
 $10,000$& $10$&$1,000,000$ &$\Delta S_A=0.167335 + 0.00957016   \frac{l}{\xi}$ & $20 \le \frac{l}{\xi} \le 100$ \\ \hline 
 $2,000$& $20$&$1,000,000$ &$\Delta S_A=0.212146 + 0.283104   \frac{l}{\xi}$ & $20 \le \frac{l}{\xi} \le 100$ \\ \hline 
            \end{tabular}
  \end{center}
  \end{table}

\subsubsection{Physical interpretation}
Here, we interpret the time evolution of $\Delta S_A$ in the CCP-type potential, physically. 

\subsubsection*{\underline{Entangled particles}}
In the window, $t< \frac{l}{2}$, $\Delta S_A$ for $l>\xi$ in the fast CCP-type quench is independent of $l$ but depends on $l$ after $t\simeq \frac{l}{2}$. As in the fast ECP-type quench,  the time evolution of $\Delta S_A$ is interpreted in terms of the propagation of entangled particles created around $t=0$. Their velocity, $v_k$, depends on $k$. The potential after $t=0$ is finite , but the maximum velocity, $v_k^{max}$, is expected to be the speed of light, $v^{max}_k \simeq \pm1$ because $m \ll 1$. If the entangled particles are created around $t=0$, the distance before $t \simeq \frac{l}{2}$ between entangled particles with $v_k^{max}$ is smaller than the subsystem size. Since the distance after $t \simeq \frac{l}{2}$ is larger than $l$, the whole region in $A$ is entangled with $B$. Therefore, $\Delta S_A$ after $t\simeq \frac{l}{2}$ depends on the subsystem size.

\subsubsection*{\underline{Minimum of $\Delta S_A$}}
$\Delta S_A$ in   CCP-type potential is characterized by $t_C$: in the fast limit, $t_C=2\xi$; in the slow limit, $t_C=2\xi_{kz}$.  $\Delta S_A$ before $t\simeq\frac{t_C}{2}$ increases monotonically. $\Delta S_A$ from $t\simeq t_C$ to $t\simeq 2t_C$ decreases monotonically, and the entropy after $t=t_C$ oscillates.

Figure \ref{fig:plot_t2xi} and \ref{fig:CCPslow_ldependence_t2xikz} show the $l$-dependence of $\Delta S_A(t=t_C)$ in both limits. As entanglement entropy in a massive theory, $\Delta S_A$ with  large $l$ is independent of $l$: $\Delta S_A(l \ge 4\xi)$ in the fast limit and $\Delta S_A(l \ge 6\xi_{kz})$ in the slow limit are independent of $l$. In the massive free theory, there is a correlation length, $\Xi_{static}$.  If the subsystem size is much larger than $\Xi_{static}$,  entanglement entropy in the $1+1$ dimensional massive free theories \cite{rt2,cm} is  given by
\begin{equation}
S_A \simeq K \log{\left(\Xi_{static}\right)},
\end{equation}
where $K$ depends on the number of the boundary of subsystem.
Therefore,  there might be an effective correlation length, $\Xi_{effective}$, in the CCP-type quenches at $t=t_C$. If $l>4\xi$, $\Delta S_A(t=t_C)$ in the fast limit is given by
\begin{equation} \label{larm}
\Delta S_A (t=t_C) \simeq K\log{\left(\Xi_{effective}\right)}- K \log{\left(\xi\right)},
\end{equation}   
where $K \log{\left(\xi\right)}$ is the entropy for the initial state.  Figure \ref{fig:plot_t2xi} shows $\Delta S_A(t=t_C)$ for $l\ge 4\xi $ approaches a negative constant. Thus, $\Xi_{effective}$ is smaller than the initial correlation length\footnote{$\Xi_{effective}$ is expected to be different from $4\xi$ but related to it.}.

If $l\ge6\xi_{kz}$, $\Delta S_A(t=t_C)$ in the slow limit is independent of $l$. Therefore, we expect  $\Delta S_A(t=t_C)$ for $l\ge 6\xi_{kz}$ to be given by (\ref{larm}). Since Figure \ref{fig:CCPslow_ldependence_t2xikz} shows $\Delta S_A$ for $l \ge 6\xi_{kz}$ approaches a positive constant, $\Xi_{effective}$ is larger than $\xi$. 

As explained above, the $l$-dependence of $\Delta S_A$ in the fast limit is interpreted in terms of the propagation of entangled particles created around $t\simeq0$. Thus, the $l$-dependence of $\Delta S_A(t=t_C)$ is expected to be interpreted in terms of entangled particles created around $t \simeq 0$\footnote{
Although the pair in the sudden ECP-type potential makes entanglement entropy increase \cite{ca1, L1, L2, MM1}, the pair in this case reduces the entropy.}. The distance at $t=2\xi$ between entangled particles with $v_k^{max}$ is $4\xi$, and it is the subsystem size where $S_A(t_c)$ becomes $K\log{\left(\Xi_{effective}\right)}$.  We expect  $\Xi_{effective}$ to be related to this distance.

If we apply the entangled particle interpretation to $\Delta S_A(t=t_C,l\ge6\xi_{kz})$ in the slow limit, entangled particles should be created at $t \simeq -t_{kz}$ when adiabaticity breaks down.
\subsubsection*{\underline{The periodicity of entanglement oscillation}}
Entanglement entropy in the CCP-type quench oscillates. 
The periodicity of oscillation of $\Delta S_A$ in late time, $\pi \xi$, is independent of whether we take the fast or slow limits. 
As in Appendix \ref{apb}, $f_k(t)$ in the window, $t \gg \delta t$ is given by
\begin{equation}
f_k(t) \simeq \mathcal{A}_k e^{i \omega_k t}+ \mathcal{B}_k e^{-i \omega_k t},
\end{equation}
where $f_k$ is a superposition of right-moving and left-moving waves, and the amplitudes, $\mathcal{A}_k$ and $\mathcal{B}_k$, depend on $k$.
Their dispersion relations are given by $\omega_k =\sqrt{4\sin^2{\left(\frac{k}{2}\right)}+m^2}$, which is consistent with the fact that the Hamiltonian might be well approximated by massive one because the CCP-type mass in  late time changes slowly.
As in Appendix \ref{apc} the spectra of two point functions, $X_k, P_k, D_k$ in $t \gg \delta t$ are given by
\begin{equation}\label{ltp}
\begin{split}
&X_k, P_k \simeq \mathcal{C}^{x,p}_k + \mathcal{D}^{x,p}_k \cos{\left(2 \omega_0 t+ \Theta^{x, p}_k\right)}, \\
&D_k\simeq \mathcal{D}^{d}_k \cos{\left(2 \omega_0 t+ \Theta^{d}_k\right)}, \\
\end{split}
\end{equation}
where $\mathcal{C}^i_k $, $\mathcal{D}^i_k$ and $\Theta^i_k$ are independent of $t$. Their spectra oscillate due to the time-dependent terms in (\ref{ltp}) which come from a coherence between the waves. Thus, we expect the entanglement oscillation in $t \gg \delta t$ to come from the coherence between the right-moving and left-moving waves. 

As explained above, the entangled pair interpretation does not seem to determine $t_{*}$. However, we apply the interpretation to $\Delta S_A$ in very late time. 
If we interpret the time evolution of $\Delta S_A$ in very late time in terms of the propagation of entangled particles with the momentum-dependent velocity, $v_k$, the particles with small $v_k$ contribute dominantly to $\Delta S_A$ in very late time.
Thus, we expect  the modes around $k=0$ to contribute dominantly to $\Delta S_A$ in the late time  because the modes around $k=0, \pm \pi$ are slow modes\footnote{We do not consider the modes around $k=\pm \pi$ because they suffer from the discretization effect.}.
The spectrum of two point functions with $k=0$ in very late time, $X_{k=0}, P_{k=0} $ and $D_{k=0}$,  oscillate with $\pi \xi$ which is consistent with  the periodicity of entanglement oscillation. Therefore, the periodicity of entanglement oscillation in very late time comes from the periodicity of the coherent oscillation of zero modes.

\subsubsection*{\underline{$l$-dependence of $\Delta S_A$}}
$\Delta S_A$ in the fast-CCP quench is independent of $l$ before $t\simeq\frac{l}{2}$, if the subsystem size is larger than $\xi$. After $t\simeq\frac{l}{2}$, $\Delta S_A$ depends on the subsystem size. $\Delta S_A$ in the late time is proportional to $l$, though its proportionality coefficient depends on $\delta t $ and $\xi$. If we keep $\xi$ a constant, taking the fast limit, the proportionality coefficient and an expectation value of number operator upon a volume, $ \frac{N}{\mathcal{V}} $, decrease when $\omega$ decreases. Here, $\mathcal{V}$ is the spatial volume of total space. The $\omega$-dependence of $\frac{N}{\mathcal{V}}$ is shown in Figure \ref{fig:num2}. Thus, the $\omega$-dependence of proportionality coefficient in the fast limit is consistent with $\frac{N}{\mathcal{V}}$. 

$\Delta S_A$ even in the slow limit can be fitted by the function in (\ref{fitslow}), if $t$ is large enough.
Its proportionality coefficient decreases when $\omega$ increases with fixed $\xi$, which is consistent with the $\omega$-dependence of $\frac{N}{\mathcal{V}}$. 
Therefore, the $l$-dependence of the late-time $\Delta S_A$ in the CCP-type potentials is determined by how dense particles in the late time are.  
\begin{figure}[htbp]
\begin{center}
 \begin{tabular}{c}
 \begin{minipage}{0.33\hsize}
  \begin{center}
   \includegraphics[width=40mm]{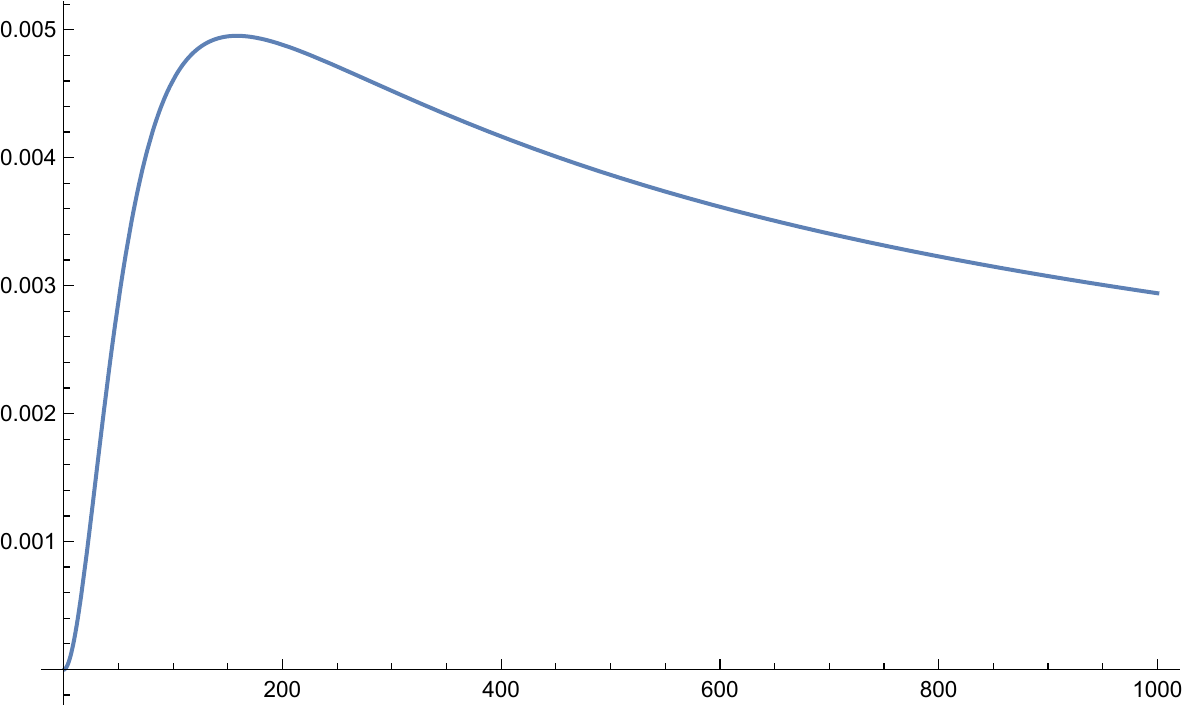}
     \put(-10,-10){$\delta t$}
    \put(-130,60){$\frac{N}{\mathcal{V}}$}
  \end{center}
 \end{minipage}
 \begin{minipage}{0.33\hsize}
  \begin{center}
   \includegraphics[width=40mm]{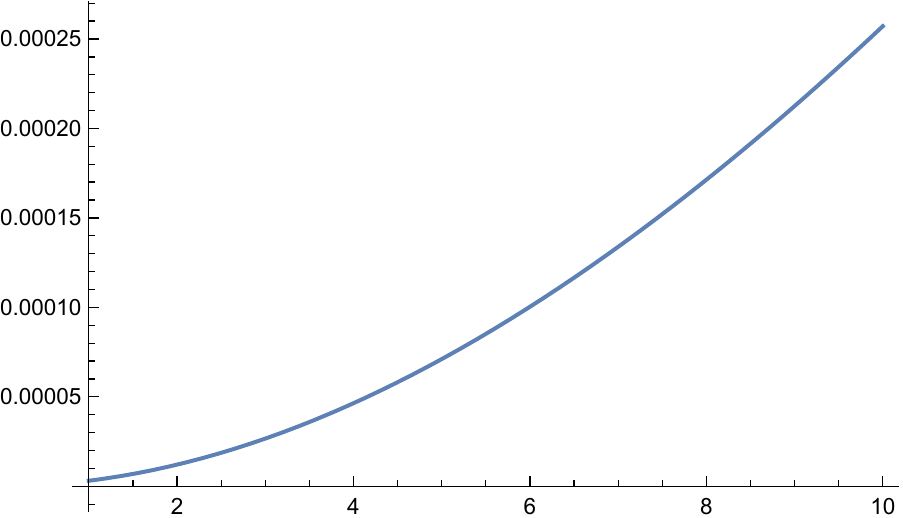}
        \put(-10,-10){$\delta t$}
    \put(-130,60){$\frac{N}{\mathcal{V}}$}
  \end{center}
 \end{minipage}
 \begin{minipage}{0.33\hsize}
  \begin{center}
   \includegraphics[width=40mm]{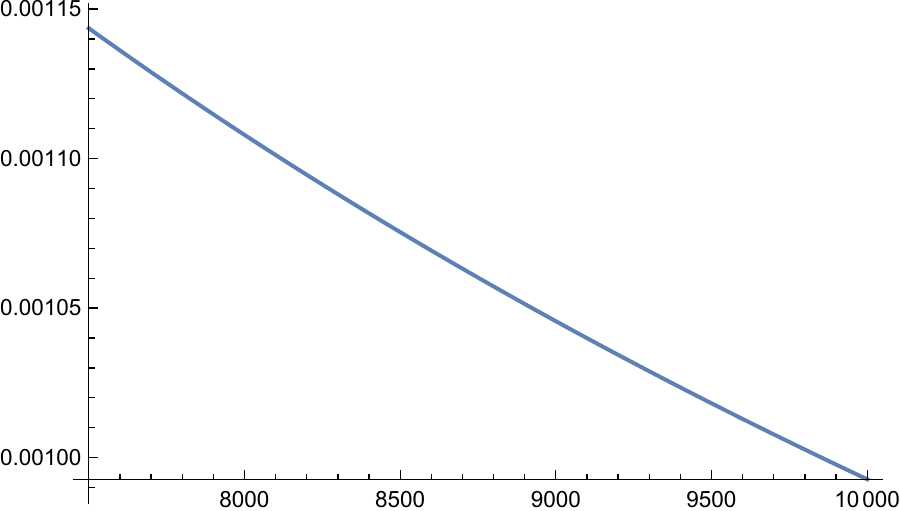}
        \put(-10,-10){$\delta t$}
    \put(-130,60){$\frac{N}{\mathcal{V}}$}
  \end{center}
 \end{minipage}
 \end{tabular}
\caption{The $\omega$-dependence of $\frac{N}{\mathcal{V}}$ with $m=\frac{1}{100}$.   In the left panel, the range of $\delta t$ is $1\le\delta t\le 1,000$. The middle panel is for small $\delta t$. The right one is for large $\delta t$.\label{fig:num2} }
\end{center}
\end{figure}


\section{Summary and Future Directions}
In this paper, we have studied the time evolution of $\Delta S_A$ in smooth quenches by taking the extreme limits, the fast and slow limits.
\subsubsection*{ECP-type quenches}
The early-time $\Delta S_A$ in the ECP-type mass is fitted by a linear function of $t$, and the late-time $\Delta S_A$ is proportional to the subsystem size. Its proportionality coefficient depends on the quenches. The coefficients in the fast and slow quenches are proportional to the initial mass, $m=\frac{1}{\xi}$, and the Kibble--Zurek energy, $E_{kz}$, respectively. Therefore, we expect  the effective temperature in the fast limit to be determined by the initial correlation length, $\xi$.
On the other hand, the temperature in the slow limit is determined by the effective energy scale defined when adiabaticity breaks down, $E_{kz}$.
We found that the time evolution of $\Delta S_A$ in the ECP-type quenches is interpreted in terms of the propagation of entangled particles with the velocity, $v_k$. 
The time evolution of $\Delta S_A$ in the fast limit is interpreted in terms of the momentum-dependent propagation of entangled particles created around $t=0$.
On the other hand, the time evolution in the slow limit is interpreted in terms of the propagation of entangled particles created when adiabaticity breaks down. 

\subsubsection*{CCP-type quenches}
We found $\Delta S_A$ in the CCP-type quenches oscillates after a characteristic time, $t_C$: $t_C\simeq 2\xi$ in the fast quench; $t_C\simeq 2\xi_{kz}$ in the slow quench.

$\Delta S_A$ in the fast CCP quench is interpreted in terms of the propagation of entangled particles created around $t=0$. Their velocity, $v_k$, depends on the momenta.

$\Delta S_A$ at $t=t_C$ approaches a constant if the subsystem size is large enough. 
The fast-quenched $\Delta S_A(t=t_C)$ in the window, $l>2 t_C$, is a negative constant, which is interpreted as entanglement entropy in the theory with an effective correlation length  $\Xi_{eff} (<\xi)$.  $\Xi_{eff} $ in the fast limit is related to the distance between the entangled particles created at $t\simeq0$.
The slow-quenched $\Delta S_A(t=t_C)$ in $l>3 t_C$ is a positive constant interpreted as the entropy in the theory with an effective correlation length $\Xi_{eff} (>\xi)$. $\Xi_{eff} $ in the slow limit is related to the distance between the particles created at $t\simeq-t_{kz}$.

After $t=t_C$, $\Delta S_A$ in the CCP-type mass oscillates with a periodicity.
The periodicity in the window, $t\gg \delta t$, is expected to come from the periodicity of oscillation of  $X_k, P_k$ and $D_k$. 
Moreover, the periodicity in vary late time, $\pi \xi$, comes from the periodicity of oscillation of  $X_{k=0}, P_{k=0}$ and $D_{k=0}$. 

 The late-time $\Delta S_A$ in both limits is fitted by the linear function of $l$,
$\Delta S_A \sim T_{eff} \cdot l $, where $T_{eff}$ depends on $\xi$ and $ \delta t$.
The $\delta t$ and $\xi$-dependence of $T_{eff}$ is consistent with the  dependence of the density of particles in the late time, $\frac{N}{\mathcal{V}}$.

\subsection*{Future Directions}
We will comment on a few of future directions.
\begin{itemize}
\item \underline{The time $\Delta S_A$ starts to oscillate}: We found that $t=t_C$ in CCP quenches plays an important role. However, we were not able to find why they determine the time $\Delta S_A$ starts to oscillate. It is interesting to study why these scale determine the initial time of entanglement oscillation.  

\item \underline{Physics in the slow CCP}: We could not find how the $l$-dependence of late-time $\Delta S_A$ in the slow CCP-type potential are interpreted. It is one of interesting future directions to study how they are interpreted.

\item \underline{$\Delta S_A $ in interacting field theories}: In this paper, we studied the time evolution of $\Delta S_A$ in the free field theories. It is important to study how an interaction changes the results in this paper.

\item \underline{Periodic potential}: The potentials in this paper can be replaced with periodic potentials. $\Delta S_A$ in the periodic potential might be related to the dynamics in Floquet   system (time crystal). Thus, we think that it is interesting to study $\Delta S_A$ in the periodic potentials.

\item \underline{Other measures}: It is interesting to study the dynamics of quantum entanglement in the smooth quenches by computing other measures such as logarithmic negativity, mutual information and so on. Some of authors in this paper have been studying the time evolution of logarithmic negativity and mutual information in \cite{oln}.
\end{itemize}

\section*{Acknowledgement }
We would like to thank Ryu Shinsei, Thomas Faulkner, Hiroyuki Fujita and Norihiro Iizuka for comments and useful discussions and especially to Sumit Das for the collaboration in the first stage, comments, useful discussions and helping to make our statement clear. 
The work of M.~Nishida was supported by Basic Science Research Program through the National Research Foundation of Korea (NRF) funded by the Ministry of Science, ICT \& Future Planning (NRF- 2017R1A2B4004810) and GIST Research Institute (GRI) grant funded by the GIST in 2017.
AT was fully supported by Heng-Tong Ding.
The work of AT was supported in part by NSFC under grant no. 11535012.

\newpage


\appendix

\newpage
\section{Calculation\label{apa}}
\subsection{CCP for Scalar Quench}
We solve the following equation,
\begin{align} 
\frac{d^2 f_k(t)}{d t^2}+(4{\rm sin}^2[k/2]+m_0^2 {\rm tanh}^2[t/\delta t])f_k (t)=0.
\end{align}
By defining
\begin{align}
z&=-{\rm sinh}^2[t/\delta t],\quad f_{k}(z)= (1-z)^{\alpha}\psi(z),
\end{align}
we find
\footnotesize
\begin{align}
z(1-z)\frac{d^2 \psi(z)}{d z^2}+\biggl(\frac{1}{2} -(2\alpha+1)z\biggl)\frac{d \psi(z)}{d z}
 -\biggl((\delta t)^2{\rm sin}^2[k/2] + \frac{\alpha}{2}\biggr)\psi(z)
+\biggl(\alpha (\alpha -1) +\frac{1}{2}\alpha +\frac{(m_0\delta t)^2}{4} \biggr)\frac{z \psi(z)}{1-z}=0
\label{DE1}.
\end{align}
\normalsize
The solution is given by the hypergeometric function. In terms of $f_k(t)$, we find
\footnotesize
\begin{align}
f_{k}(t) =({\rm cosh}[t/\delta t])^{2\alpha} \biggl[
 A ~ _2 F_1 \biggl(a,b;\frac{1}{2} ;-{\rm sinh}^2[t/\delta t] \biggr)
+B~{\rm sinh}[t/\delta t] _2 F_1 \biggl(a+\frac{1}{2},b+\frac{1}{2};\frac{3}{2} ;-{\rm sinh}^2[t/\delta t] \biggr)
\biggl],
\end{align}
\normalsize
where
\begin{align}
&a = \alpha -\frac{{\rm i} \omega_0 \delta t}{2},~b = \alpha +\frac{{\rm i} \omega_0 \delta t}{2},
\\
&\alpha = \frac{1+\sqrt{1-4(m_0\delta t)^2}}{4},
\\
&\omega^2_0 = 4{\rm sin}^2[k/2] +m_0^2 ,
\end{align}
and $A$ and $B$ are the coefficients. These coefficients can be determined by two conditions,
\begin{align}
&{\rm i}\biggl( f_k^{*}(t) \frac{d f_k(t)}{d t} -\frac{ f_k^{*}(t)}{d t}f_k(t) \biggr)=1,
\label{NC}
\\
&f_k(t) \sim \frac{{\rm e}^{-{\rm i}\omega_0 t}}{\sqrt{2\omega_0 }} ~~(t\to-\infty).
\label{AC}
\end{align}
Note that the hypergeometric function can be written as
\begin{align}
_2 F_1 (a,b;c;z)
&=
\frac{\Gamma(c)\Gamma(b-a)}{\Gamma(b)\Gamma(c-a)}(-z)^{-a} ~_2 F_1 (a,a-c+1;a-b+1;\frac{1}{z})
\nonumber \\
&\qquad+\frac{\Gamma(c)\Gamma(a-b)}{\Gamma(a)\Gamma(c-b)}(-z)^{-b} ~_2 F_1 (b,b-c+1;b-a+1;\frac{1}{z}).
\end{align}
Thus by defining
\begin{align}
E_{1/2}=\frac{\Gamma(1/2)\Gamma(b-a)}{\Gamma(b)\Gamma(1/2-a)}
,~
E_{3/2}=\frac{\Gamma(3/2)\Gamma(b-a)}{\Gamma(1/2+b)\Gamma(1-a)}
,~
E'_{c}=E_c (a\leftrightarrow b),
\end{align}
we have
\footnotesize
\begin{align} \label{sol1}
f_k(t)
&=
({\rm cosh}[t/\delta t])^{2\alpha} \biggl\{
A\biggl[
 E_{1/2} |{\rm sinh}^2[t/\delta t]|^{-a}~_2 F_1\biggl(a,a+\frac{1}{2};a-b+1;-\frac{1}{{\rm sinh}^2[t/\delta t]}\biggr)
 \nonumber \\
&\qquad +
 E'_{1/2}|{\rm sinh}^2[t/\delta t]|^{-b}~_2 F_1\biggl(b,b+\frac{1}{2};b-a+1;-\frac{1}{{\rm sinh}^2[t/\delta t]}\biggr)
\biggl]
 \nonumber \\
&\qquad +
B {\rm sinh}[t/\delta t] 
\biggl[
E_{3/2}  |{\rm sinh}^2[t/\delta t]|^{-a-1/2} ~_2 F_1\biggl(a+\frac{1}{2},a;a-b+1;-\frac{1}{{\rm sinh}^2[t/\delta t]}\biggr)
 \nonumber \\
&\qquad +
E'_{3/2}  |{\rm sinh}^2[t/\delta t]|^{-b-1/2} ~_2 F_1\biggl(b+\frac{1}{2},b;b-a+1;-\frac{1}{{\rm sinh}^2[t/\delta t]}\biggr)
\biggr]
\biggr\}.
\end{align}
\normalsize
In the limit $t\to-\infty$, $f_k(t)$ reduces to
\begin{align}
f_k(t)
&\sim 
(AE_{1/2} -BE_{3/2}) {\rm e}^{-{\rm i}\omega_0 t}+(AE'_{1/2}-BE'_{3/2}){\rm e}^{{\rm i}\omega_0 t}.
\end{align}
In order to satisfy $f_{k}(t)\sim{\rm e}^{-{\rm i}\omega_0 t}$, we impose
\begin{align}
B=\frac{E'_{1/2}}{E'_{3/2}}A.
\end{align}
Next we determine the coefficient $A$ from the normalization condition \eqref{NC}.  This condition does not depend on the time $t$, and we consider $t\to-\infty$. In this situation the hypergeometric function is expanded as the following series expansion,
\begin{align}
~_2 F_1(\alpha, \beta; \gamma;z)
=
\frac{\Gamma(\gamma)}{\Gamma(\alpha)\Gamma(\beta)}
\sum_{n=0}^{\infty}
\frac{\Gamma(\alpha+n)\Gamma(\beta+n)}{\Gamma(\gamma+n)}\frac{z^n}{n!}.
\end{align}
In our case
\begin{align}
&\frac{d}{d t}~_2 F_1\biggl(\alpha, \beta; \gamma;-\frac{1}{{\rm sinh}^2[t/\delta t]}\biggr)
\nonumber \\
&=\frac{\Gamma(\gamma)}{\Gamma(\alpha)\Gamma(\beta)}
\sum_{n=1}^{\infty}
\frac{\Gamma(\alpha+n)\Gamma(\beta+n)}{\Gamma(\gamma+n)}(-{\rm sinh}^2[t/\delta t])^{-n-1}
\frac{2{\rm sinh}[t/\delta t]{\rm cosh}[t/\delta t]}{(n-1)!}
\nonumber \\&
\rightarrow0~~(t\to-\infty).
\end{align}
In the calculation of $\frac{d f_k(t)}{d t}$, the terms that the derivative acts on $_2 F_1\biggl(\alpha, \beta; \gamma;-\frac{1}{{\rm sinh}^2[t/\delta t]}\biggr)$ become zero.
Then \eqref{NC} reduces to
\begin{align}
&{\rm i}\Biggl[
\Biggl\{
-\frac{2\alpha |A|^2}{\delta t} \biggl(\frac{E_{1/2}E'_{3/2}-E_{3/2}E'_{1/2}}{E'_{3/2}}\biggr)^2
-\frac{2a^* |A|^2}{\delta t} \biggl(\frac{E_{1/2}E'_{3/2}-E_{3/2}E'_{1/2}}{E'_{3/2}}\biggr)^2
\Biggr\}
\nonumber \\
&\qquad-
\Biggl\{
\frac{2\alpha |A|^2}{\delta t} \biggl(\frac{E_{1/2}E'_{3/2}-E_{3/2}E'_{1/2}}{E'_{3/2}}\biggr)^2
+\frac{2a |A|^2}{\delta t} \biggl(\frac{E_{1/2}E'_{3/2}-E_{3/2}E'_{1/2}}{E'_{3/2}}\biggr)^2
\Biggr\}
\Biggr]=1
\nonumber \\
&\Rightarrow
|A| = \frac{1}{\sqrt{2\omega_{0}}}\frac{E'_{3/2}}{E_{1/2}E'_{3/2}-E_{3/2}E'_{1/2}}.
\end{align}
Finally, the phase factor can be determined from \eqref{AC}. In the limit of $t\to-\infty$, we have
\begin{align}
(\text{phase of }f_{k})= 2^{-{\rm i} \omega_0  \delta t} {\rm Arg}[A]
\Rightarrow
A= \frac{2^{{\rm i} \omega_0  \delta t}}{\sqrt{2\omega_{0}}}\frac{E'_{3/2}}{E_{1/2}E'_{3/2}-E_{3/2}E'_{1/2}}.
\end{align}
 By shifting $4{\rm sin}^2[k/2] \to 4{\rm sin}^2[k/2] + M_0^2$, we can obtain the result with the mass term,
\begin{align}
m^2(t)=M_0^2 + m_0^2 {\rm tanh}^2 [t/\delta t].
\end{align}

In the slow CCP limit, it takes a long time to compute $f_k(t)$ numerically. In order to avoid this difficulty, we use adiabatic approximation at large $|k|$ \cite{sm5}. Approximation functions in our numerical computations are
\begin{align}\label{apsol}
f_k(t)&\sim\frac{{\rm e}^{-{\rm i}\omega(t) t}}{\sqrt{2\omega(t) }},\notag\\
\dot{f}_k(t)&\sim\frac{{-i\omega(t)\rm e}^{-{\rm i}\omega(t) t}}{\sqrt{2\omega(t) }},\notag\\
\omega^2(t)&= 4{\rm sin}^2[k/2] +m^2(t).
\end{align}
An error from this approximation in the computations is examined in Appendix \ref{sec:error_examination}.

\subsection{ECP for Scalar Quench}
Next we consider the ECP-type protocol for scalar quenches. $f_k(t)$ in the ECP-type protocol  satisfies the following differential equation,
\begin{align}
&\frac{d^2 f_k(t)}{d t^2}+\biggl(4{\rm sin}^2[k/2]+\frac{m^2}{2}(1- {\rm tanh}[t/\delta t])\biggr)f_k (t)=0.
\end{align}
The calculation is almost the same as the ECP case.  By defining
\begin{align}
z=\frac{1+{\rm tanh}[t/\delta t]}{2},~~{\rm tanh}[t/\delta t] = 2z-1,\quad f_k(z)&=z^{\beta}(1-z)^{\alpha},\psi(z),
\end{align}
we obtain the following differential equation,
\footnotesize
\begin{align}
z(1-z)\frac{d^2 \psi(z)}{dz^2} +
\bigl( 1+2\beta -\bigl((\alpha +\beta+1)+(\alpha+\beta)+1\bigr)z\bigr)\frac{d \psi(z)}{dz}
-\bigl((\alpha +\beta +1)(\alpha +\beta) \bigr)\psi(z) =0,
\end{align}
\normalsize
where 
\begin{align}
\alpha = -{\rm i}|(\delta t) {\rm sin}[k/2]|,~\beta ={\rm i}\sqrt{ (\delta t)^2 {\rm sin}^2[k/2] + \frac{m^2 (\delta t)^2}{4} }.
\label{diffeq}
\end{align}
The solution can be again expressed by the hypergeometric function,
\footnotesize
\begin{align}\label{sol2}
f_k(z) &=A \biggl( \frac{1+{\rm tanh}[t/\delta t]}{2}\biggr)^{\beta} \biggl( \frac{1-{\rm tanh}[t/\delta t]}{2}\biggr)^{\alpha}
~_2 F_1(\alpha + \beta +1 , \alpha +\beta; 2\beta+1;(1+{\rm tanh}[t/\delta t])/2)
\nonumber \\
&\qquad
+B \biggl( \frac{1+{\rm tanh}[t/\delta t]}{2}\biggr)^{-\beta} \biggl( \frac{1-{\rm tanh}[t/\delta t]}{2}\biggr)^{-\alpha}
~_2 F_1(-\alpha - \beta +1 , -\alpha -\beta; -2\beta+1;(1+{\rm tanh}[t/\delta t])/2).
\end{align}
\normalsize
Again we impose the conditions  \eqref{NC} and
\begin{align}
f_k (t) \sim \frac{{\rm e}^{-2\beta t/\delta t}}{\sqrt{4(-{\rm i})\beta/\delta t}}~~(t\to-\infty),
\end{align}
which is  similar to \eqref{AC}\footnote{Since $\alpha$ and $\beta$ depend on $\delta t$, the boundary condition is slightly different.}. 
One can fix the coefficients under these condition,
\begin{align}
A=0,\quad B=\frac{1}{\sqrt{4(-{\rm i})\beta/\delta t}}.
\end{align}
We can obtain the result with the mass term
\begin{align}
m^2(t) = M^2 + m^2 {\rm tanh}[t/\delta t],
\end{align}
by shifting
\begin{align}
& 4{\rm sin}^2[k/2] \to 4{\rm sin}^2[k/2] +M^2+m'^2 ,\\
& m^2 \to -2m^2,
\end{align}
and taking the limit $m'\to m$.

\section{Number operator in the late time \label{apb}}
$f_k(t)$ in the late time is given by
\begin{equation} \label{lfk}
\lim_{t \rightarrow \infty}f_k(t)=V~ 2^{-i \omega_0 \delta t}e^{i \omega_0 t}+W~ 2^{i \omega_0 \delta t}e^{-i \omega_0 t},
\end{equation}
where $V=A E_{\frac{1}{2}}+B E_{\frac{3}{2}}, W=A E'_{\frac{1}{2}}+B E'_{\frac{3}{2}}$.
$X_k(t)$ in the late time is given by
\begin{equation}
\begin{split}
&\lim_{t \rightarrow \infty}X_K(t)=\lim_{t \rightarrow \infty}f_k(t)a_k+\lim_{t \rightarrow \infty}f^*_k(t) a^{\dagger}_{-k} \\
&=\left(V~ 2^{-i \omega_0 \delta t}e^{i \omega_0 t}+W~ 2^{i \omega_0 \delta t}e^{-i \omega_0 t}\right) a_k+\left(V^*~ 2^{i \omega_0 \delta t}e^{-i \omega_0 t}+W^{*}~ 2^{-i \omega_0 \delta t}e^{i \omega_0 t}\right) a^{\dagger}_{-k} \\
&=\frac{1}{N_0}\left[b_k e^{-i \omega_0 t}+b_{-k}^{\dagger} e^{i \omega_0 t} \right],\\
\end{split}
\end{equation}
where 
\begin{equation}
b_k=N_0~2 ^{i\omega_0 \delta t}\left(W~a_k+V^*~ a^{\dagger}_{-k}\right).
\end{equation}
$\left[b_k, b^{\dagger}_k\right]$ is given by
\begin{equation}
\begin{split}
&\left[b_k, b^{\dagger}_k\right]=\left|N_0\right|^2\left(\left|W\right|^2-\left|V\right|^2\right) \\
&=\left|N_0\right|^2\frac{1}{2\omega_0}\cdot\left[-\left|\frac{E'_{\frac{3}{2}}E_{\frac{1}{2}}+E_{\frac{3}{2}}E'_{\frac{1}{2}}}{E'_{\frac{3}{2}}E_{\frac{1}{2}}-E_{\frac{3}{2}}E'_{\frac{1}{2}}}\right|^2+4\left|\frac{E'_{\frac{3}{2}}E'_{\frac{1}{2}}}{E'_{\frac{3}{2}}E_{\frac{1}{2}}-E_{\frac{3}{2}}E'_{\frac{1}{2}}}\right|^2\right],
\end{split}
\end{equation}
where 
\begin{equation}
\begin{split}
E_{\frac{3}{2}}E'_{\frac{1}{2}}&=\Gamma{\left(\frac{3}{2}\right)}\Gamma{\left(\frac{1}{2}\right)}\times \frac{1}{\Gamma{(1-a)}\Gamma(a)}\times \frac{1}{\Gamma{(b+\frac{1}{2})}\Gamma{(\frac{1}{2}-b)}}\times \left|\Gamma{(i \omega_0 \delta t)}\right|^2 \\
&=\Gamma{\left(\frac{3}{2}\right)}\Gamma{\left(\frac{1}{2}\right)}\times \frac{\sin{(\pi a)}}{\pi}\times \frac{\cos{(\pi b)}}{\pi}\times \left|\Gamma{(i \omega_0 \delta t)}\right|^2, \\
E'_{\frac{3}{2}}E_{\frac{1}{2}}&=\Gamma{\left(\frac{3}{2}\right)}\Gamma{\left(\frac{1}{2}\right)}\times \frac{1}{\Gamma{(1-b)}\Gamma(b)}\times \frac{1}{\Gamma{(a+\frac{1}{2})}\Gamma{(\frac{1}{2}-a)}}\times \left|\Gamma{(i \omega_0 \delta t)}\right|^2 \\
&=\Gamma{\left(\frac{3}{2}\right)}\Gamma{\left(\frac{1}{2}\right)}\times \frac{\sin{(\pi b)}}{\pi}\times \frac{\cos{(\pi a)}}{\pi}\times \left|\Gamma{(i \omega_0 \delta t)}\right|^2, \\
E'_{\frac{3}{2}}E_{\frac{1}{2}}-E_{\frac{3}{2}}E'_{\frac{1}{2}}&=\frac{\Gamma{\left(\frac{3}{2}\right)}\Gamma{\left(\frac{1}{2}\right)}}{\pi^2}\times \left(\sin{(\pi a)}\cos{(\pi b)}-\cos{(\pi a)}\sin{(\pi b)}\right)\times \left|\Gamma{(i \omega_0 \delta t)}\right|^2  \\
&=\frac{\Gamma{\left(\frac{3}{2}\right)}\Gamma{\left(\frac{1}{2}\right)}}{\pi^2}\times \sin{\left[\pi\left( a-b\right)\right]}\times \left|\Gamma{(i \omega_0 \delta t)}\right|^2, \\
E'_{\frac{3}{2}}E_{\frac{1}{2}}+E_{\frac{3}{2}}E'_{\frac{1}{2}}&=\frac{\Gamma{\left(\frac{3}{2}\right)}\Gamma{\left(\frac{1}{2}\right)}}{\pi^2}\times \left(\sin{(\pi a)}\cos{(\pi b)}+\cos{(\pi a)}\sin{(\pi b)}\right)\times \left|\Gamma{(i \omega_0 \delta t)}\right|^2  \\
&=\frac{\Gamma{\left(\frac{3}{2}\right)}\Gamma{\left(\frac{1}{2}\right)}}{\pi^2}\times \sin{\left[\pi\left( a+b\right)\right]}\times \left|\Gamma{(i \omega_0 \delta t)}\right|^2,  \\
V&=\frac{2^{i \omega_0 \delta t}}{\sqrt{2\omega_0}}\left(\frac{\sin{\left[\pi (a+b)\right]}}{\sin{[\pi(a-b)]}}\right), \\
E'_{\frac{3}{2}}E'_{\frac{1}{2}}&=\Gamma{\left(\frac{3}{2}\right)}\Gamma{\left(\frac{1}{2}\right)}\Gamma{(i \omega_0 \delta t)}^2 \times \frac{1}{\Gamma{(a)}\Gamma{\left(\frac{1}{2}-b\right)}\Gamma{\left(a+\frac{1}{2}\right)}\Gamma{(1-b)}}.
\end{split}
\end{equation}

If we assume that $\omega \ll 1$ (the fast limit), then $a^*=b$.
The absolute value of $E'_{\frac{3}{2}}E'_{\frac{1}{2}}$ is given by
\begin{equation}\label{cross}
\begin{split}
\left| E'_{\frac{3}{2}}E'_{\frac{1}{2}}\right|^2&=\left(\Gamma{\left(\frac{3}{2}\right)}\Gamma{\left(\frac{1}{2}\right)}\right)^2\times \left|\Gamma{(i \omega_0 \delta t)}\right|^4 \times \frac{1}{\Gamma{(a)}\Gamma{(1-a)}}\times \frac{1}{\Gamma{(a+\frac{1}{2})}\Gamma{(-a+\frac{1}{2})}} \\
&\times \frac{1}{\Gamma{(b)}\Gamma{(1-b)}}\times \frac{1}{\Gamma{(b+\frac{1}{2})}\Gamma{(-b+\frac{1}{2})}}. \\
\end{split}
\end{equation}
If we take the  limit $\omega \gg 1$, then $\alpha=\frac{1+i \sqrt{4\omega^2-1}}{4}$.
Conjugates of the parameters are given by
\begin{equation}
\begin{split}
a^*=\frac{1}{2}-a, ~ b^*= \frac{1}{2}- b. 
\end{split}
\end{equation}
Therefore,  the absolute value of $E'_{\frac{3}{2}}E'_{\frac{1}{2}}$ is the same as (\ref{cross}).
In both limits, $\left|W\right|^2$ is given by
\begin{equation}
\left|W\right|^2=4\frac{1}{2\omega_0}\left|\frac{E'_{\frac{3}{2}}E'_{\frac{1}{2}}}{E'_{\frac{3}{2}}E_{\frac{1}{2}}-E_{\frac{3}{2}}E'_{\frac{1}{2}}}\right|^2=\frac{4}{2\omega_0}\times\frac{\sin{\pi a}\cos{\pi a}\sin{\pi b}\cos{\pi b}}{\left|\sin{(\pi(a-b))}\right|^2},
\end{equation} 
and the conjugates of $\sin{\pi(a-b)}$ and $\sin{\pi(a+b)}$ are given by
\begin{equation}
\sin{\pi (a-b)}^*=\sin{\pi (b-a)}, \sin{\pi(a+b)}^*=\sin{\pi(a+b)}.
\end{equation}
Then, $\left|V\right|^2$ is given by
\begin{equation}
\begin{split}
\left|V\right|^2&=\frac{1}{2\omega_0}\frac{(\sin{\pi a}\cos{\pi b})^2+(\sin{\pi b}\cos{\pi a})^2+2(\sin{\pi a}\cos{\pi a}\sin{\pi b}\cos{\pi b})}{\left|\sin{\left[\pi(a-b)\right]}\right|^2}, \\
\end{split}
\end{equation}
and $\left|W\right|^2-\left|V\right|^2$ is given by
\begin{equation}
\begin{split}
\left|W\right|^2-\left|V\right|^2&=\frac{1}{2\omega_0}\frac{-(\sin{(\pi a)}\cos{(\pi b)}-\cos{(\pi a)}\sin{(\pi b)})^2}{\left|\sin{(\pi(a-b))}\right|^2} \\
&=\frac{1}{2\omega_0}\frac{\sin{(\pi(a-b))}^2}{\sin{(\pi(a-b))}^2} =\frac{1}{2\omega_0}.
\end{split}
\end{equation}
Therefore, we obtain $N_0=\sqrt{2\omega_0}$.

A number operator $N$ is given by
\begin{align}
N&=\int^{\pi}_{-\pi}dk \left\langle 0\right |_{in} b^{\dagger}_k b_k \left|0 \right \rangle_{in}=\int^{\pi}_{-\pi}dk N_0^2 \left\langle 0\right |_{in} \left(W^* a^{\dagger}_k+V a_{-k}\right) \left(W a_k+V^* a^{\dagger}_{-k}\right) \left|0 \right \rangle_{in}\notag\\
&=\int^{\pi}_{-\pi}dk 2 \omega_0 \times \left|V\right|^2 \times \left\langle 0\right |_{in} a_{-k} a^{\dagger}_{-k} \left|0 \right \rangle_{in}=\mathcal{V}\int^{\pi}_{-\pi}dk \left(\frac{\cos{\left(\frac{\pi \sqrt{1-4\omega^2}}{2}\right)}}{\sinh{(\pi \omega_0 \delta t)}}\right)^2,
\end{align}
where $\mathcal{V}=\left\langle 0\right |_{in} a_{-k} a^{\dagger}_{-k} \left|0 \right \rangle_{in}$.
\section{Late-time expansion \label{apc}}
Since $f_k(t)$ in the CCP-type potential at $t\gg \delta t$ is (\ref{lfk}), the spectra of $X, P$ and $D$ in $t\gg \delta t$ are given by
\begin{equation} 
\begin{split} \label{lsp}
&\left|f_k(t)\right|^2 \simeq \left|V\right|^2+\left|W\right|^2+2 \left(Re\Omega \cdot \cos{(2\omega_0 t)}+Im \Omega \sin{(2\omega_0 t)}\right), \\
&\left|\dot{f}_k(t)\right|^2 \simeq \omega_0^2\left[\left|V\right|^2+\left|W\right|^2-2 \left(Re\Omega \cdot \cos{(2\omega_0 t)}+Im \Omega \sin{(2\omega_0 t)}\right)\right], \\
&Ref^{*}_k(t)\dot{f}_k(t) \simeq \omega_0[-2Re \Omega \cdot  \sin{(2\omega_0 t)}+2Im \Omega \cdot  \cos{(2\omega_0 t)}],
\end{split}
\end{equation}
where $\Omega = WV^*2^{i \omega_0 \delta t}$. Time-dependent terms in (\ref{lsp}) with $k=0$ oscillate with a periodicity, $\pi \xi$, which is consistent with the periodicity of entanglement oscillation in very late time. 


\section{Details of numerical calculation} \label{sec:appendix_numerical_calculation}
\newcommand{\ourprec}{\varepsilon}
In this appendix, we explain details of our numerical calculation.
We compute the entanglement entropy based on the correlator method with numerical integration.
As already we have mentioned in the main text, the 
time-dependent correlators for the fundamental field and its conjugate momentum are expressed in the momentum space,
\begin{align} \label{nap}
X_{ab}(t)  &=\int^{\pi}_{-\pi} \frac{dk}{2\pi}\left|f_k(t)\right|^2\cos{\left(k\left|a-b\right|\right)}, \\
P_{ab}(t) &=\int^{\pi}_{-\pi} \frac{dk}{2\pi}\left|\dot{f}_k(t)\right|^2\cos{\left(k\left|a-b\right|\right)}, \\
D_{ab}(t) &=\int^{\pi}_{-\pi} \frac{dk}{2\pi}\text{Re}\left[\dot{f}^*_k(t)f_k(t)\right]\cos{\left(k\left|a-b\right|\right)}, 
\end{align}
where $a,b \in [1,l]$ are  a lattice coordinate, $l$ is the subsystem size and $t\in { \boldsymbol R}$ is the time. 
%
The function, $f_{k}(t)$, depends on the protocols. The function, $f_k(t)$, in the CCP-type mass is in (\ref{sol1}), and $f_k(t)$ in the ECP-type mass is in (\ref{sol2}).
The integrations in (\ref{nap}) are performed numerically.
As a consequence, $\left|f_k(t)\right|^2$, $\left|\dot{f}_k(t)\right|^2$ and 
$\text{Re}\left[\dot{f}^*_k(t)f_k(t)\right]$ are calculated by Mathematica numerically to keep enough significant digits.
The numerical integration is performed by replacing the integrals with summations as in the definition of the Riemannian integral.
We perform the following replacement:
\begin{align}
\int_{-\pi}^{\pi} dk g(k) \to \sum_{k_j=-\pi}^{\pi} \ourprec g(k_j),
\end{align}
where $g(k)$ is a function of $k$, and $\ourprec$ is a small positive real number
and $k_j = -\pi + j\ourprec$.
Note that, $\ourprec$ is independent of the lattice spacing, $\epsilon$. 
The integrants contain $\cos (k |a-b|)$, and the cosine causes cancellations between some modes.
If the argument takes value around $2\pi \ourprec l \sim 1$, in this case such expected cancellations are not perfect in the summation.
In this sense, $\ourprec $ must be taken to be smaller than $(2\pi l)^{-1}$,
and our case satisfies this condition.

In order to check how dependent our numerical computations are on $\ourprec$, we perform calculations for several $\ourprec$.
Our results do not depend on $\ourprec$ so much.

After the numerical integration, we calculate eigenvalues $\nu_j(t)$ of the following $2l\times 2l$ non-hermitian matrix, $\mathcal{M}$, numerically,
\begin{align}
\mathcal{M}
&= i J \Gamma, \\
\Gamma
&=\begin{bmatrix}
X_{ab}(t) & D_{ab}(t) \\
D_{ab}(t) & P_{ab}(t) \\
\end{bmatrix}, 
\end{align}
where $J$ is the symplectic matrix.
As a consequence of the Williamson's theorem,  
the eigenvalues $\nu_j(t)$ appear in pairs, $\nu_j = \pm \gamma_1(t), \pm \gamma_2(t), \cdots, $
and they are real.
Note that,  $\gamma_j(t) \geq \frac{1}{2}$ in the analytic computation, but in the numerical computation 
some of them are slightly smaller than $\frac{1}{2}$ due to  numerical errors. We just use only eigenvalues which satisfy $\gamma_j(t)> \frac{1}{2}$
so that we keep entanglement entropy finite. 
%
Finally, we compute the time evolution and size-dependence of entanglement entropy using $\gamma_j(t)$,
\begin{align}
S_A=\sum_j \left[ (\frac{1}{2}-\gamma_j)\log(\gamma_j-\frac{1}{2}) + (\frac{1}{2}+\gamma_j)\log(\gamma_j+\frac{1}{2}) \right].
\end{align}

\section{Error examination} \label{sec:error_examination}
Here, we analyze a systematic error which would comes from the numerical integration and the approximation where we replace  the exact spectra $f_k(t)$ for $|k|>|k_*|$  in (\ref{nap})  with the adiabatic solutions in (\ref{apsol}). We use this approximation when we compute $\Delta S_A$ in the slow CCP-type mass.
In the numerical computations, we have to choose $\ourprec$ and $|k_*|$
(see Appendix \ref{sec:appendix_numerical_calculation} for the notation $\ourprec$).
In order to avoid the contamination of numerical error,
we need to study how  our computation results depend on them.

Figure \ref{fig:prec1} and \ref{fig:prec2} show the $\ourprec$ and $|k_*|$-dependence of $\Delta S_A$ in the CCP-type quenches with different $(\xi, \delta t, l)$ 
in the range $-10<\frac{t}{\xi_{kz}}<20$ and $20<\frac{t}{\xi_{kz}}<100$.
%
Red points are for $\ourprec=10^{-5}$ with $|k_*|=0.05$, and black points are for $\ourprec=10^{-6}$ with $|k_*|=0.05$.
The red and black points appear to be on top of each other. Thus, we can conclude that our results in this range is not dependent on $\ourprec$ so much. 
%
Blue points are for  $\ourprec=10^{-5}$ with  $|k_*|=0.3$.
The red and blue points appear to be on top of each other. Thus, we can conclude that our approximation does not affect to the results in this $t$ range so much.

%

\begin{figure}[htbp]
 \begin{minipage}{0.22\hsize}
  \begin{center}
    \includegraphics[clip,width=3cm]{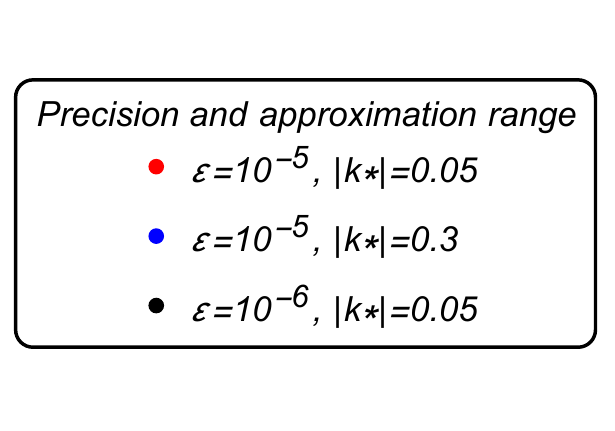}
\end{center}
 \end{minipage}
 \begin{minipage}{0.38\hsize}
  \begin{center}
   \includegraphics[width=60mm]{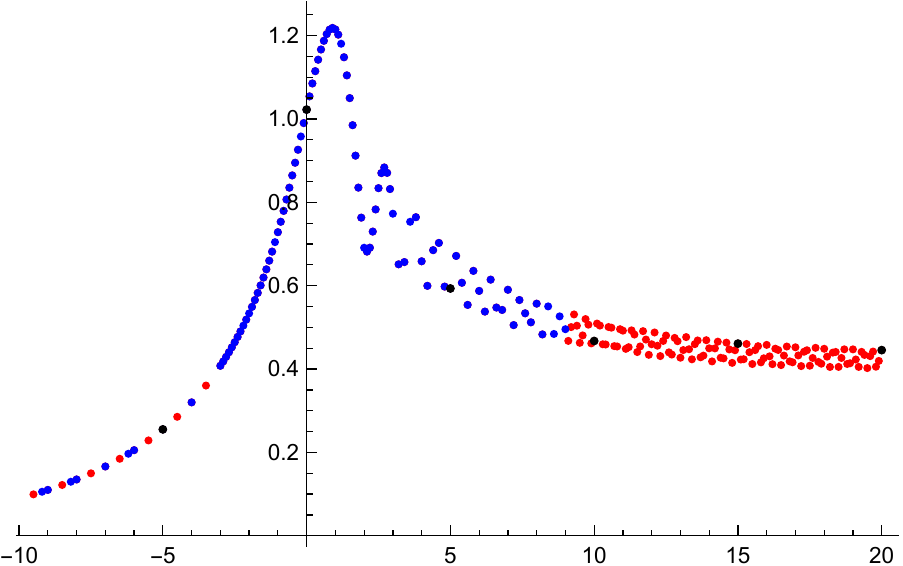}
       \put(-15,-10){$t/\xi_{kz}$}
   \put(-165,100){$\Delta S_A$}
  \end{center}
 \end{minipage}
 \begin{minipage}{0.38\hsize}
  \begin{center}
   \includegraphics[width=60mm]{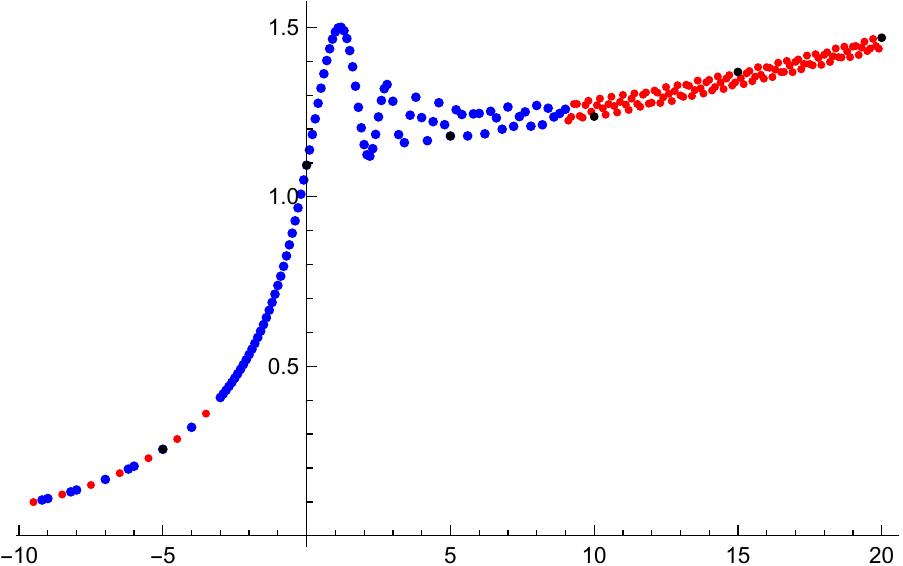}
         \put(-15,-10){$t/\xi_{kz}$}
   \put(-165,100){$\Delta S_A$}
  \end{center}
 \end{minipage}
\caption{
The left and right panels show $\Delta S_A$ with $(\xi, \delta t, l)=(10, 1,000, 100)$ and  $\Delta S_A$ with $(\xi, \delta t, l)=(10, 1,000, 2,000)$ in the CCP-type mass, respectively.
The time range is $-10 \le \frac{t}{\xi_{kz}}\le 20$.
}
\label{fig:prec1}
\end{figure}

\begin{figure}[htbp]
 \begin{minipage}{0.22\hsize}
  \begin{center}
    \includegraphics[clip,width=3cm]{fig/legend16.pdf}
\end{center}
 \end{minipage}
 \begin{minipage}{0.38\hsize}
  \begin{center}
   \includegraphics[width=60mm]{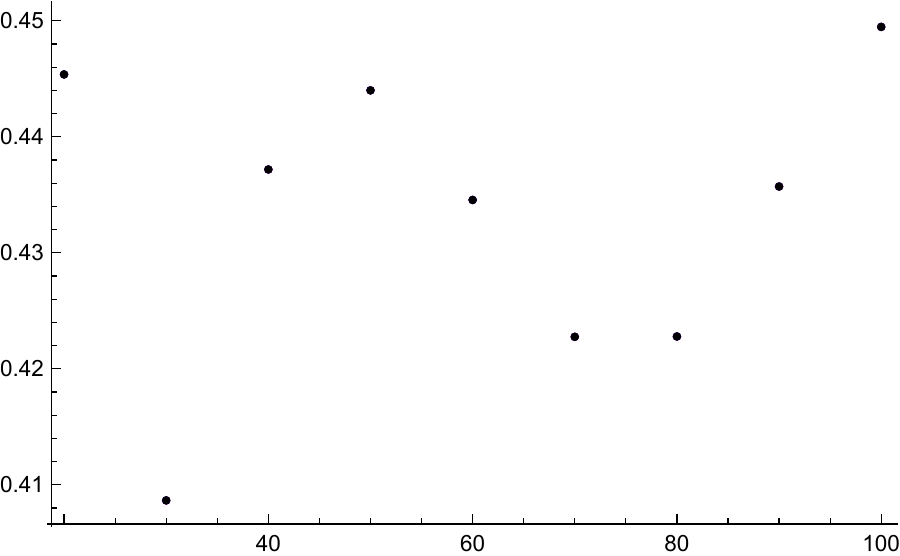}
      \put(-15,-10){$t/\xi_{kz}$}
   \put(-180,110){$\Delta S_A$}
  \end{center}
 \end{minipage}
 \begin{minipage}{0.38\hsize}
  \begin{center}
   \includegraphics[width=60mm]{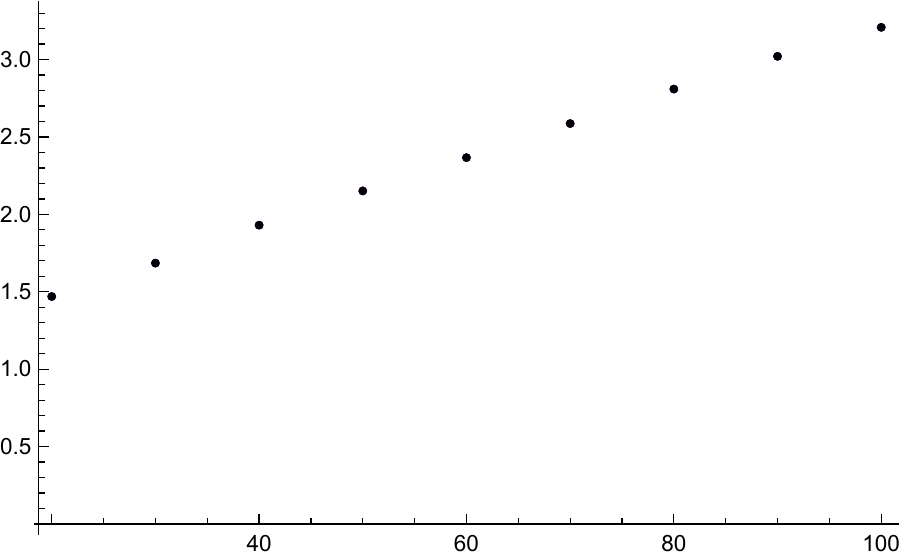}
       \put(-15,-10){$t/\xi_{kz}$}
   \put(-180,110){$\Delta S_A$}
  \end{center}
 \end{minipage}
\caption{
The left and right panels show $\Delta S_A$ with $(\xi, \delta t, l)=(10, 1,000, 100)$ and $\Delta S_A$ with $(\xi, \delta t, l)=(10, 1,000, 2,000)$ in the CCP-type mass, respectively.
 The time range is $20 \le \frac{t}{\xi_{kz}} \le 100$.
In both panels, $\Delta S_A$ with
$(\ourprec=10^{-5}, |k_*|=0.05)$, $(\ourprec=10^{-5}, |k_*|=0.3)$ and $(\ourprec=10^{-6}, |k_*|=0.05)$ are plotted and lie on the same curve.
}
\label{fig:prec2}
\end{figure}

\clearpage
\bibliographystyle{Common/utphys}
\bibliography{Common/Ref}

\end{document}